\documentclass[prd,nofootinbib,english,onecolumn]{revtex4}
\usepackage{amsmath}
\usepackage{amsfonts,color}
\usepackage{amssymb,float}
\usepackage[utf8]{inputenc}
\usepackage{color}
\usepackage[unicode=true,pdfusetitle,
 bookmarks=true,bookmarksnumbered=true,bookmarksopen=true,bookmarksopenlevel=3,
 breaklinks=false,pdfborder={0 0 1},backref=false,colorlinks=true]
 {hyperref}
\usepackage{enumitem}
\usepackage{graphicx}
\usepackage{appendix}
\usepackage{hyperref}
\usepackage{orcidlink}

\makeatletter

\begin{document}

\title{Is a Bose--Einstein Condensate a good candidate for Dark Matter? A test with Galaxy Rotation Curves}

\author{El\'ias Castellanos \orcidlink{0000-0002-7615-5004}}
\email{ecastellanos@mctp.mx}
\affiliation{Mesoamerican Centre for Theoretical Physics, Universidad Aut\'onoma de Chiapas, Carretera Zapata Km. 4, Real del Bosque (Ter\'an), 29040, Tuxtla Guti\'errez, Chiapas, M\'exico.}

\author{Celia Escamilla-Rivera \orcidlink{0000-0002-8929-250X}}
\email{celia.escamilla@nucleares.unam.mx}
\affiliation{Instituto de Ciencias Nucleares, Universidad Nacional Aut\'onoma de M\'exico, 
Circuito Exterior C.U., A.P. 70-543, M\'exico D.F. 04510, M\'exico.}

\author{Jorge Mastache %\orcidlink{0000-0002-2709-6319}
}
\email{jhmastache@mctp.mx}
\affiliation{Consejo Nacional de Ciencia y Tecnolog\'ia, Av. Insurgentes Sur 1582, Col Cr\'edito Constructor, Del. Benito Ju\'arez, CP 03940. Mexico,}
\affiliation{Mesoamerican Centre for Theoretical Physics, Universidad Aut\'onoma de Chiapas, Carretera Zapata Km. 4, Real del Bosque (Ter\'an), 29040, Tuxtla Guti\'errez, Chiapas, M\'exico.}

\begin{abstract}
We analyze the rotation curves that correspond to a Bose--Einstein Condensate (BEC) type halo surrounding a Schwarzschild--type black hole to confront predictions of the model upon observations of galaxy rotation curves. We model the halo as a Bose--Einstein condensate in terms of a massive scalar field that satisfies a Klein--Gordon equation with a self--interaction term. We also assume that the bosonic cloud is not self--gravitating. To model the halo, we apply a simple form of the Thomas--Fermi approximation that allows us to extract relevant results with a simple and concise procedure. Using galaxy data from a subsample of SPARC data base, we find the best fits of the BEC model by using the Thomas--Fermi approximation and perform a Bayesian statistics analysis to compare the obtained BEC's scenarios with the Navarro--Frenk--White (NFW) model as pivot model.  We find that in the centre of galaxies we must have a supermassive compact central object, i.e., supermassive black hole, in the range of $\log_{10} M/M_\odot = 11.08 \pm 0.43$ which condensate a boson cloud with average particle mass $M_\Phi = (3.47 \pm 1.43 )\times10^{-23}$ eV and a self--interaction coupling constant $\log_{10} (\lambda \; [{\rm pc}^{-1}]) = -91.09 \pm 0.74 $, i.e., the system behaves as a weakly interacting BEC. We compare the BEC model with NFW concluding that in general the BEC model using the Thomas--Fermi approximation is strong enough compared with the NFW fittings. Moreover, we show that BECs still well--fit the galaxy rotation curves and, more importantly, could lead to an understanding of the dark matter nature from first principles.
\end{abstract}

\maketitle

%%%%%%%%%%%%%%%%%%%%%%%%%%%%%%%%%%%%%%%%%%%%
%%%%%%%%%%%%%%%%%%%%%%%%%%%%%%%%%%%%%%%%%%%%

\section{Introduction}
\label{sec:intro}

The relation between scalar fields and (relativistic) Bose--Einstein condensates (BEC's) has already a long history, see for instance \cite{dol,wei,MG,ET, ET1,JB,LP} and references therein. The aforementioned relation remains a very important topic that must be still fully understood. 

On one hand, which is in the core of the current  astrophysics and cosmology, scalar fields seem to be a relevant candidate to describe dark matter  (DM) \cite{DM,DM1,DM2,DM3,DM4}. The current state-of-the-art in cosmology and astrophysics constrains the amount of dark matter in the universe around 26\% of the total energy density. Evidence of dark matter can be found in the following observations, e.g., from the kinematic of galaxies and clusters  \cite{Rubin:1980zd, Zwicky:1933gu}, 
the physics of the cosmic microwave background radiation (CMBR) and baryon acoustic oscillations (BAO), \cite{Aghanim:2018eyx, Eisenstein:2005su} as well as observations from Supernovae Type Ia (SNIa) and Gravitational Lensing (GL). \cite{Astier:2005qq, Kaiser:1992ps}.

In this line of thought, dark matter could consist of some type of generic scalar field(--particles) of spin--zero, for instance, Weakly Interacting Massive Particles (WIMPs), axions and others that depend on the model under consideration. Although these particles have been not yet observed, scalar fields interpreted as dark matter open up a very interesting model to confront observations. On the other hand, the theory of relativistic bosonic gases and its transitions, under certain circumstances as BEC's open the door to the interpretation of dark matter in the form of a condensate of generic bosonic particles \cite{BoehmerHarko2007,Ure,TR}. Furthermore, there are physical models related to the above ideas in the literature, such as the so--called hairy wigs models, which describe the dynamical behavior of scalar field configurations surrounding black holes \cite{B,QU}. In these works, among others, it was found that these systems can form a stable structure for enough time, make them plausible candidates to describe dark matter galactic halos.  Even more, since it is generally accepted that almost all the galaxies host a supermassive black hole at the center, and together with the assumption that dark matter is some kind of scalar field, this leads to the analysis of the existence of bound or quasi--bound scalar field configurations surrounding these compact objects. In fact, in the case of a Schwarzschild black hole and a massive scalar field without self--interactions, it was found that such quasi--bound states exist \cite{B1}.

Furthermore, when self--interactions are present for these scalar field configurations surrounding black holes (or hairy wigs models) and when quasi--bound states exist, the system can be also analyzed from the Bose--Einstein condensation point of view \cite{NOS,NOS1,NOS2}. It is quite interesting that in the {aforementioned} scenario a Gross--Pitaevskii like--equation can be deduced from the corresponding Klein--Gordon equation.
In other words, it is possible to study the system by using the formalism behind the Bose--Einstein condensation to extract relevant information. 
More precisely, in the formalism developed in refs.\,\cite{NOS,NOS1,NOS2} it was shown that for a spherically symmetric test--scalar field configurations surrounding  different types of black holes, the system can be interpreted as a trapped BEC. In other words, the corresponding Klein--Gordon equation can be rewritten in a form of a Gross--Pitaevskii equation, which describes the dynamics of usual laboratory BEC's, under certain circumstances. Additionally, in the three quoted references, it was analyzed the limit of validity of the so--called Thomas--Fermi approximation which is capable to give us an approximated solution (in principle, by using a simple algebraic procedure) for the system under consideration. Here it is important to mention that in the standard theory of condensates (for instance, ref.\,\cite{pethick2002}) the Thomas--Fermi approximation is an algebraic procedure that is useful for exploring relevant properties associated with the system when interactions are present. Moreover, the Thomas--Fermi approximation is valid when the kinetic energy is negligible with respect to the potential and interaction ones. Then, the kinetic energy can be neglected from the very beginning in the corresponding Gross--Pitaevskii equation. As a consequence, the non--linear differential equation becomes an algebraic equation for the order parameter, which in principle, is easy to solve. Additionally, the Thomas--Fermi approximation is valid when the corresponding scattering length, which describes the interaction among the particles within the system, is much smaller than the mean inter--particle spacing for sufficient large clouds. In other words, when the system is diluted enough and contains a large number of particles. Finally, we must add that the Thomas--Fermi approximation fails for trapped condensates near the edge of the cloud, due to the divergent behavior of the kinetic energy (i.e. the total kinetic energy per unit area diverges on the boundary of the system \cite{pethick2002}). The last assertion can be used to define the validity of the Thomas--Fermi approximation and sets the limit in which we can extract information from the system. Therefore, in the case of the scalar field configurations viewed as BEC's surrounding black holes, the divergent behavior of the kinetic energy on the predicted boundary of the system can be used to test the region in which the Thomas--Fermi approximation is valid. In other words, we are not able to extract information beyond the region in which the approximation is valid.
Deeper research is needed to support the description of these systems as dark matter. In addition, according to the results obtained in refs.\,\cite{NOS,NOS1,NOS2}, observable features related to the \emph{scalar cloud} within the Thomas--Fermi approximation, for discriminating between different types of black holes are tiny.

The main goal of the present manuscript is to identify the particle density distribution obtained from the Thomas--Fermi approximation (or the BEC density distribution) as a DM density profile surrounding galaxies (i.e., we restrict our analysis to the region in which the Thomas--Fermi approximation is valid). Moreover, we will be comparing the BEC density distribution with the so--called  Navarro--Frenk--White (NFW) density profile characterized by a cusp central density, to analyze the consequences upon the galaxy rotation curves of some set of galaxies. 

Galaxy rotation curves cannot be explained by luminous matter alone and we have to appeal to an extra matter component, i.e. DM or BEC--DM in our scenario, to explain the observations \cite{Salucci:2018hqu,Salucci:2018eie}. Several DM density profiles have been proposed, we can categorize the profiles by its central behavior. Profiles whose densities grow with a power law of $\rho \sim r^{-1}$ are known as cuspy profiles, i.e. Navarro--Frenk--White \cite{Navarro:1995iw, Moore:1999gc}. Profiles whose density tends to a constant value at the center of the galaxy are know as core profiles, i.e. isothermal, Burket profiles \cite{Gunn:1972sv, Burkert:1995yz}. Most cuspy profiles come from numerical simulations while the later are phenomenological proven, but must of them do not offer a clear explanation for the DM fundamental nature. Moreover, there is a tension between cores and cusp profiles because, on one hand, cuspy profiles are the one predicted from numerical simulations, on the other hand, observations seem to prefer cored profiles, the so called core--cusp problem \cite{deBlok:2009sp}.

We aim to constrain the parameters of the BEC--DM model by  using a total of 20 high resolution, circular galaxies. The mass density of the BEC is given in terms of the mass of the bosonic particles $M_{\Phi}$, a frequency $\omega$, the coupling constant $\lambda$ (which describes the interparticle interaction within the system) and, the only astronomical parameter, the mass of the black hole $M$. Except for the mass of the black hole, the profile is only given by the underlying particle model, the BEC indeed, which we expect to be equal for all galaxies, leaving the mass of the black hole as the only astronomical parameter that may vary from one galaxy to other.

To set a statistical study behind the models obtained, we will use Bayesian inference to update knowledge about unknown parameters, e.g., the mass boson, with information from SPARC data and performing the comparison with a pivot model (NFW model). 

The outline of the paper is as follows: in Sec.\,\ref{sec:gp} we summarize the results found in previous works \cite{NOS,NOS1,NOS2} in which the density distribution is obtained through the Thomas--Fermi approximation upon a bosonic cloud in the form of a BEC surrounding a black hole. Such as density is interpreted in the present work as the galactic dark matter halo. In Sec.\,\ref{sec:rotation_curve} we describe the method to analyze the corresponding rotation curves, starting by imposing physical bounds given by the characteristics related to the bosonic particle. In Sec. \ref{sec:sample} we describe the mass model and SPARC sampler. In Sec.\,\ref{sec:data}, we present the results for the rotation curves and perform the statistics for each scenario. Finally, in Sec.\,\ref{con} we report the main results and a discussion of the predictions obtained in the present work.

%%%%%%%%%%%%%%%%%%%%%%%%%%%%%%%%%%%%%%%%%%%%
%%%%%%%%%%%%%%%%%%%%%%%%%%%%%%%%%%%%%%%%%%%%
\section{Background theory}
\label{sec:gp}

In this section, we summarize some important results obtained in refs. \,\cite{NOS,NOS1,NOS2}. In the aforementioned references a Gross--Pitaevskii like--equation is deduced from the corresponding Klein--Gordon equation in spherically symmetric and static black hole spacetimes. Also, a self--interacting scalar potential is assumed, allowing to link the system with the BEC's point of view. It is important to mention that both equations contain the same information related to the system. After the deduction of the Gross--Pitaevskii like--equation we apply the Thomas--Fermi approximation, which allows us to deduce with a very simple procedure, the corresponding density profile that we assume as a galactic dark matter halo. 

In order to deduce the density profile we consider a test scalar field(--particle) in a spherically symmetric and static spacetime where the metric which in standard spherical coordinates is given by
\begin{equation}\label{eq:g}
ds^{2}
= -f(r) c^2 dt^{2}+\frac{1}{f(r)}dr^{2}+r^{2}(d \theta^{2}+\sin^{2}\theta
d\phi^{2}) \, .
\end{equation}

Thus, the Klein--Gordon equation for a complex test--scalar field $\Phi$ with a
scalar potential $V(\Phi)$ in a spacetime with metric $g_{\mu \nu}$ can be written as follows:
\begin{equation}
\label{eq:KG0}
\frac{1}{\sqrt{-g}} \partial_{\mu} 
\Bigl(\sqrt{-g} g^{\mu \nu} \partial_{\nu}\Phi \Bigr)
-\frac{d\,V(\Phi \Phi ^*)}{d\,\Phi^{*}}=0,
\end{equation}
where the star is complex conjugation and $g$ is the determinant of the metric.
If we want to link the system with a weakly interacting Bose--Einstein condensate of some generic bosonic particles, we assume a scalar potential of the form
\begin{equation}
\label{eq:MEX}
V(\Phi \Phi^*)=\mu^{2}\Phi\Phi^{*}+\frac{\lambda}{2}(\Phi\Phi^{*})^{2},
\end{equation}
where $\mu$ is the scalar mass parameter which is related to the mass of the bosonic particles $M_{\Phi}$ through the inverse of the Compton wavelength of the particles $\mu = \dfrac{M_{\Phi} c}{\hbar}$. Additionally, $\lambda$ is the self--interaction coupling constant which is interpreted as the scattering length of the particles.

We can find solutions of the Klein--Gordon equation by using the following ansatz 
\begin{equation}\label{eq:omega}
\Phi (t,r) =
e^{i\,\omega\,t}\,\frac{u(r)}{r},
\end{equation}
where we assume that the frequency $\omega$ is \emph{real}. Notice that in general the function $u(r)$ is a complex--valued classical function that can be interpreted as the macroscopical wave function of the system or the order parameter as in standard theory of condensates.
By using the ansatz (\ref{eq:omega}) the corresponding Klein--Gordon equation reduces to a Gross--Pitaevskii--like 
equation of the form
\begin{equation}
\Bigg(-  \dfrac{d^2}{dr_*^2}  + V_{\rm eff} (r) +
\lambda_{\rm eff} (r) \,\dfrac{|u(r)|^2}{r^2} \Bigg)\,u (r)
= \dfrac{\omega^{2}}{c^2} \,u (r) \, ,
\label{eq:KGGP}
\end{equation}
where we have also introducing the $r_*$ coordinate defined as $d\,r_*=\frac{d\,r}{f(r)}$, i.e., the so--called tortoise coordinate. 

Notice that in equation (\ref{eq:KGGP}) we define the effective trapping potential as
\begin{equation}
V_{\rm eff}(r)= f(r)\,\,\left(\mu^2 + \frac{f'(r)}{r} \right)
\label{eq:Veff}
\end{equation}
where the prime indicates derivatives respect to the $r$ coordinate, together with an effective self--interaction parameter with the following functional form

\begin{equation}
\lambda_{\rm eff} (r) =\lambda\,f(r) \, .
\label{eq:lameff}
\end{equation}
Here is important to mention that the effective potential in equation (\ref{eq:Veff}) is caused by the curvature of the spacetime itself, together with the contribution of the mass parameter $\mu$. In other words, the effective potential $V_{eff}$ allows the bosonic cloud to admit the existence of quasi--bound states. Additionally, $\lambda$ is \emph{modulated} by the influence of the spacetime geometry, i.e., the interactions show a position-dependent behavior as was shown in refs.\,\cite{NOS,NOS1,NOS2}. Finally, we must mention that the term $\omega^{2}/c^2$ can be also identified with effective chemical potential. 

As was mentioned in the introduction, the Thomas--Fermi approximation assumes that the kinetic energy 
is negligibly small in comparison to the potential energy and the self-interaction energy. Then, we can neglect the kinetic energy in equation (\ref{eq:KGGP}) from the very beginning, with the result of an algebraic equation, from which we can obtain the density of particles $\rho(r) \equiv \dfrac{|u(r)|^2}{r^2}$.

 The solution for the Gross--Pitaevskii--like 
equation (\ref{eq:KGGP}) within the Thomas--Fermi approximation is then given by 
\begin{equation}
\label{TF}
\dfrac{|u(r)|^2}{r^2} \equiv \rho(r)= \left( \frac{\omega^{2}}{c^2}-V_{\rm eff} (r) \right) \dfrac{1}{\lambda_{\rm eff} (r)} \, .
\end{equation}
Notice that the above equation is well defined as long as the right-hand side
is positive. The value of $\rho(r)$ is zero outside the region delimited by the equation $V_{\mathrm{eff}} (r) = \omega ^2/c^2$. Moreover, the equation $V_{\mathrm{eff}} (r) = \omega ^2/c^2$ sets the size of the cloud within de Thomas--Fermi approximation. The region in which  the condensate lies is a spherical shell of inner radius $r_{min}$ and outer radius $r_{max}$, where $r_{min}$ and $r_{max}$ are  precisely the solutions of the equation $V_{\mathrm{eff}} (r) = \omega ^2/c^2$. Additionally, as was point it out in \cite{NOS2}, the Thomas--Fermi approximation becomes arbitrarily good if $\lambda \, N$ (with $N$ the total number of particles within the condensate) becomes sufficiently big. In the present case, $\lambda \, N$ increases if the $\omega /c$ is chosen very close to the mass parameter 
$\mu$. Then we expect that the Thomas--Fermi approximation becomes quite good if the corresponding parameters are chosen properly.

For simplicity, let us assume that the function $f(r)$ in the metric (\ref{eq:g}) in ordinary units is given by:

\begin{equation}
f(r)=\Bigg(1- \frac{\alpha}{r}\Bigg),
\end{equation}
where $\alpha \equiv 2 G M / c^2$ for the Schwarzschild metric, where $G$ is the gravitational constant and $M$ the mass of the black hole. 

Let us remark that the model under consideration comes from \textit{first principles} in the sense that we only assume a scalar distribution in a form of  some kind of Bose--Einstein condensate composed of generic bosonic particles surrounding a black hole. Then, we can obtain the density of particles or the density profile that we interpreted as a galactic dark matter halo by using the so--called Thomas--Fermi approximation.

%%%%%%%%%%%%%%%%%%%%%%%%%%%%%%%%%%%%%%%%%%%%
%%%%%%%%%%%%%%%%%%%%%%%%%%%%%%%%%%%%%%%%%%%%

\section{Rotation Curves}
\label{sec:rotation_curve}
In this section we analyze if it is possible to obtain a bosonic halo as replacement for the dark matter. We first do an analysis of the equations exposed in Sec.\,\ref{sec:gp} in order to found bounds to the parameters. We then explain the mass model, to finally explore the statistics of the parameters.

\begin{figure}
  \centering
  \includegraphics[width=0.47\textwidth]{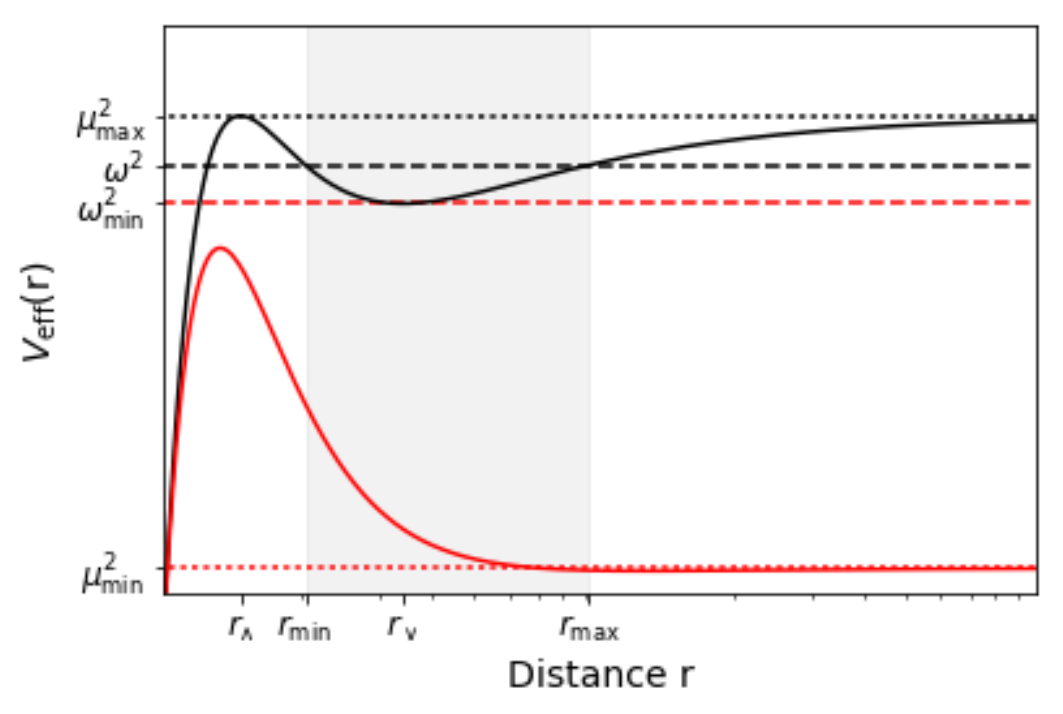}
   \includegraphics[width=0.47\textwidth]{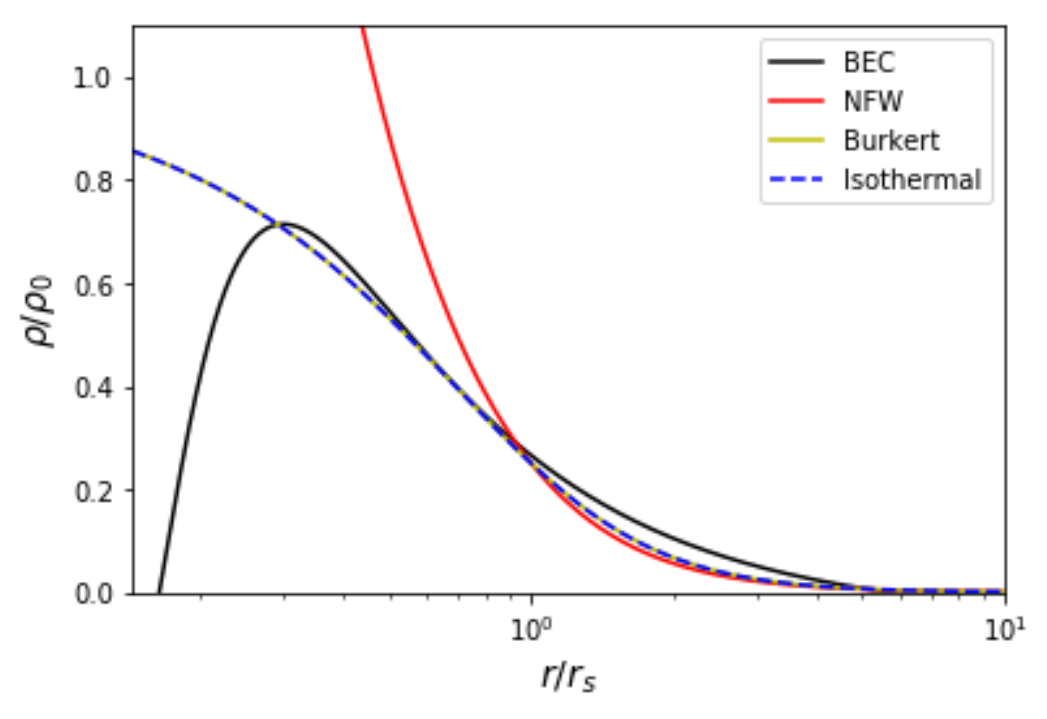}
  \caption{\footnotesize{\textit{Left:} Plot of the potential $V_{\rm eff}$, Eq.\,(\ref{eq:Veff}). Solid lines are two examples for different values of mass parameter $\mu$, the black lines is given $\mu_{\rm max} = V_{\rm eff}(r_{\wedge})$) and the red line is given $\mu_{\rm min}$. The $\omega$ parameter defines the the region where the Thomas-Fermi approximation is valid, $r_{\rm max} > r > r_{\rm min}$ and could take the value $\omega_{\rm max} (= \mu_{\rm max}) > \omega > \omega_{\rm min} (= V_{\rm eff}(r_{\vee}))$.  \textit{Right:} Plot of the profile density of BEC (solid black), NFW (solid red), Burkert (solid yellow) and isothermal (dashed blue).    }}
  \label{fig:PotentialBounds}
\end{figure}

%%%%%%%%%%%%%%%%%%%%%%%%%%%%%%%%%%%%%%%%%%%%
%%%%%%%%%%%%%%%%%%%%%%%%%%%%%%%%%%%%%%%%%%%%

\subsection{Physical bounds}
The halo model that we are proposing consist in  four free parameters, the mass of the black hole, $M$, the mass parameter of the scalar field $\mu$,  the frequency of the field, $\omega$ and the self--interaction parameter, $\lambda$, that we assume as the corresponding scattering length. Nevertheless, the mass of the black hole is the only galactic parameter that can change from one galaxy to another. The other three parameters must be the same since we are assuming the same particle forming galactic halos.

The density highly depends on the effective potencial $V_{\rm eff}(r)$, equation\,\eqref{eq:Veff}.  And this function will actually define the value of the parameters that are valid for our model.  First, assume a $\mu = 0$ and notice that $f^\prime = (1 - f)/r$, therefore, $V_{\rm eff}(r) = (1 - \frac{\alpha}{r})\frac{\alpha}{r^3}$. For small $r$ the potential will increase as $1 - \frac{\alpha}{r}$, will reach a maximum and decrease proportional to $r^{-3}$. The region that meet $r \propto \alpha$ are going to be the interesting cases, because its going to be of the order where the potential reaches its upper bound value and is also the minimum radius where the approximation is valid, for instance, for $r = 1$ pc, the mass of the black hole should be of the order $ \mathcal{O} (10^{13}) \, \textup{M}_\odot $.

We are particularly interested in computing the maximum of the potential because it will define the maximum value that the mass parameter, $\mu$, can take and the validity region where the density can be computed. To compute the maximum we solve for $\frac{{\rm d}V_{\rm eff}}{{\rm d}r} = 0$, this give us
\begin{equation}
\label{eq:V_deriv}
  (f-1)\left(4 f - \mu^2  r^2-1\right) = 0
\end{equation} 
Where we have used $f^{\prime \prime} = 2 (f-1)/r^2$. The trivial solution is when $f = 1$, this can only be met if $r \rightarrow \infty$, this is, when $V_{\rm eff} \rightarrow \mu^2$. Computing the other roots for equation\,\eqref{eq:V_deriv}, will define the radius where the potential is minimum, $r_\vee$, and maximum, $r_\wedge$.

A special case is to compute the maximum value of the mass parameter $\mu_{\rm max}$, which is defined when $V_{\rm eff}(r_\wedge) = \mu^2_{\rm max}$ at the radius $r_\wedge$, from this condition we obtain that $\mu_{\rm max}^2 = f(r_\wedge)/r_\wedge^2$. Using equation\,\eqref{eq:V_deriv} is easy to compute that $ f(r_\wedge) = 1/3$. Therefor $r_\wedge = 3 G M / c^2$. For instance, for a black hole of $M_{\rm bh} = 10^{13} \textup{M}_\odot$, the radii where the potential is maximum is $r_\wedge \sim 1.4$ pc. Is important to notice that the maximum value of $\mu_{\rm max}$ is given only in terms of the mass of the black hole.

We take the ansatz $V_{\rm eff}(r_\vee)  = 0.95 \, \mu^2_{\rm min}$ in order to compute $\mu_{\rm min}$, which range (0.7 -- 0.85) $\mu_{\rm max}$. Taking $V_{\rm eff}(r_\vee)  = \mu^2_{\rm min}$ makes no physical sense because there will be no potential well where the condensate could form.

The next physical parameter to analyze is the frequency of the field $\omega$. This parameter is particularly important because the region where the Tomas--Fermi approximation is valid is given by the condition $V_{\rm eff}(r) = \omega^2/c^{2}$. The region $r_{\rm max} > r > r_{\rm min} $ where the condensate could form is given by the roots of the polynomial $(\mu^2 - \omega^2)r^4 - \alpha \mu^2 r^3 - \alpha r + \alpha^2 = 0$. The extreme case where $\omega \rightarrow \mu$ gives a region where $r_{\rm max} \rightarrow \infty$. The closer the value of $\omega$ is to $\mu$ the larger the potential well of the condensate. For instance, assuming a black hole of $M = 10^{13} \textup{M}_\odot$ and $\mu = \mu_{\rm max}$, if $\omega$ is just 10\% less than $\mu$ then $r_{\rm max} \sim 4$ pc, which will make the model not testable for currents observations. However, if the percentile difference is of the order $\mathcal{O} (10^{-4})$ then $r_{\rm max} \sim 50$ kpc, enough extension for a Milky Way galaxy size. If we assume that all galaxies are surrounded for a dark matter halo then the frequency parameter is the most constrained of all four parameters.

The maximum radius at which the approximation is valid, at least the maximum observed radius, also depends on the value of $\omega$, the smaller the value of $\omega$ the smaller is the region. We find out that if $1 - \mu/\omega < 10^{-5}$ the maximum radius is of the order of $r_{\rm max} \sim \mathcal{O}(10)$ kpc. Therefore we fix the value of $\omega$ such that $1 - \mu/\omega = 10^{-7}$ in order to make sure that the maximum observed galaxy rotation curve is well inside the region where the Thomas--Fermi  approximation is valid.

The scattering length (or the self--interacting parameter $\lambda$) acts as a weight parameter, this should be or the order of $\mathcal{O} (10^{-70})$ 1/m to fit current observational velocities. The smaller the value of $\lambda$ the higher velocities we can compute, this will also mean that the scalar field is behaving almost as an ideal condensate. Heuristically speaking, we found that the bounds should be the order of $10^{-50} > \lambda [ {\rm m^{-1}} ] > 10^{-90}$ .

%%%%%%%%%%%%%%%%%%%%%%%%%%%%%%%%%%%%%%%%%%%%
%%%%%%%%%%%%%%%%%%%%%%%%%%%%%%%%%%%%%%%%%%%%

\subsection{Bose--Einstein Condensate: Thomas--Fermi  profile}
\label{modelo}
We assume that galaxies with a central black hole could possibly create enough potential well in order to create a condensate cloud of  almost non--interactive particles that could affect the dynamics of the galaxy. In the present work, we estimate the parameters of the scalar field(--particle) from rotation curves. It is worth pointing out that this in principle is an \emph{universal profile} extracted from a particle model in which only the astronomical parameter is the mass of the black hole, other parameters are inherent of the particle which means should be the same for all galaxies.

The density profile is simple enough in order to compute the mass as function of the radius, the integral over the volume gives,

\begin{eqnarray}\label{eq:mass_exact}
  M(r) &=& \frac{ \mu}{\lambda } \left[ \alpha ^2 \omega ^2 r + \frac{1}{2} \alpha \omega ^2 r^2 - \frac{1}{3}  \left( \mu ^2-\omega ^2 \right) r^3 \right. \nonumber \\ && \left.
   - \alpha  \log(r) 
  + \alpha ^3 \omega ^2 \log (r-\alpha )   \right]  
\end{eqnarray}

This expression should be evaluated between $r_{\rm min}$ and $r_{\rm max}$. Dimensional analysis of the expression for the mass let us conclude that the $r^3$ term is negligible because $\mu \sim \omega$, therefore the two most relevant terms are the ones proportional to $\log{r}$ and $r^2$, that dominates at small and large radii, respectively. And because $\alpha \gg \omega$ we can approximate the mass as,
\begin{equation}\label{eq:mass_aprox}
  M(r) \approx \frac{ \mu \alpha}{\lambda } \left[ \frac{1}{2} \omega ^2 r^2 -   \log(r)  \right]  ,
\end{equation}
from the last expression we clear see that $r_{\rm min}$ can be approximately computed when $M(r)=0$, which gives $r^2_{\rm min} \approx -\mathcal{W}(-\omega^2)/\omega^2$, where $\mathcal{W}$ is the Lambert function. The value of $r_{\rm max}$ can also be approximated when the term proportional to $r^3$ gets bigger to the $r^2$ term in equation\,(\ref{eq:mass_exact}), which would lead to a negative values for the mass. Thus, $r_{\rm max} \sim 3 \alpha / 2(1--\mu^2/\omega^2)$, given the bound we fix for $\omega$, this would lead to $r_{\rm max} \sim \mathcal{O} (\alpha 10^{14})$. Then, the region where the Thomas--Fermi approximation is valid is given by 
\begin{equation}
  \frac{3}{2}\frac{\alpha}{ 1 -\mu^2/\omega^2} \gtrsim r \gtrsim \frac{[-\mathcal{W}(-\omega^2)]^{1/2}}{\omega}
\end{equation}
Notice that if we do not have observation for $r \sim \mathcal{O}(r_{\rm min})$ then the parameters $\mu$, $\omega$ and $\lambda$ will be degenerate for the fitting, this is, we can find any combination of $\mu \omega^2/\lambda = {\rm cte}$ to valid for one galaxy. We can break the degeneracy with $\alpha$  because $r_{\max}$ highly depends on its value.

We assume circular rotation velocities of test particles in the plane of the galaxy and also spherical DM halos and BEC distributions. For this distribution of matter the circular velocity at radius $r$ is given by $V_H^2(r) = G M(r)/r$. From this last expression and equation\,\eqref{eq:mass_bec} we see that in the case of the BEC at large radii the velocity grows as $v \propto r^{1/2}$

\begin{figure}
  \centering
     \includegraphics[width=0.45\textwidth]{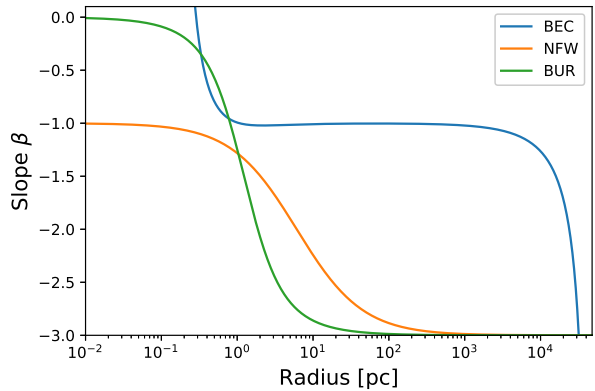}
  \caption{\footnotesize The slope for different profiles. For BEC (blue line) notice that the density is bounded between some $r_{\rm min}$ and $r_{\rm max}$, in this region the slope is almost constant with value $\beta =  -1$. For NFW (orange line) we notice its cuspy behavior close to the galactic centre and then dilutes as $\rho \propto r^{-3}$, while core profiles as Burkert (green line) reach constant value in the centre $\beta = 0$ and $\rho\propto r^{-3}$ at large radii.}
  \label{fig:slope}
\end{figure}

A parameter that could help us understand the behavior of the BEC is the slope of the density profile, which  is given by
\begin{equation}\label{eq:mass_bec}
  \beta \equiv - \frac{d\, \log \rho}{d\, \log r} = \frac{ \alpha  r \omega ^2 - 3 \alpha  f^2/r}
            {f \left(\left(\mu ^2 r^2+\frac{\alpha }{r}\right) f -  r^2 \omega ^2  \right)}
\end{equation}
and takes the values $\beta=-1$ for most of the bounded region $r_{\rm max} \gtrsim r \gtrsim r_{\rm min}$,  see Fig.\,\ref{fig:slope}.

Fittings to observational data have shown that the NFW halo profile is not a good description for rotation curves, and it is generally preferred a core dominated halo model \cite{vandenBosch:2000rza, deBlok:2009sp}. In the present work, our objective is to test the realization of the BEC model, equation\,\eqref{eq:mass_exact}, through rotation curves of a sample of high-resolution galaxies. Form Fig.\ref{fig:PotentialBounds} notice that we may have no presence of BEC at the center of the galaxy (i.e. for $r<r_{\rm min}$), which make sense if we have a central super-massive object as a black hole. Thus, this profile is either core or cuspy, we may call it null central profile. The extension of the condensate reaches up to $r_{\rm max}$, which is a finite value close to the observational limit of the galaxy, in contrast with the usual notion in the DM approach, where halos may extend up to $r_{200}$, where the density of the halo is $\rho_{dm} = 200 \rho_c$, where $\rho_c$ is the critical density of the Universe. This particular feature makes some question-related stability and galaxy formation jump right away, and it should be something that needs to be explored in the future.

%%%%%%%%%%%%%%%%%%%%%%%%%%%%%%%%%%%%%%%%%%%%
%%%%%%%%%%%%%%%%%%%%%%%%%%%%%%%%%%%%%%%%%%%%

\section{Mass Model and Data Sample}
\label{sec:sample}

The SPARC catalog has an observing data sample of 175 galaxies from $HI/H\alpha$ studies with a large range of luminosities and Hubble types \cite{Lelli:2016zqa}. The curves for some galaxies are limited to a few data points, therefore significantly increasing the uncertainties on the parameters, making difficult to draw conclusions from this approach. Therefore, we limit our sample to 20 galaxies, not because of the type of the galaxy but because of its observational characteristics and leave the analysis of the complete sample after this work. Our sample consists on galaxies with a large number of observational data points (above 30 data points), preferably with observations close to the galactic center, their rotation curve is smooth, with no  relevant wiggles and extended to large radii, have none or small bulge. These characteristics provide a good estimate of the parameters.

We focus or analysis to constraint the parameters of the BEC model. The total rotational velocity is computed taken the values for the gas and the stars from the mass model of the SPARC catalog, The SPARC database already offers robust mass models for the complete sample of galaxies using Spitzer $3.6 \mu m$ photometry. In most cases, the stellar disk can be well described by a single exponential disk, and their photometry was fitted by an exponential disk model. The bulge and the stellar disk model is given by the catalog.

The mass model include the four main components of a galaxy: the budge (when is present), $V_b$, the gas disk, $V_g$, the stellar disk, $V_{\star}$, and the BEC halo, $V_{\rm bec}$. The total gravitational potential of the galaxy is the sum of each component of the galaxy, thus the observed rotation velocity is,
\begin{equation}\label{eq:Vel_tot}
	V_{\rm tot}^2 = V_g^2 +  \Upsilon_b V_b^2 + \Upsilon_{\star} V_{\star}^2 + V_{\rm bec}^2 \, ,
\end{equation}
where $\Upsilon_{\star}$ and $\Upsilon_b$ are the mass-to-light ratio of the star and bulge disk, respectively. SPARC database gives the $\Upsilon_{\star} = 1 M_{\odot}/L_{\odot}$ at $3.6 \mu m$ with significant uncertainty. It is not surprising that the stellar and bulge model in SPARC underestimates (or overestimates) the luminosity, therefore, we could have stellar contributions up above the total rotation curve. We toke the mass-to-light ratio of the star, $\Upsilon_{\star}$, as a free parameter. Some assumptions have to be made respect to $\Upsilon_b$ in order to reduce the number of free parameters in the model, therefore we assume a heuristic relation between the bulge and the stellar disk so they hold the relation $\Upsilon_b = 1.4 \Upsilon_\star$ \cite{Lelli:2016zqa}. 

When fitting the BEC parameters, the $\Upsilon_\star$ remains as the major source of uncertainty. The unknown value of $\Upsilon$ provides different stellar and bulge mass contributions. Fixing the value of the mass--to--light ratio is out of the reach of this work because it is well know that is model dependent which precise value relies on extinction, star formation history, IMF, among others, but we expected to be studied in future work. Therefore, we ignore a priori any knowledge of the IMF and treat $\Upsilon_\star$ as an extra free parameter.

\subsection{DM profiles \label{MaMoDM}}
We compare the BEC distribution with the one obtained from $\Lambda$CDM simulations DM density profile characterized by a cusp central density, the NFW profile. On the other hand, observational determinations of the inner mass density distribution seem to indicate that mass density profiles of DM halos can be better described using an approximately constant-density inner core ($\rho \sim r^\alpha$ with $\alpha \ll 1$). This core has a typical size of order of a {\rm kpc} \cite{Moore:1994yx,deBlok:1996jib}, and examples of these profiles are the isothermal ($\rho_{\rm iso}$) \cite{vanAlbada:1984js} and Burkert ($\rho_{\rm bur}$) profiles \cite{Burkert:1995yz}

\begin{eqnarray}
  \rho_{\mathrm{bur}}(r)&=&\frac{\rho_{s}}{\left(1+r / r_{s}\right)\left(1+\left(r / r_{s}\right)^{2}\right)} \\
  \rho_{\mathrm{iso}}(r)&=&\frac{\rho_{s}}{1+\left(r / r_{s}\right)^{2}}
\end{eqnarray}
In contrast, a cuspy profiles such as the NFW profile takes the form
 \begin{equation}
   \rho_{{\rm nfw}} = \frac{\rho_0}{\frac{r}{r_s}\left(1+\frac{r}{r_s}\right)^2},
 \end{equation} \label{eq:rhoNFW}
where $r_s$ is the characteristic radius of the halo, and $\rho_0$ is related to the density of the Universe at the time of collapse of the DM halo. This mass distribution gives rise to a halo rotation curve
 \begin{equation}
   v_{\rm nfw} = V_{200}\sqrt{ \frac{\ln(1+ cx) - cx/(1+cx)}{x[\ln(1+c) - c/(1+c)]} }
 \end{equation}
This density profile has an inner and outer slope of -1 and -3, respectively. The inner slope implies a density cusp. The halo density can be specified in terms of a concentration parameter $c = r_{200}/r_s$ that indicates the amount of collapse that the DM halo has undergone, where the radius $r_{200}$ is defined as the radii at which the density contrast of the galaxy is 200 times greater than the critical density $\rho_{\rm cr}$, defined as $\rho_{\rm  cr} = 3 H^2/(8 \pi G)$, {\it H} being the Hubble parameter.
 
We use the observed rotation curve, stellar, and gas component as an input for the numerical code, to obtain the properties of the BEC. When fitting equation\,(\ref{eq:Vel_tot}) to the observed rotation curves, we apply a non-linear least-squares method to perform the fit, minimizing the residual sum of the $\chi^2$--test. The $\chi^2$--goodness--of--fit test, that tells us how close are the theoretical values to the observed ones. In general, the $\chi^2$-test statistics is of the form:
\begin{equation}
  \chi^2 = \sum_{i=1}^n \left(\frac{V_{{\rm obs}_i}-V_{{\rm model}_i}(r,\rho_0, r_s, r_c)}{\sigma_i}\right)^2,
\end{equation}

where $\sigma$ is the standard deviation, and $n$ is the number of observations.Comparison of the fits derived can tell us which of the DM models is preferred. More important are the differences between the reduced $\chi^2_{\rm red}=\chi^2/(n-p-1)$ values, where $n$ is the number of observations and $p$ is the number of fitted parameters. The uncertainties in the rotation velocity are reflected in the uncertainties in the model parameters.

%%%%%%%%%%%%%%%%%%%%%%%%%%%%%%%%%%%%%%%%%%%%
%%%%%%%%%%%%%%%%%%%%%%%%%%%%%%%%%%%%%%%%%%%%

\section{Statistical analysis}
\label{sec:data}

Using the SPARC sample we can now compute the profiles for our two models: NFW and BEC and perform the statistical comparison between rotation curves.

\begin{figure*}
\begin{tabular}{cc}
  \includegraphics[width=0.45\textwidth]{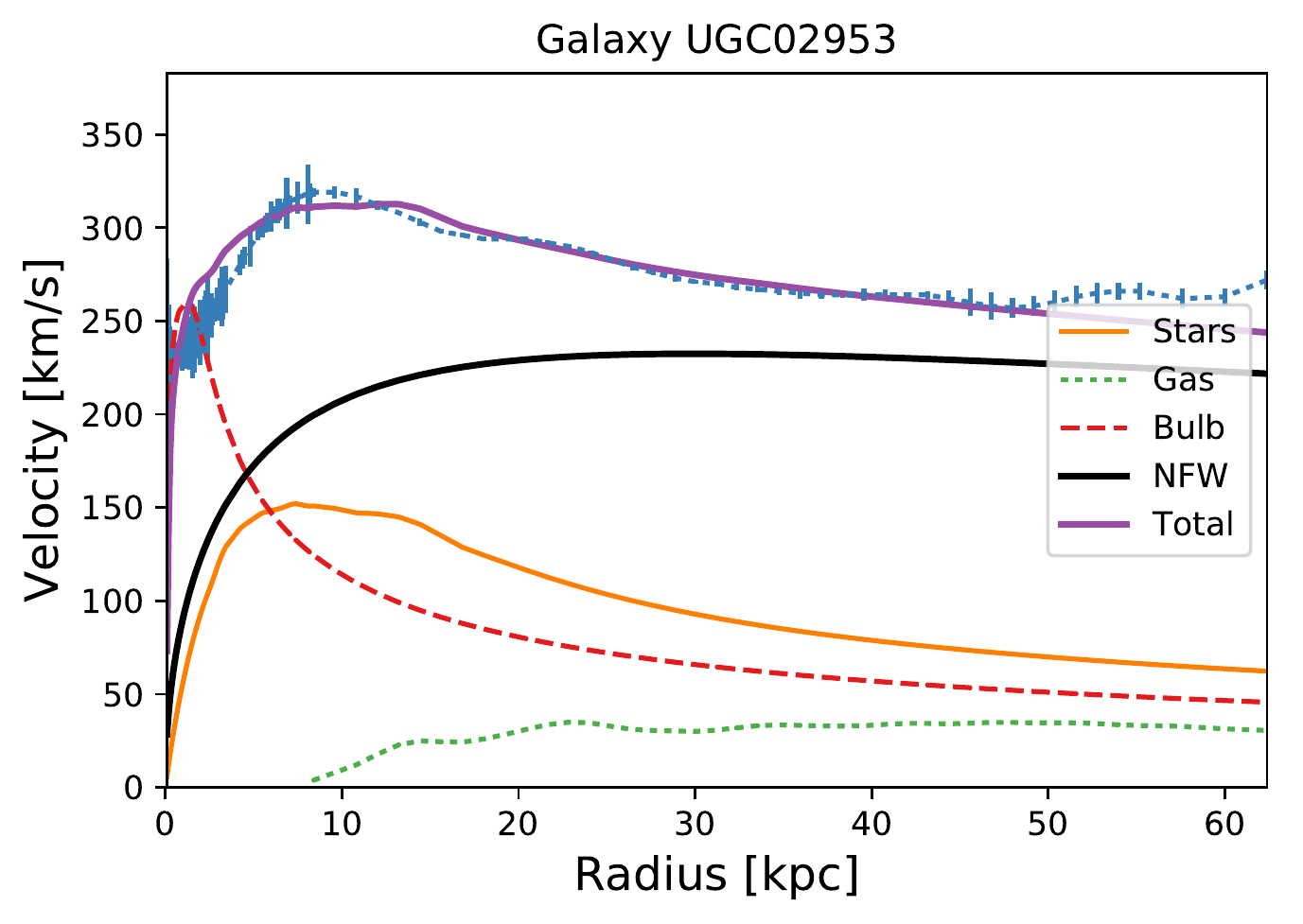}    &   
  \includegraphics[width=0.45\textwidth]{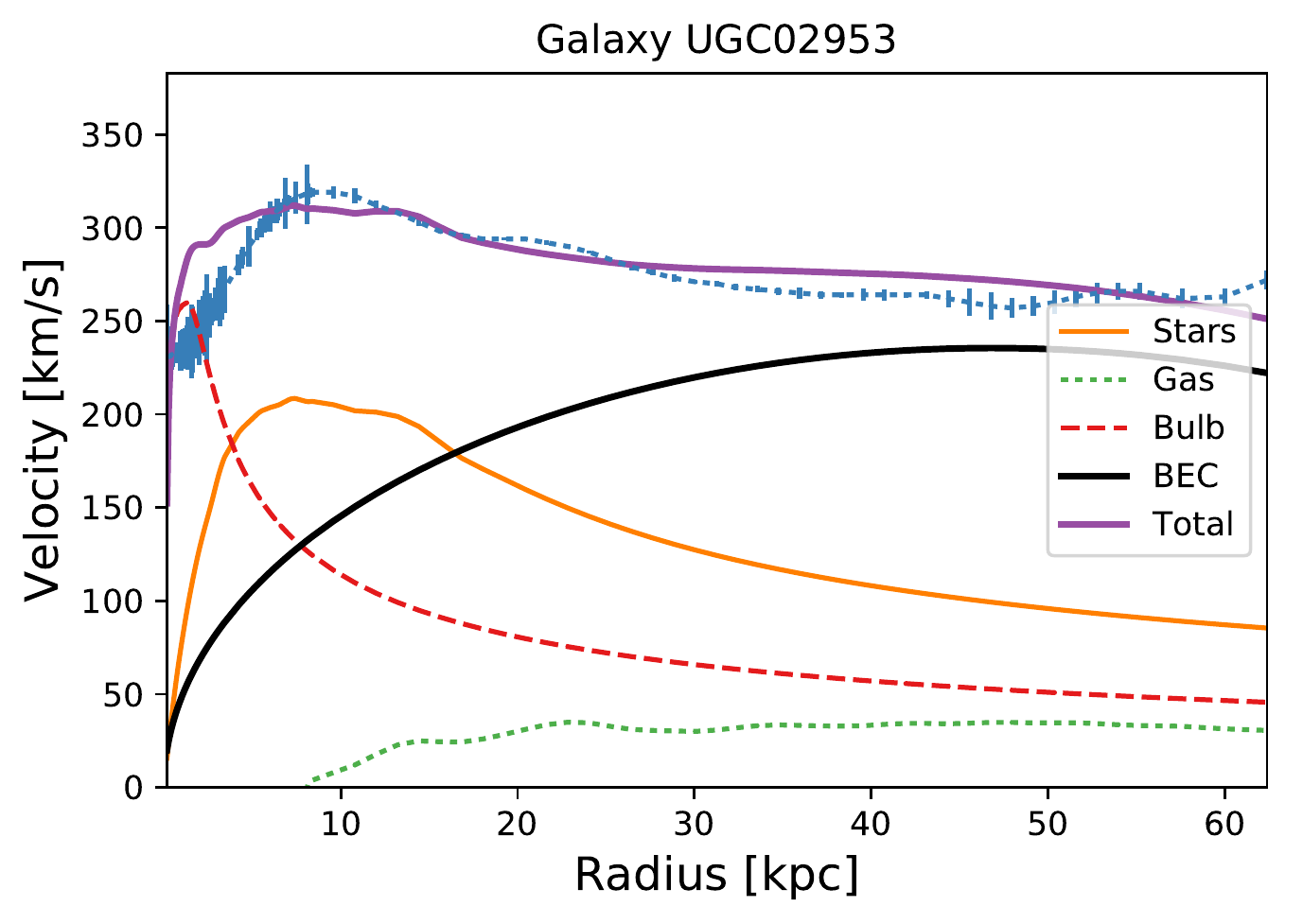}  \\
  (a) UGC02953 with NFW  & (b) UGC02953 with BEC \\[6pt]
  \includegraphics[width=0.45\textwidth]{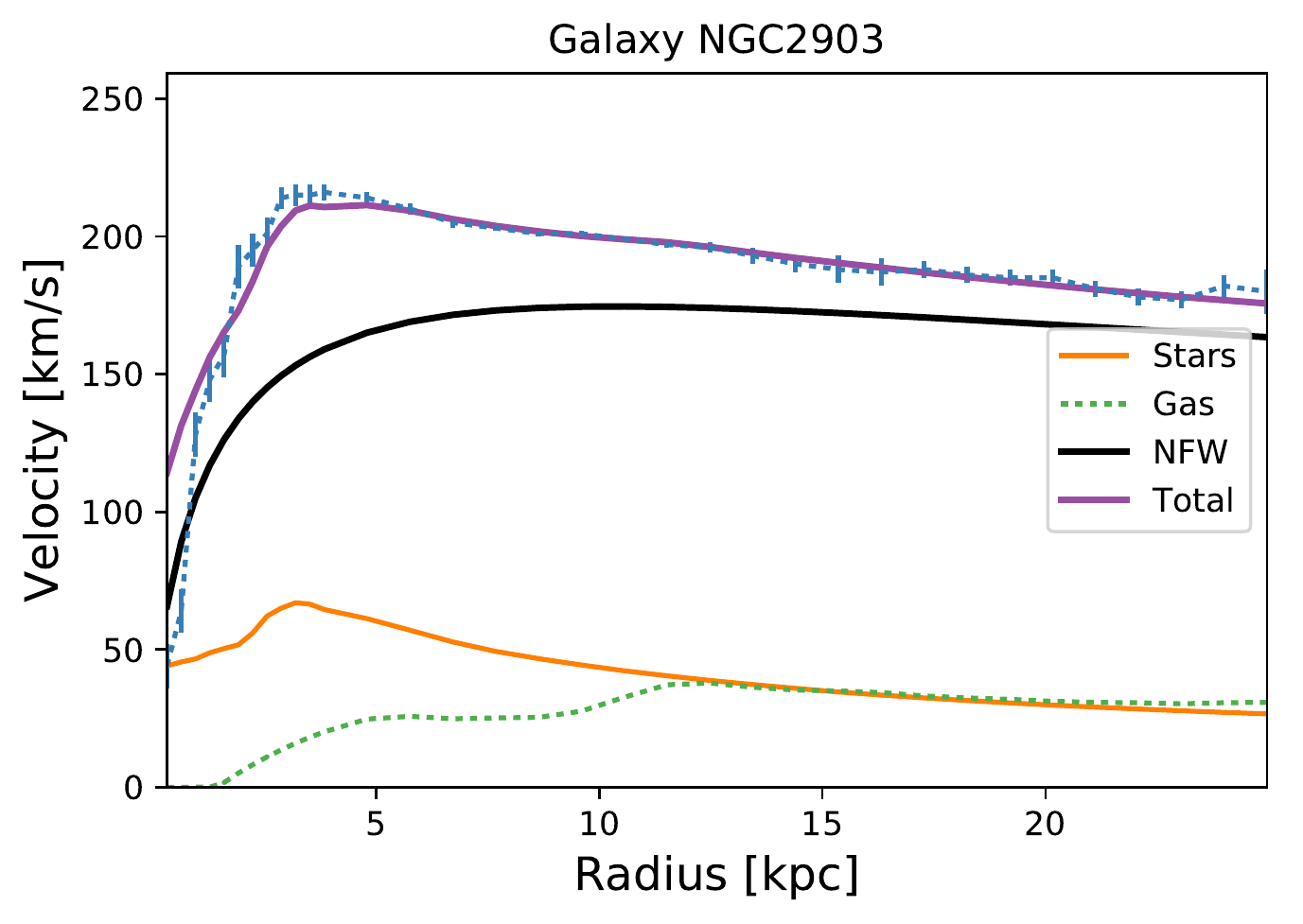} &   
  \includegraphics[width=0.45\textwidth]{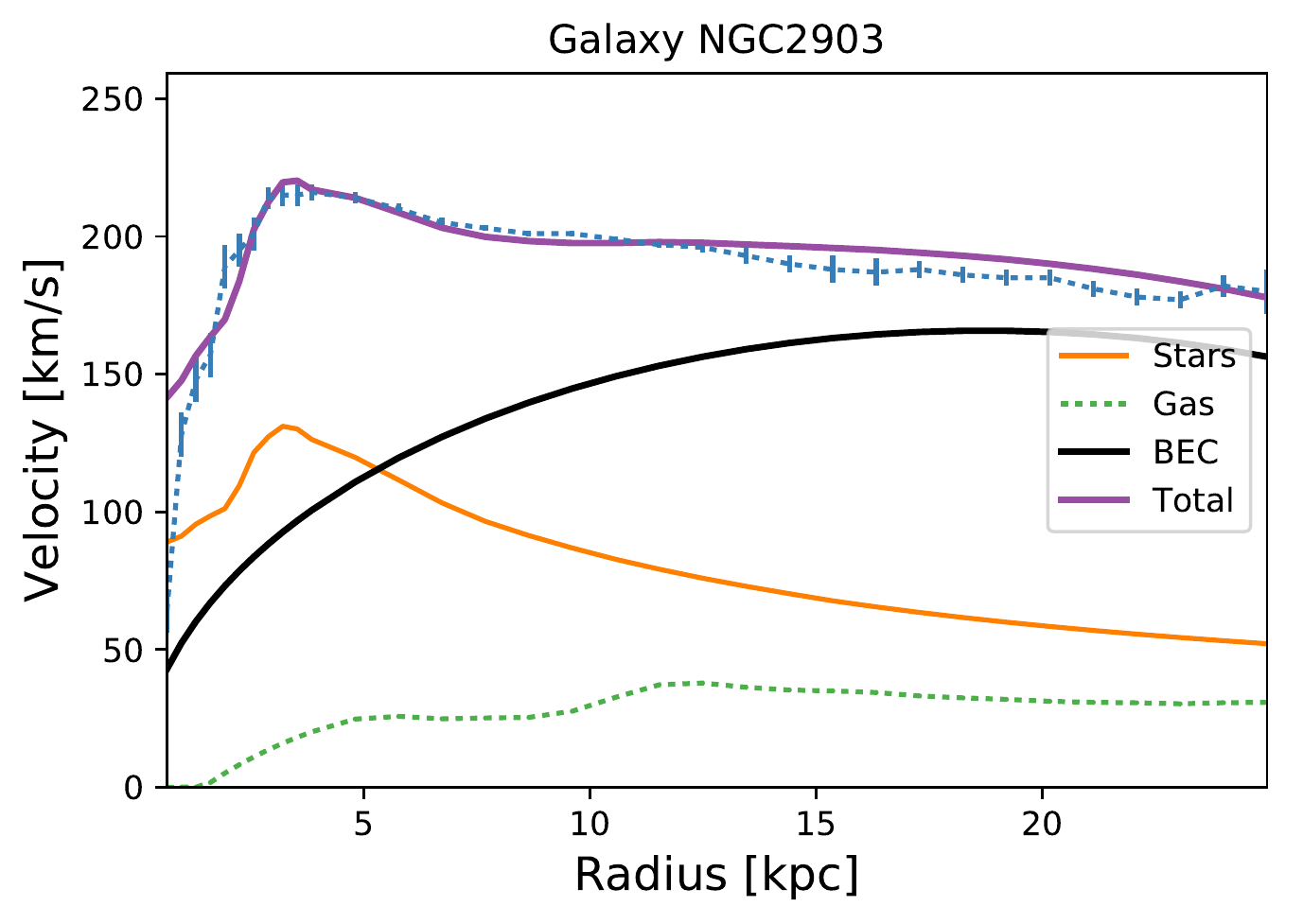} \\
  (c) NGC2903 with NFW & (d) NGC2903 with BEC \\[6pt]
\end{tabular}
\caption{Best--fit rotational curves of two galaxies of our sample. Blue dots with error--bars denote the total observed rotational velocity. The NFW/BEC component is represented by the black solid curve, figures in the left correspond to NFW while BEC is shown at the right. The red dashed curve draws the contribution of the bulge component, the orange solid curve shows the contribution of the stars, and the dotted green curve denotes the contribution of the gas (In appendix \ref{appendix:rot_curves} we use the same color code for each galaxy component).}
\label{fig:fit_example}
\end{figure*}

Using a Bayesian model selection, a methodology to describe the relationship between the galaxy model,
the astrophysical data and the prior information about the free parameters \cite{bayes-th},
we can update the prior model probability to the posterior model probability. However, when we compare
models, the \textit{evidence} is used to evaluate the model's evolution using the data available. The evidence
is given by
\begin{eqnarray}\label{eq:bayes}
\mathcal{E} =\int{\mathcal{L}(\theta) P(\theta) d\theta}
\end{eqnarray}
where $\theta$ is the vector of free parameters, which in our analysis correspond to $(M/L, \lambda, \mu, \omega)$ and $P(\theta)$
is the prior distribution of these parameters. Equation  (\ref{eq:bayes}) can be difficult to calculate due that
cannot be evaluated in closed form and
the integrations can consume to much time when the parametric space is large.
Nevertheless, even when several methods exist \cite{gregory,Trotta:2005ar}, in this work we applied a multi nested
sampling algorithm \cite{skilling} which has proven practicable in cosmology and astrophysical applications \cite{Liddle:2006kn}.

We compute the logarithm of the Bayes factor between two models $\mathcal{B}_{ij}=\mathcal{E}_{i}/\mathcal{E}_{j}$,
where the reference model ($\mathcal{E}_{i}$) with highest evidence is the NFW model. Using this method we perform the interpretation scale known as Jeffreys's scale \cite{Jeffreys:1939xee},
is given as: if
$\ln{B_{ij}}<1$ there is not significant preference for the model with the highest evidence; if $1<\ln{B_{ij}}<2.5$ the
preference is substantial; if $2.5<\ln{B_{ij}}<5$ it is strong; if $\ln{B_{ij}}>5$ it is decisive. In Table \ref{tab:bayes_factor_def} is reported the regions
related to our results. 

\begin{table}
\begin{center}
\begin{tabular}{|c|c|c|}
\hline $\ln{B_{i0}}$   		    & Strength of evidence    		&  Color region        \\
\hline >5	    			& Strong evidence for model $i$	 &Yellow    				 \\
\hline [2.5,5]            		& Moderate evidence for model $i$ 	  & Red   					 \\
\hline [1,2.5] 		    	& Weak evidence for model $i$ 	    &Blue 	\\
\hline [-1,1]  			& Inconclusive 	   				&Green \\
\hline 
\end{tabular}
\caption{Jeffreys's scale as presented in \cite{trottabayes}. The Bayes factor of each Galaxy ID's where computed in comparison to model $i$: NFW model. We used the information from Table \ref{tab:galaxies_models}.}
\label{tab:bayes_factor_def}
\end{center}
\end{table}

{\renewcommand{\tabcolsep}{4mm}
{\renewcommand{\arraystretch}{1.5}
\begin{table*}
\begin{minipage}{\textwidth}
\caption[]{Numerical fits for the BEC model. For each Galaxy we reported the values of: $\Upsilon$ relation total mass-to-light ratio on luminosity,
the reduced $\chi^2$ in terms of the degree of freedom for the BEC model $\chi^2/\text{dof}$, which correspond to 3 and 4 dof, respectively. The small $\chi^2$ values are mainly due to the large data error bars. $\log_{10}M_{\text{BH}}/M_{\odot}$ is the mass of the black hole given in solar units $M_{BH}=10^x M_{\odot}$. The free parameters for the BEC model are: $\lambda$ and $\mu$. The last column is the $\ln B$ factor in comparison with the NFW model.}
\resizebox*{\textwidth}{!}{
\begin{tabular}{ccccccccc}
\hline
\multicolumn{1}{c}
{Galaxy ID} & \multicolumn{5}{c}{BEC model} 						& \multicolumn{1}{c}{$\ln B$}\\
\hline \hline
$  $ 				                & $\Upsilon_\star^{\rm bec}$     		& $\chi^2/\text{dof}$ 	   & $\log_{10}M_{\text{BH}}/M_{\odot}$     & $\log_{10}\lambda \;{\rm pc}^{-1}$    & $\mu [\text{pc}^{-1}]$    \\
\hline
IC2574 			 	 	     & $0.21^{+1.07}_{-0.21}$  	     & 5.37 			           & $12.00 \pm 0.73$ 					   & $-89.83\pm0.78$ 		& $4.06\pm0.97 $ 	& $4.44$\\
NGC2403                                       & $0.68 \pm 0.46 $		       & 14.14 		   	   & $10.64 \pm 0.51$ 					   & $ -91.88\pm0.96 $ 		& $4.12\pm0.30 $  	& $4.42$\\
NGC2841 				     & $0.95 \pm0.16$		                 & 2.50		  	   & $11.20 \pm 0.10$ 				  	   & $-91.43\pm0.17 $ 		& $4.01\pm1.00 $   	 & $2.23$\\
NGC2903 		 		     & $0.44^{+1.39}_{-0.44} $		& 5.75 			   & $10.72 \pm 0.94$ 					   & $-92.77\pm0.42$ 		& $2.34\pm0.97$  	& $1.47$\\
NGC3198 		 		     & $ 0.73 \pm 0.04 $		        & 2.58 			   & $10.96 \pm 0.16$ 					   & $-91.30\pm0.47$ 		& $4.24\pm1.00 $ 	& $1.40$\\
NGC3521			  		     & $0.54 \pm 0.10 $		         & 0.21 			   & $11.71 \pm 1.35$ 					   & $ -89.99\pm1.49$ 		& $7.87\pm0.95 $  	& $1.61$\\
NGC4559			  		     & $0.52^{+2.16}_{-0.52}$		& 0.24 			   & $10.78 \pm0.97$ 					   & $-91.03\pm0.91$ 		& $5.89\pm1.64 $         & $1.24$\\
NGC6015                     		     & $0.94 \pm0.05 $  		        & 7.16 			   & $11.08 \pm 0.40 $ 					   & $-91.35\pm0.54$ 		& $3.60\pm1.03$          &$1.35$\\
NGC6503                   		     & $0.63 \pm0.10$		                & 5.09 			   & $10.72 \pm 0.34$ 					   & $-91.58\pm0.51 $ 		& $4.33\pm0.46 $ 	& $1.10$\\
NGC6946                       		     & $0.50 \pm0.06 $  		        & 1.91 			   & $10.77 \pm 0.61 $ 					   & $-90.59\pm0.98 $ 		& $ 9.79\pm1.54 $ 	   & $2.06$\\
NGC7331			 		     & $0.38 \pm 0.03 $  		        & 0.90 			   & $11.05 \pm0.25$ 					   & $-90.36\pm0.46$ 		& $10.20\pm0.94 $ 	   & $1.37$\\
NGC7793			 		     & $0.71^{+1.18}_{-0.71} $  		& 0.78 			   & $10.70 \pm7.54$ 					   & $-90.79\pm42.99$ 		& $7.54\pm4.65 $          & $1.39$\\
UGC02953		 		     & $0.67^{+1.08}_{-0.67}$  		& 14.86 			   & $11.12 \pm1.00$ 					   & $-91.58\pm1.59$ 		& $4.01\pm0.45$           & $1.54$\\
UGC03205                                      & $0.25 \pm 0.16 $                      &  5.10                          & $10.78\pm0.47$                                                & $ -91.32\pm0.19 $ &        $3.74\pm1.19$                & 2.54$$\\
UGC03580			              & $0.81 \pm 0.07$  		        & 3.11 			   & $10.78 \pm 0.47$ 		                            & $-91.57\pm0.84 $ 		& $4.23\pm0.98$ 	& $1.92$\\
UGC05253		 		     & $0.52^{+0.98}_{-0.52} $  		& 6.56 			   & $11.05 \pm0.95 $ 					   & $-91.54\pm1.62 $ 		& $4.30\pm1.01$ 	& $1.47$\\
UGC07524		 		     & $1.11^{+2.64}_{-1.11} $  		& 0.59 			   & $10.54 \pm 1.05 $ 					   & $ -90.66\pm1.86 $ 		& $8.42\pm1.02$         & $1.96$\\
UGC08699				     & $ 0.56 \pm 0.01 $  		        & 1.44 			   & $ 10.73 \pm 0.25 $ 					   & $ -92.16\pm0.34 $ 		& $3.51\pm0.05 $ 	& $1.29$\\
UGC09133		 		     & $0.58^{+0.97}_{0.58} $  		& 23.93 			   & $11.35 \pm 2.36$ 					   & $ -90.65\pm2.61 $ 		& $5.21\pm1.60$ 	& $1.57$\\
UGC11455				     & $0.54 \pm0.08$  		         & 3.60 			   & $11.67 \pm1.12$ 					   & $ -89.78\pm1.24 $ 		& $8.56\pm1.00 $ 	& $2.95$\\
UGC11914				     & $0.64^{+0.84}_{-0.64} $  		& 2.16 			   & $12.01\pm0.99$ 					   &          $ -90.83\pm2.16 $ 		& $3.96\pm0.95 $        & $0.92$\\
\hline
\hline\\
\end{tabular}\label{tab:galaxies_models1}}
\end{minipage}
\end{table*}}}

{\renewcommand{\tabcolsep}{8mm}
{\renewcommand{\arraystretch}{1.9}
\begin{table*}
\begin{minipage}{\textwidth}
\caption[]{Numerical fits for the NFW model. For each Galaxy ID we reported the values of: $\Upsilon$ relation total mass-to-light ratio on luminosity,
the reduced $\chi^2$ in terms of the degree of freedom $\chi^2/\text{dof}$, which correspond to 3 and 4 dof, respectively. Also we report $c$ and $v_{200}$. The small $\chi^2$ values are mainly due to the large data error bars.}
\resizebox*{\textwidth}{!}{
\begin{tabular}{ccccc}
\hline
\multicolumn{1}{c}
{Galaxy ID} & \multicolumn{4}{c}{NFW model} \\
\hline \hline
$  $ 				&       $\Upsilon_\star^{\rm nfw}$ 	&         $\chi^2/\text{dof}$ 	     &  $c$     		&                                 $v_{200}$  \\
\hline
IC2574 			&       $0.10^{+0.33}_{-0.10} $ 	&     $32.10$ 		     &  $0.84\pm0.05$  		                   & $180.01\pm 0.19$\\
NGC2403 		&      0.29 $\pm$ 0.03		     & $8.76$		     & 17.18 $\pm$ 0.69                                 & 85.93 $\pm$ 1.13\\
NGC2841 		&     $0.83 \pm 0.50$ &                     $1.671$ 		     & 8.76 $\pm$ 8.67		                   & 209.50 $\pm$ 34.14\\
NGC2903 		&     $0.22 \pm0.22$ &                       $5.429$ 		     & 31.58 $\pm$ 4.81 		                    & 105.93 $\pm$ 20.12 \\
NGC3198 		&     $0.45^{+0.77}_{-0.45}$&            $0.734$		      & $10.75^{+22.19}_{-10.75}$                 &  109.47 $\pm$ 47.41\\
NGC3521			&    $0.54 \pm0.02$ &                         $0.214$ 		     & 3.22 $\pm$ 0.72 		                   & 366.35 $\pm$ 45.48\\
NGC4559			&     $0.12^{+0.48}_{-0.12}$ &             $0.178$ 		     & 14.87 $\pm$ 10.61		                  & 83.60 $\pm$ 10.65 \\
NGC6015			&     $0.92\pm0.01$ &                          $7.585$ 		     & 3.72 $\pm$ 0.63  		                  & 165.53 $\pm$ 24.57 \\
NGC6503			&    $0.39 \pm0.28$ &                          $1.289$ 		     & 14.97 $\pm$ 2.49		                  & 79.85 $\pm$ 1.84 \\
NGC6946			&    $0.36\pm0.31$ &                            $1.576$              & 14.95 $\pm$ 12.87  		                  & 105.42 $\pm$ 14.08  \\
NGC7331			&   $0.36\pm0.09$ &                             $0.726$ 	    & $6.87^{+27.64}_{-6.87}$  		         & 196.37 $\pm$ 82.58 \\
NGC7793			&   $0.60\pm0.10$ &                            $0.840$ 		     & 9.57 $\pm$ 3.72  		                 & 75.74 $\pm$ 17.87   \\
UGC02953		&   $0.49\pm0.06 $ &                            $6.007$              & 17.39 $\pm$ 1.59  		                  & 168.08 $\pm$ 14.64\\
UGC03205		&   $0.79\pm0.53$	                       & $3.065$ 		     & $4.16^{+16.74}_{-4.16}$  		          & 203.70 $\pm$ 93.95	\\
UGC03580                  & 0.17 $\pm$ 0.04                            & 3.30                   & 9.17 $\pm$ 1.02                                 &99.38 $\pm$ 12.33 \\
UGC05253		&    $0.44\pm0.01$ &                          $2.922$ 		     & 12.53 $\pm$ 1.19  		                 & 157.52 $\pm$ 6.88  \\
UGC07524		&    $0.10^{+1.13}_{-0.10}$ &             $0.682$ 		     & 6.53 $\pm$ 2.27  		                  & 80.47 $\pm$ 7.57   \\
UGC08699		&    $0.52\pm0.07$ &                           $0.878$ 		     & $13.09^{+26.16}_{-13.09}$  		 & 120.99 $\pm$ 32.18 \\
UGC09133		&    $0.46\pm0.02$ &                           $8.397$ 		     & 10.25 $\pm$ 0.28 		                 & 168.91 $\pm$ 3.75  \\
UGC11455		&    $0.52^{+1.00}_{-0.52} $ &             $4.684$ 		     & 2.77 $\pm$ 0.80                                   & 341.90 $\pm$ 1.00 \\
UGC11914		&   $0.64\pm0.39$ &                            $2.264$ 		     & 3.86 $\pm$ 3.27           		          & 500.00 $\pm$ 99.54\\
\hline
\hline\\
\end{tabular}\label{tab:galaxies_models}}
\end{minipage}
\end{table*}}}

\begin{figure*}
  \centering
   \includegraphics[width=0.45\textwidth]{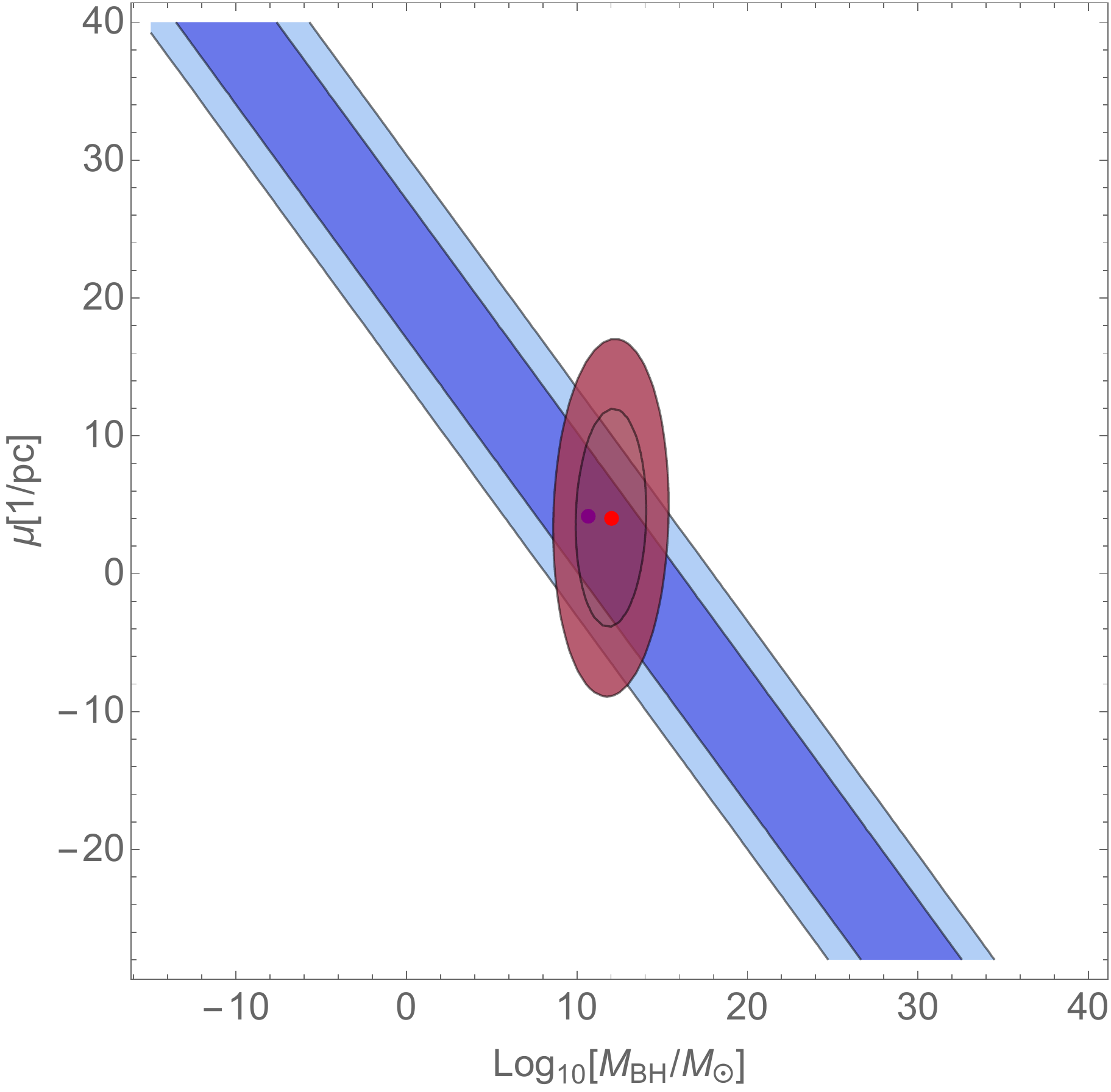}
   \includegraphics[width=0.44\textwidth]{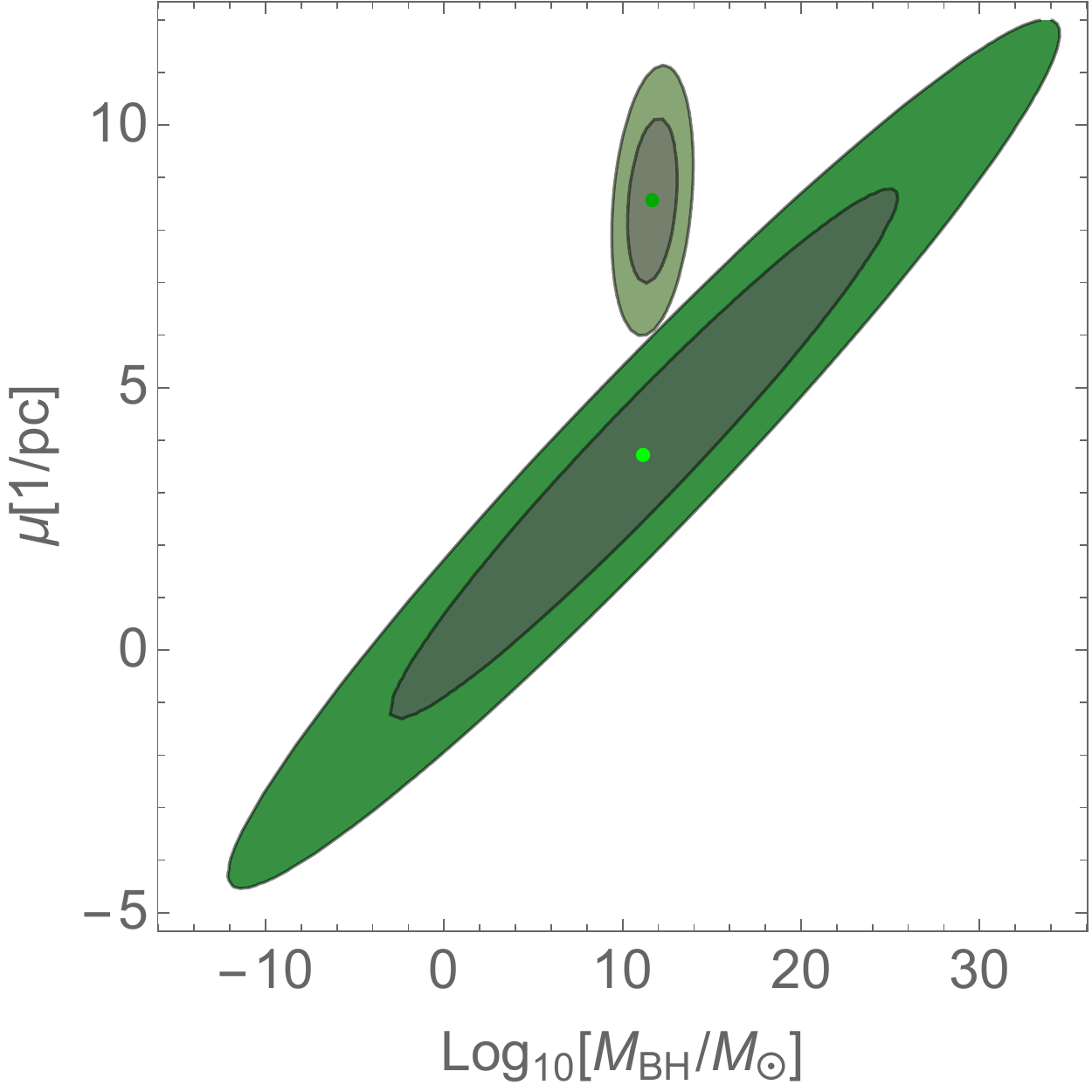}
    \caption{\textit{Left:} Contour regions for IC2574 (blue) and NGC2403 (red) galaxies, which are in the moderate (red) region. \textit{Right:} Contour regions for UGC07524 (green) and UGC11455 (darker green) galaxies, which are in the weak (blue) region.}
  \label{fig:bayes_allgalaxies}
\end{figure*}

\begin{figure*}
  \centering
  \includegraphics[width=0.7\textwidth]{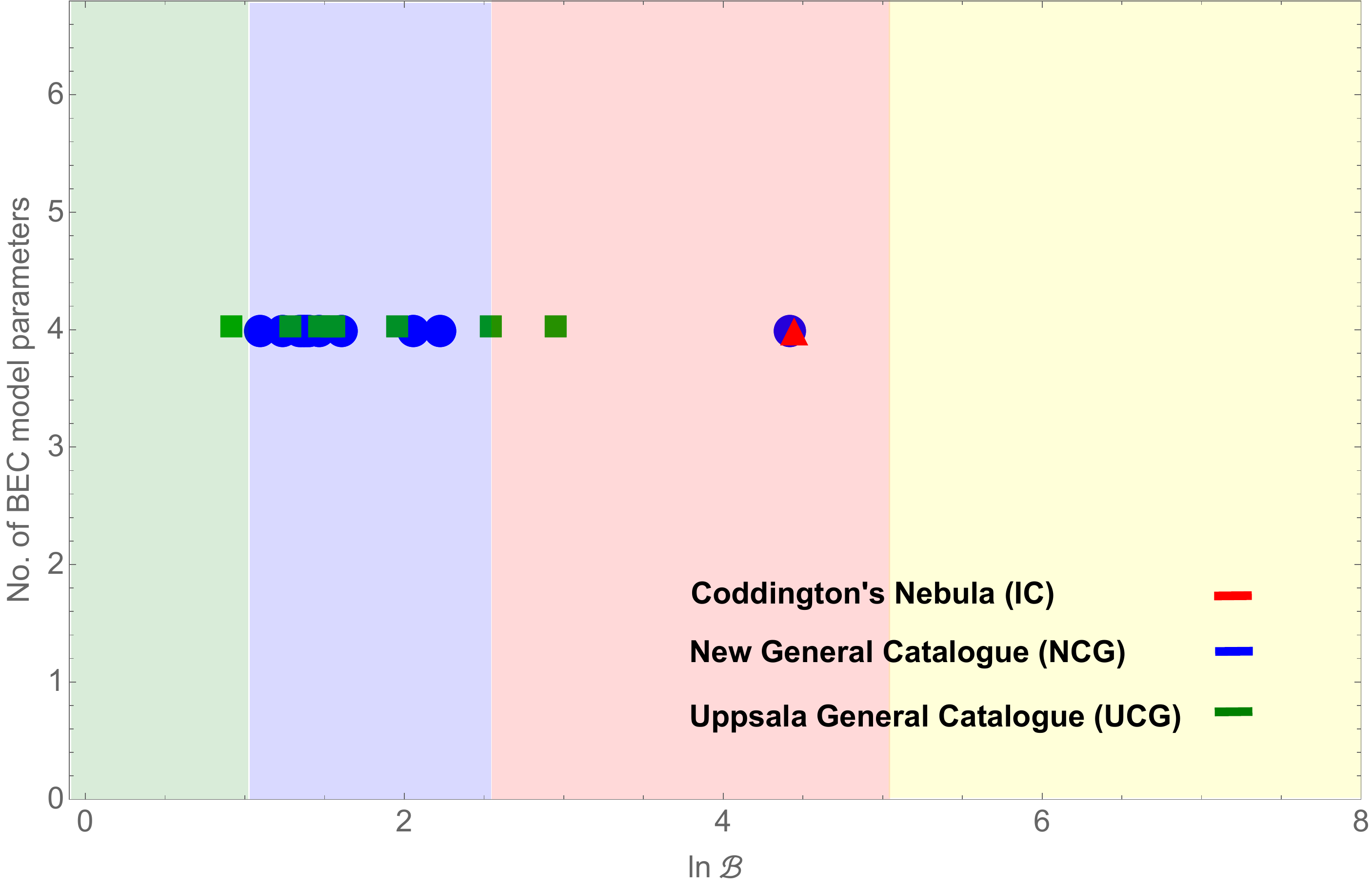}
    \caption{Bayes regions for the Galaxy ID's reported in Table \ref{tab:galaxies_models}.}
  \label{fig:bayes_allgalaxies1}
\end{figure*}

%%%%%%%%%%%%%%%%%%%%%%%%%%%%%%%%%%%%%%%%%%%%
%%%%%%%%%%%%%%%%%%%%%%%%%%%%%%%%%%%%%%%%%%%%

\section{Results and Conclusions}
\label{con}

According to the results obtained in the present work, it seems to be that the interpretation of dark matter as some kind of BEC is, in fact, a good model to describe the kinematics of the galaxy rotation curves. Even more, we have shown that the corresponding Thomas--Fermi approximation can give insights in a simple way that match with the observations of rotation curves in galaxies. From the simple analysis, we notice that density is proportional $\rho \propto r^{-1}$ for all radii, in contrast with most DM profiles that behave as $\rho \propto r^{-3}$ at large radii. The mass of the BEC behaves as $m_{\rm bec} \propto r^2$, and therefore $v_{\rm bec} \propto r^{1/2}$, this coincides with NFW in the limit $r \ll r_{\rm max}$ where $v_{\rm nfw}\propto r^{1/2}$, with $r_{\rm max} \sim 2.163 r_s$.

We can see in Table  \ref{tab:galaxies_models1} the fittings of our SPARC sample for the BEC model, we also report the corresponding fittings for the NFW model in Table \ref{tab:galaxies_models}. We have obtained an average $\mu = 5.43 \pm 2.24 \ {\rm pc}^{-1}$ which corresponds to an average boson mass of $M_\Phi = (3.47 \pm 1.43 )\times10^{-23}$ eV.  Previous analysis has found masses for ultralight dark matter (see \cite{Chavanis:2018pkx} for a compilation of results), which constrain the mass of the boson in the order of $10^{-22}$ eV. Notice that the latter result agrees with the one obtained in the present work. For the self--interacting parameter we get the values $\log_{10} (\lambda \; [{\rm pc}^{-1}]) = -91.09 \pm 0.74 $. In other words, the halo viewed as a Bose--Einstein condensate can be interpreted almost as an ideal condensate of some generic bosons, which also agree with the results reported in \cite{NOS2}. The mass of the black holes at the galaxy centre is given by $\log_{10} M/M_\odot = 11.08 \pm 0.43$. This kind of scenarios, e.g  Abell~85 and Holm 15A, has been reported in \cite{abell85, mehrgan} with a core that host supermassive black holes of mass $10^9$--$10^{11}$ $M_{\odot}$.  Although recent observation of supermassive black holes in galaxies supports our fitting for the mass of the black hole, we could also think of the possibility that the black hole may not be the only option to trap the boson cloud. In future works, we could use a weak field approximation to also include the mass of stars, bulge, and gas in the center of the galaxy, even more, the self-gravitating effect of the bosonic halo or the effects caused by the  rotation of the halo as a contribution to the gravitational potential and therefore reduce the contribution of the black hole as the only source of the potential.

Though we leave the mass--to--light ratio of starts, $\Upsilon_\star$, as free parameter, we find that the average value for the BEC is $\Upsilon_\star^{\rm bec} = 0.61 \pm 0.21 \; M_\odot/L_\odot$, while the NFW fit well with an average value of $\Upsilon_\star^{\rm nfw} = 0.44 \pm 0.23 \; M_\odot/L_\odot$. This is, NFW suppress the contribution of star 17\% more comparing with BEC. In other words, NFW prefers a less contribution of the stars to the total rotation curve. This can be clearly seen in Figure\,\ref{fig:fit_example}, for both galaxies NGC 2903 and UGC 2953. The average value of $\Upsilon_\star^{\rm nfw}$ along the big uncertainties of both concentration parameter $c = 10.33 \pm 6.82$ and $v_{200} = 171.65 \pm {106.79}$ km/s let us to conclude that for some galaxies the NFW profile represent an unphysical halo. 

The value of the mass--to--light ratio of bulge and starts add the biggest uncertainties, and make our conclusions dependent on the value of $\Upsilon_\star$. Notice how the $\mu$ and $M_{BH}/M_{\odot}$ are quite correlated for the four galaxies that are in a moderate statistical region in comparison to NFW model.

Notice that in general our results fits as well as the NFW profile, but objectively we can not say the observations prefer the NFW or the BEC model. In three galaxies (UGC07524, UGC11455, IC2574) the BEC fits better than NFW. In five galaxies (UGC02953 NGC3198, NGC6503, UGC09133, UGC05253) the BEC model seems to be that does not fit better than the NFW model. However, we have to keep in mind that we only use the Thomas--Fermi approximation related to the scalar halo. In the rest of the galaxies, our fit equally competes with NFW. On average the ratio $\chi^2_{\rm bec}/\chi^2_{\rm nfw} = 1.49$.
Additionally, from the Bayesian point of view, we notice a moderate preference for IC2574 galaxy in comparison to NFW model using the rest of the binned sample. Also, we should mention that NGC2403 galaxy shows a moderate preference for NFW, but this result disagrees with the $\chi$--square fit since the covariance errors for this galaxy is minor in comparison to the UGC catalog showing a strong correlation between the astrophysical parameters of the model.  c.f. with Figure \ref{fig:bayes_allgalaxies}. After a careful analysis, we realize that exists a statistical anomaly in NGC2403 due to the larger density of data ($\chi^2/\text{dof}$) in comparison to UGC07524, since from the frequentist point of view these densities produces a better fit but the bayesian analysis remains lower.

 Indeed, the BEC model on average fits quite well and is not a bad fitting at all given the simplicity of the model is something that must be taken into account. 
The BEC model in the Thomas--Fermi approximation is strong enough as the NFW model to describe galaxy rotation curves as we can see from Fig.\,\ref{fig:bayes_allgalaxies1}. More important, we cannot ignore that the BEC model comes from first principles and that makes use the Thomas--Fermi approximation and assume that is only a compact central object that condensates the relativistic boson cloud. This model can be extended and generalized to include bulge and central stars contribution that may diminish the mass of the central black hole, or the--self gravitating nature of the halo and also with rotation (and perhaps charge). The good fitting of the BEC model using galaxy rotation curves may be indicating that the nature of dark matter may be close to be unveiled. 

The fit in this model helps us to set physical bounds to the parameters of the Bose--Einstein dark matter approach. These bounds can be used to extend the analysis in bigger samples of galaxies. Finally, more general scenarios can be taken into account in order to  study the corresponding rotation curves, for instance,  in charged and rotating black holes spacetimes  and perhaps as discrimination criteria of the nature of supermassive black holes. 

%%%%%%%%%%%%%%%%%%%%%%%%%%%%%%%%%%%%%%%%%%%%
%%%%%%%%%%%%%%%%%%%%%%%%%%%%%%%%%%%%%%%%%%%%

\begin{acknowledgments}
EC acknowledges MCTP/UNACH for financial support. 
CE-R is supported by the \textit{Royal Astronomical Society} as FRAS 10147, \textit{PAPIIT} Project IA100220 and ICN-UNAM projects. This article is also based upon work from COST action CA18108, supported by COST (European Cooperation in Science and Technology).
JM acknowledges CONACYT--MCTP/UNACH for financial support. This work is partially supported by CONACyT.
\end{acknowledgments}

%%%%%%%%%%%%%%%%%%%%%%%%%%%%%%%%%%%%%%%%%%%%
%%%%%%%%%%%%%%%%%%%%%%%%%%%%%%%%%%%%%%%%%%%%
\begin{appendix}

\section{Images: Galaxy Rotation Curves}\label{appendix:rot_curves}

Here we show the rest of the images in which we compare the BEC model with NFW.

\begin{figure}
  \includegraphics[width=0.45\textwidth]{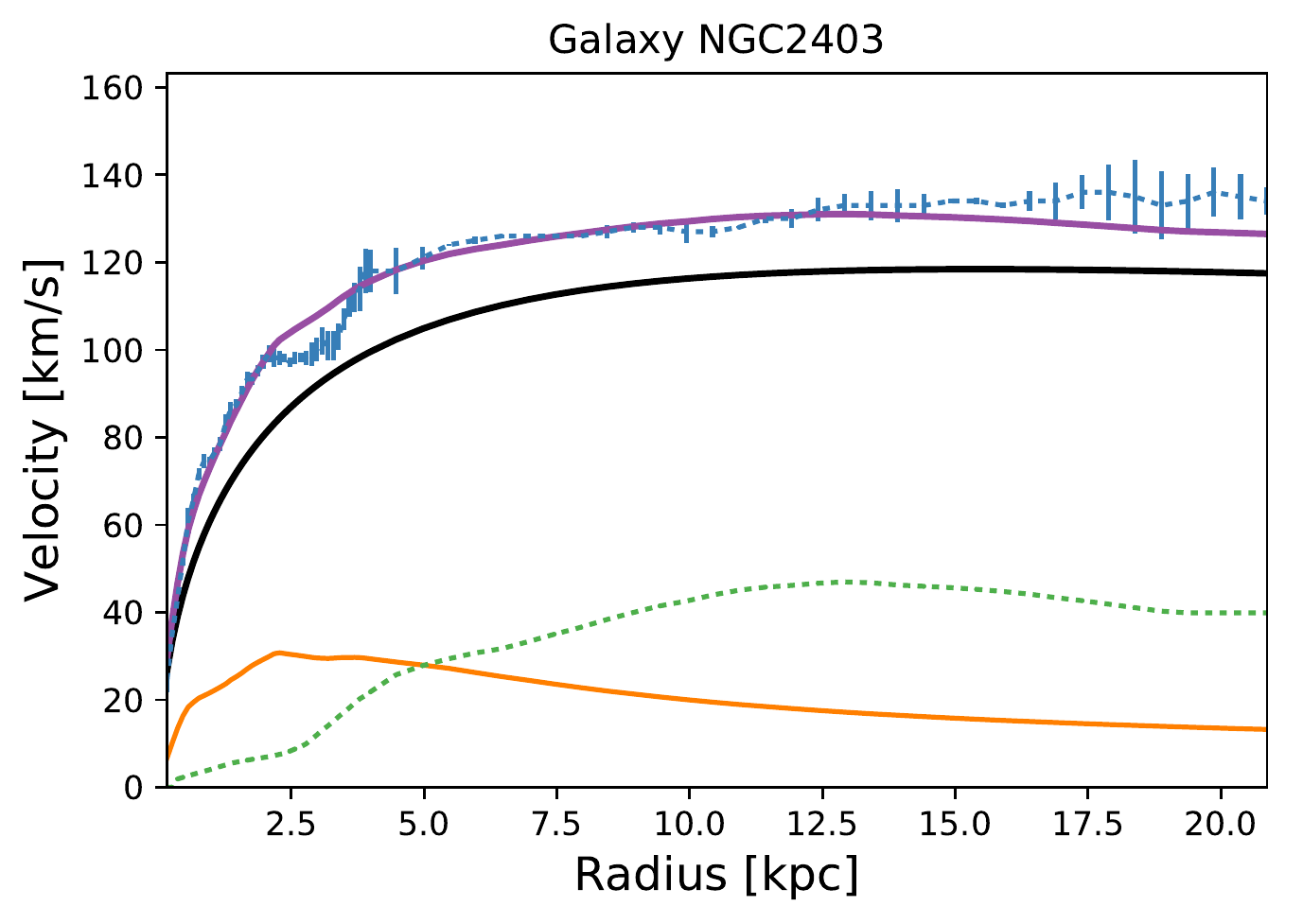}   %&   
  \includegraphics[width=0.45\textwidth]{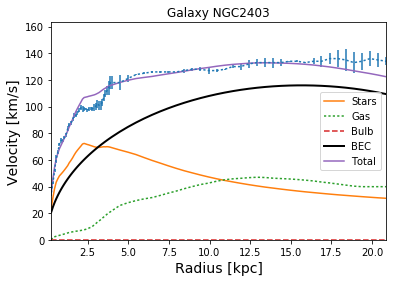}   \\
  \includegraphics[width=0.45\textwidth]{Galaxy_NGC2903_NFW.pdf} %&   
  \includegraphics[width=0.45\textwidth]{Galaxy_NGC2903.pdf} %\\
\caption{Best--fit rotational curves. NFW profiles on the left. BEC on the right. Legends are like in Fig.\,\,\ref{fig:fit_example}.}
\end{figure}

\begin{figure}
  \includegraphics[width=0.45\textwidth]{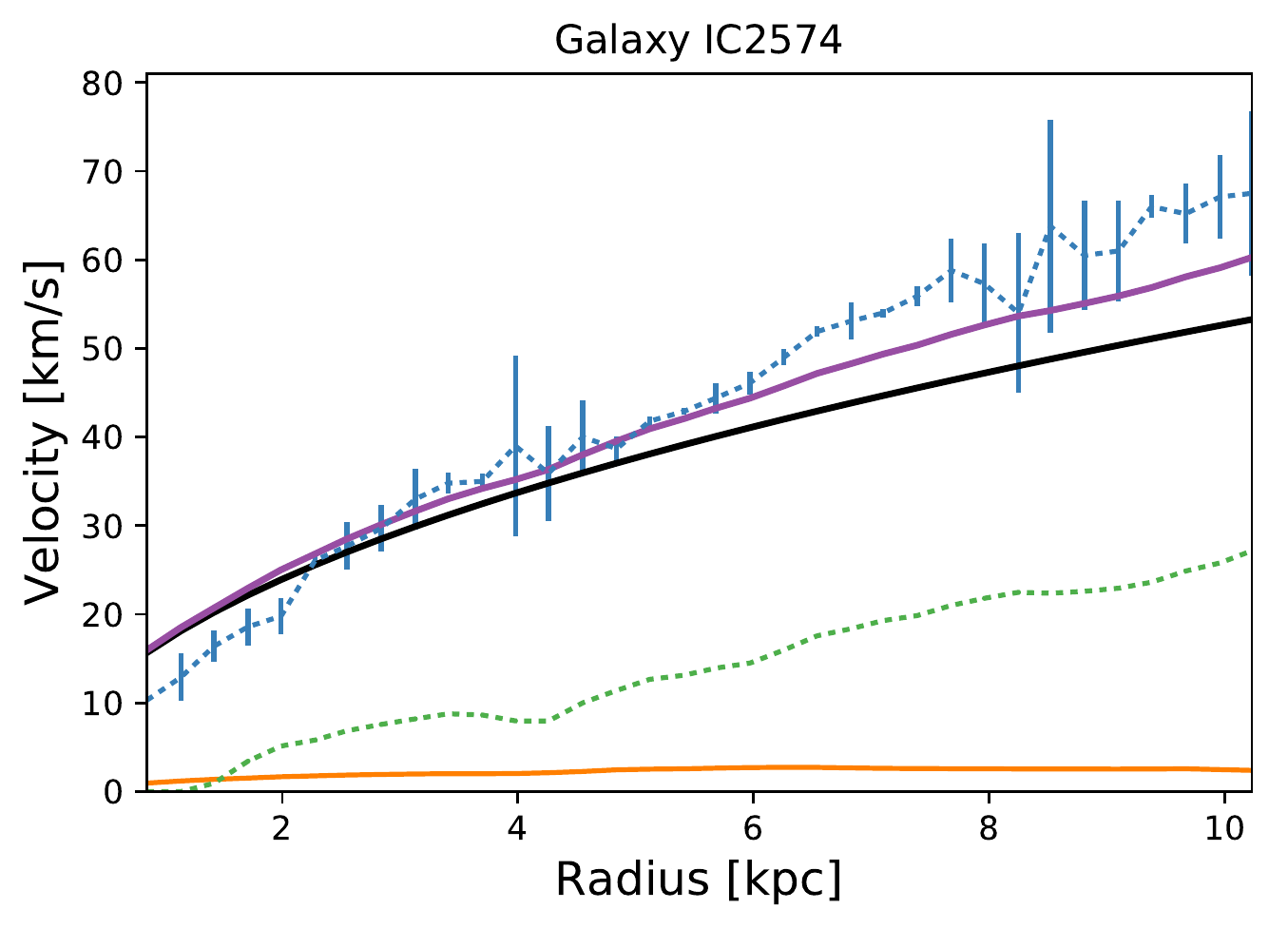}   %&   
   \includegraphics[width=0.45\textwidth]{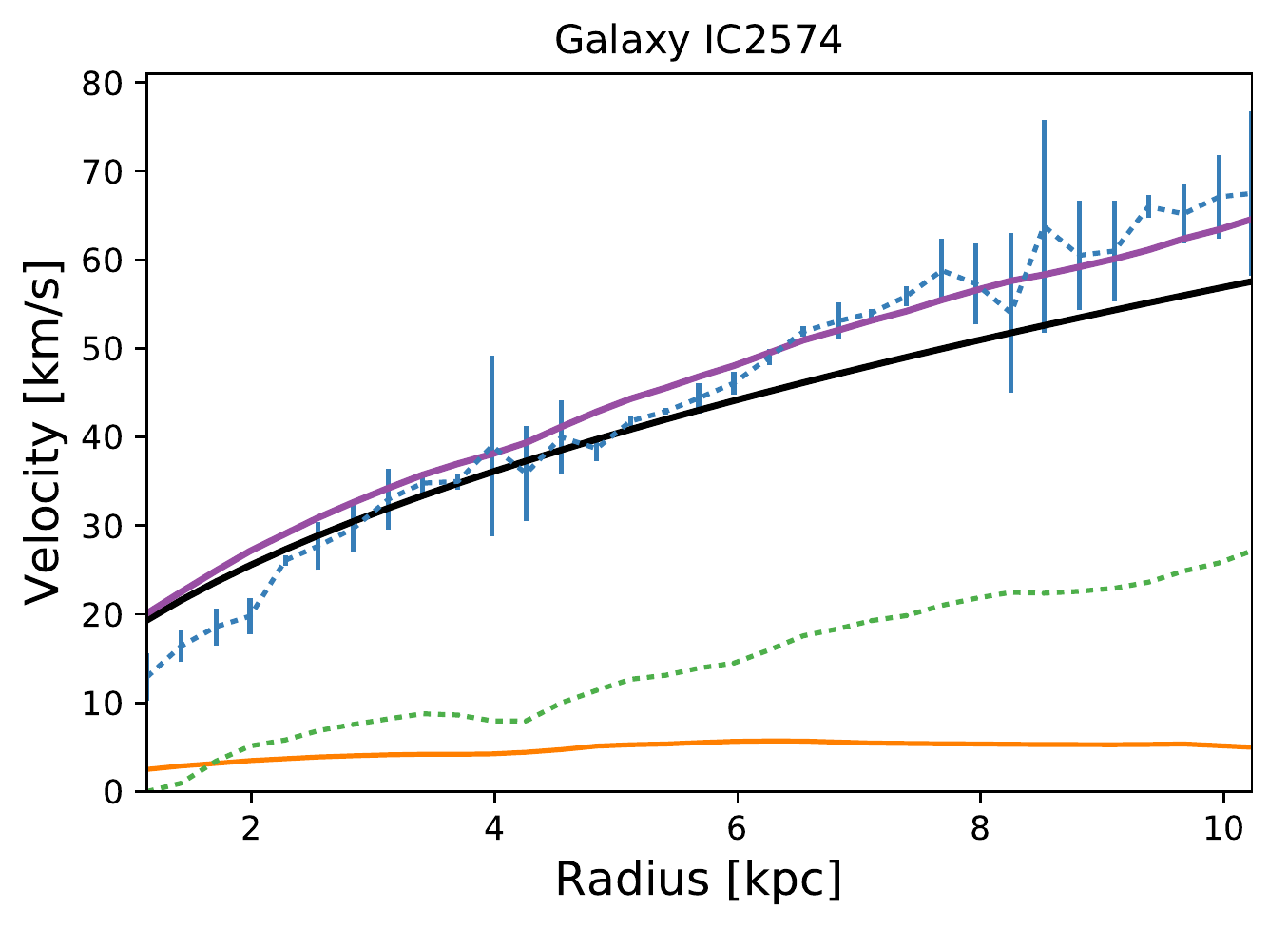}   %\\
\caption{Best--fit rotational curves. NFW profiles on the left. BEC on the right. Legends are like in Fig.\,\,\ref{fig:fit_example}.}
\end{figure}

\begin{figure}
 \includegraphics[width=0.45\textwidth]{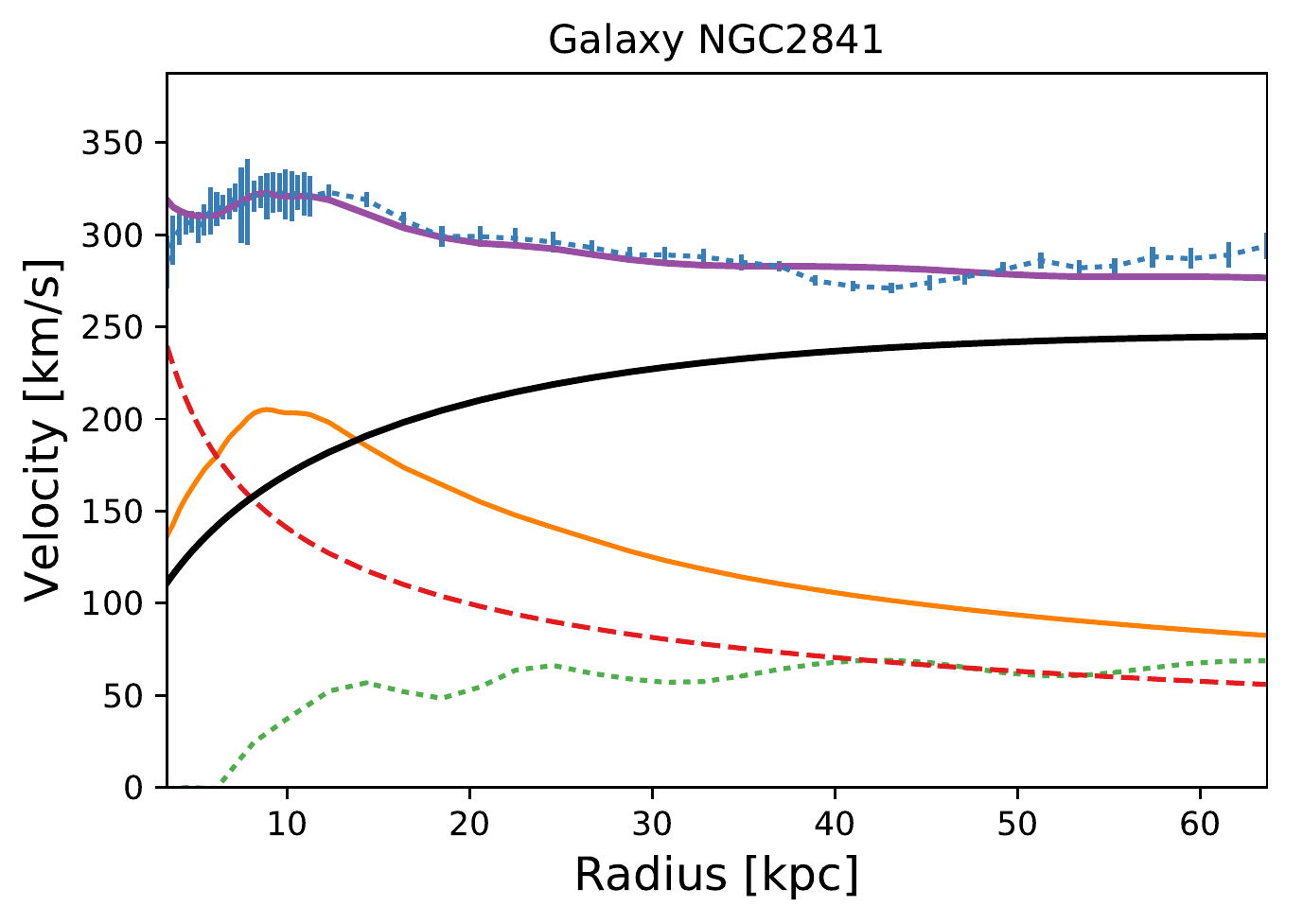}  %&   
  \includegraphics[width=0.45\textwidth]{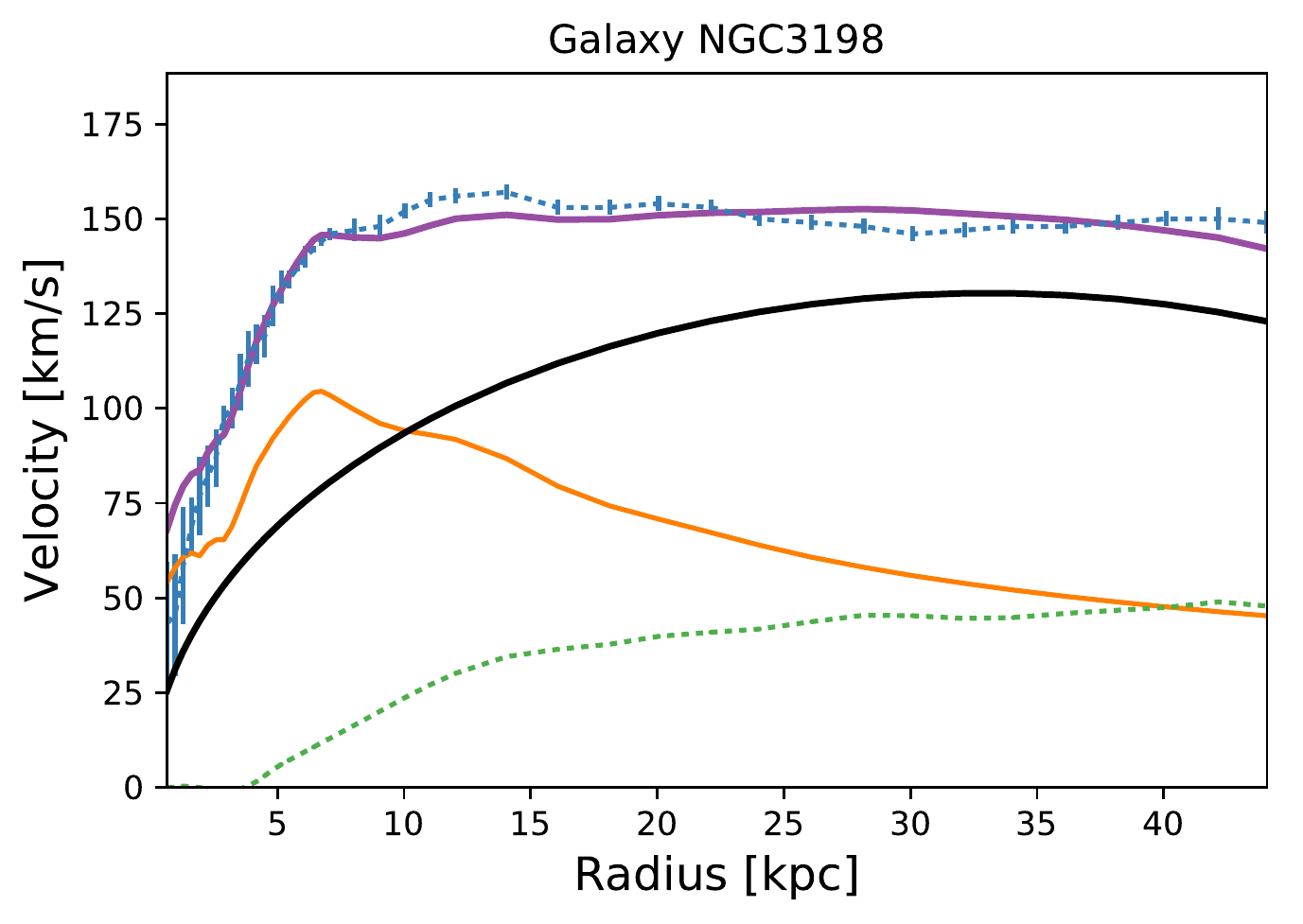} % \\
  \includegraphics[width=0.45\textwidth]{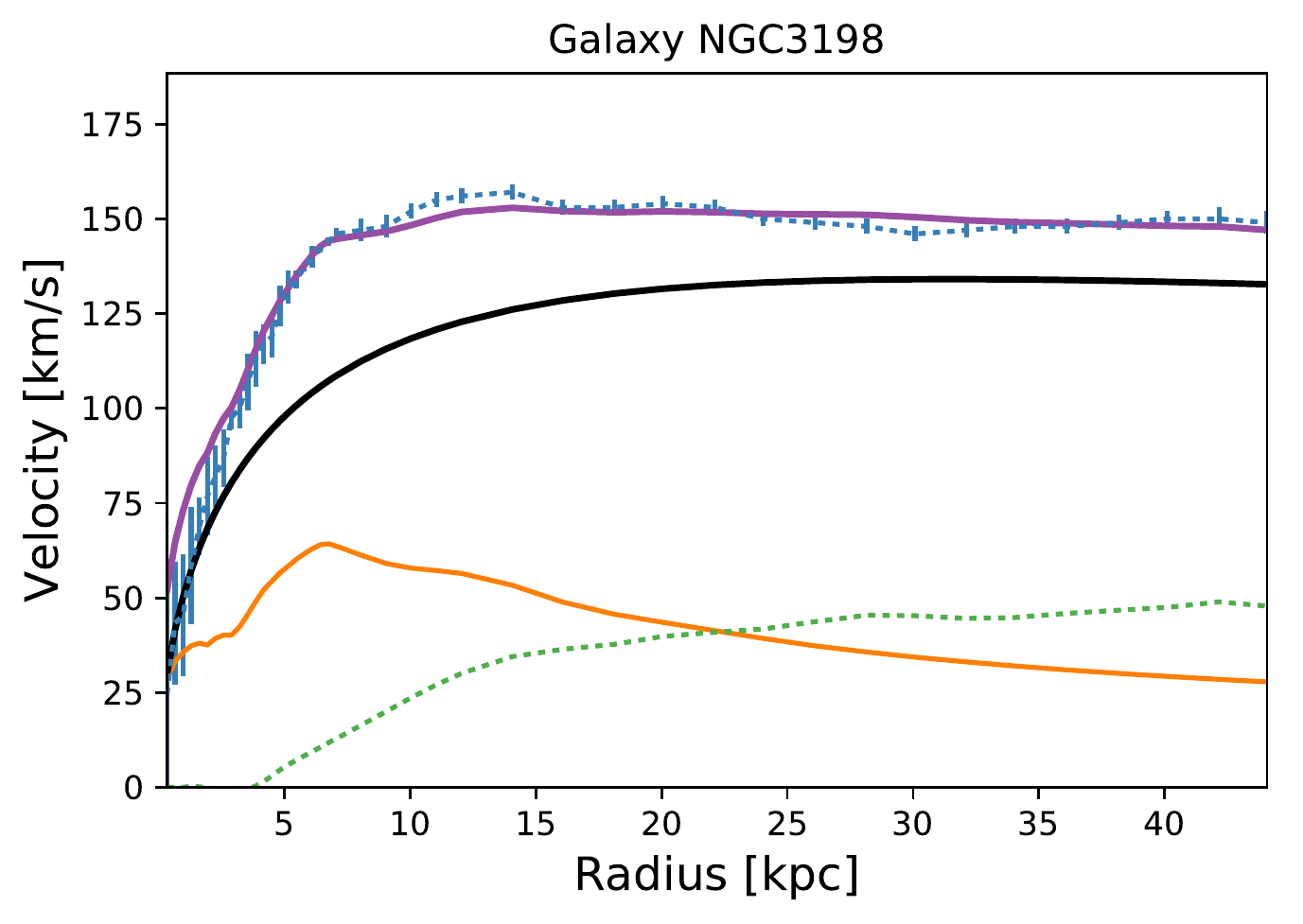}   %&   
  \includegraphics[width=0.45\textwidth]{Galaxy_NGC3198.pdf}  % \\
\caption{Best--fit rotational curves. NFW profiles on the left. BEC on the right. Legends are like in Fig.\,\,\ref{fig:fit_example}.}
\end{figure}

\begin{figure}
  \includegraphics[width=0.45\textwidth]{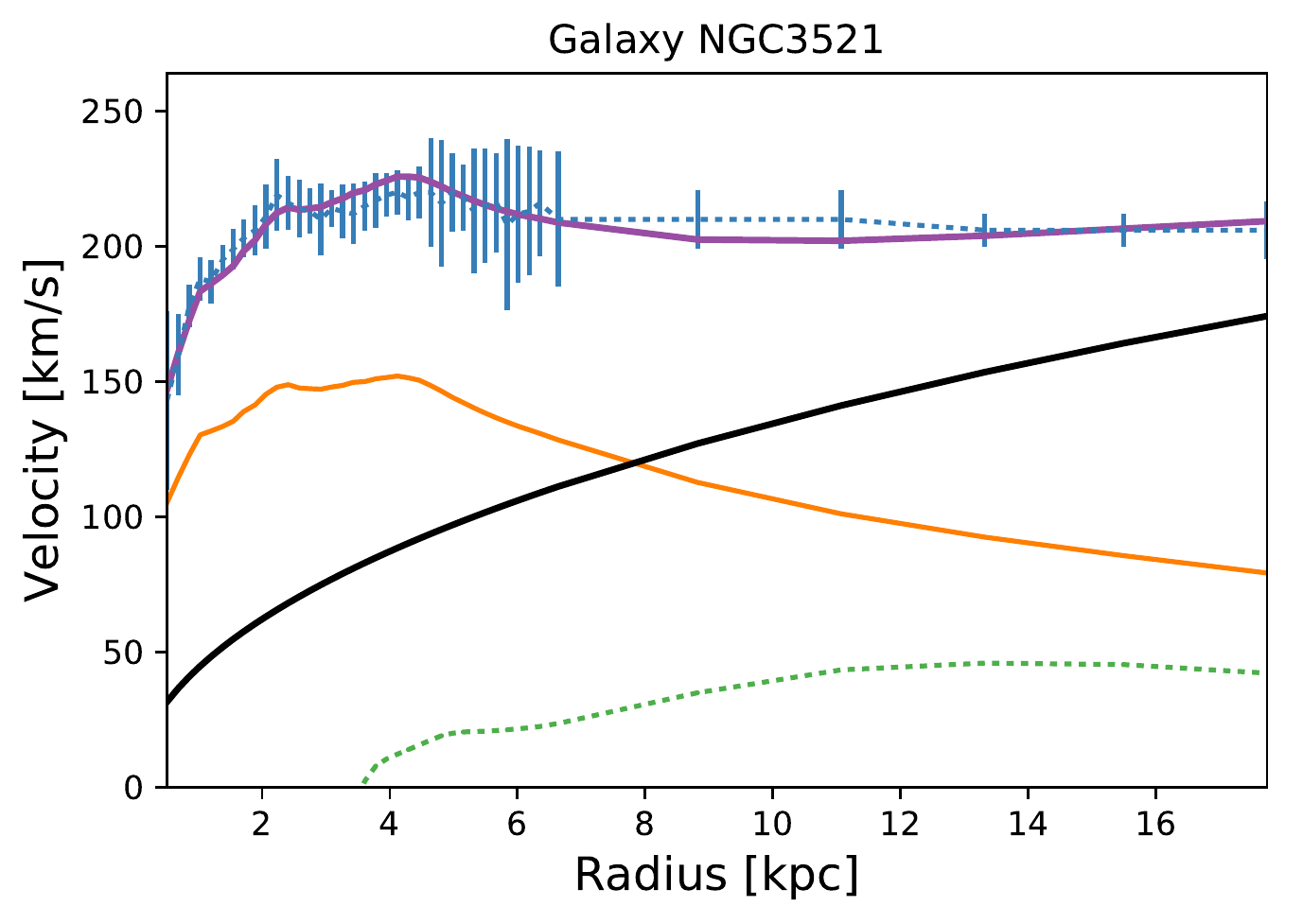}   %&   
   \includegraphics[width=0.45\textwidth]{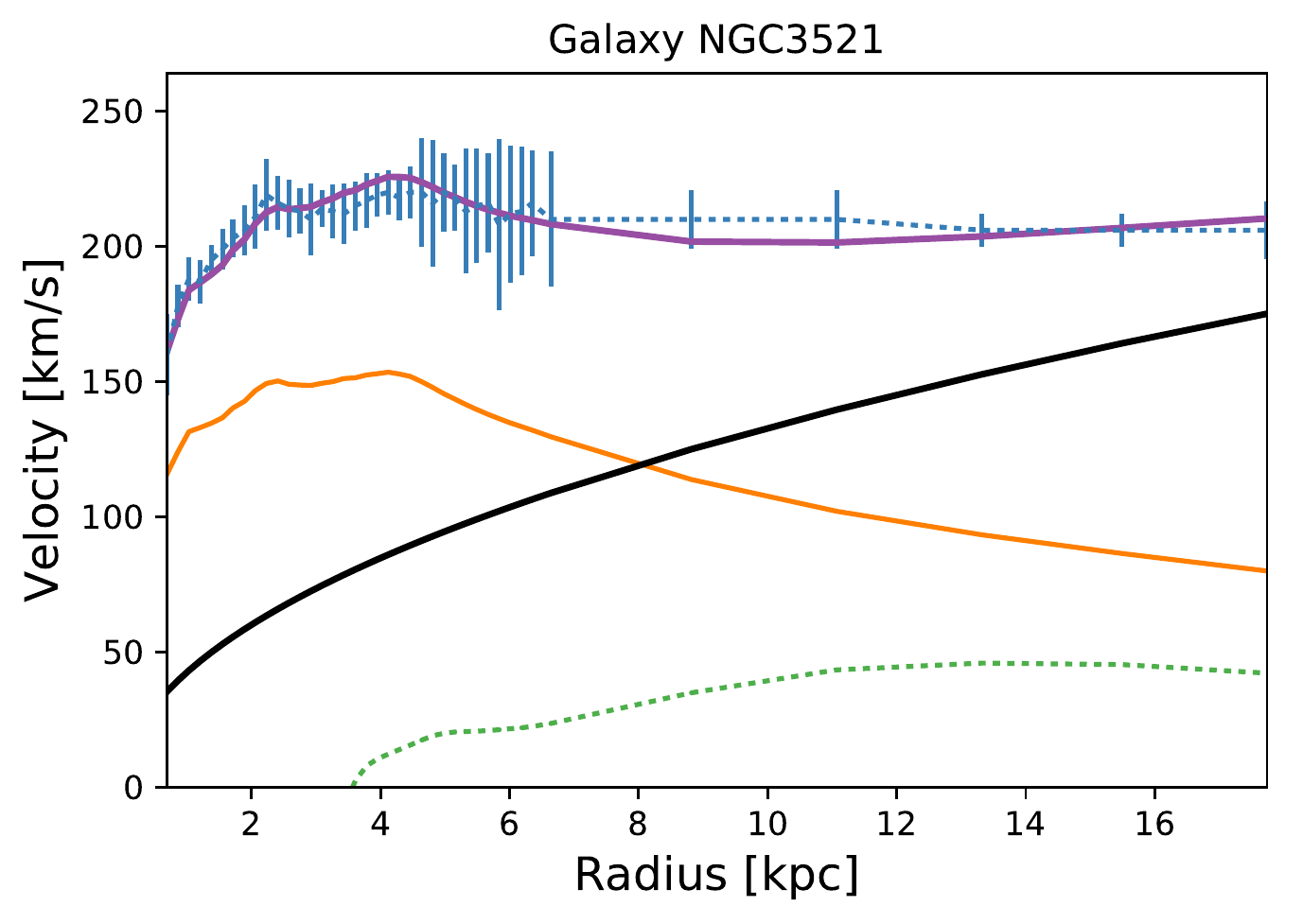}   \\
  \includegraphics[width=0.45\textwidth]{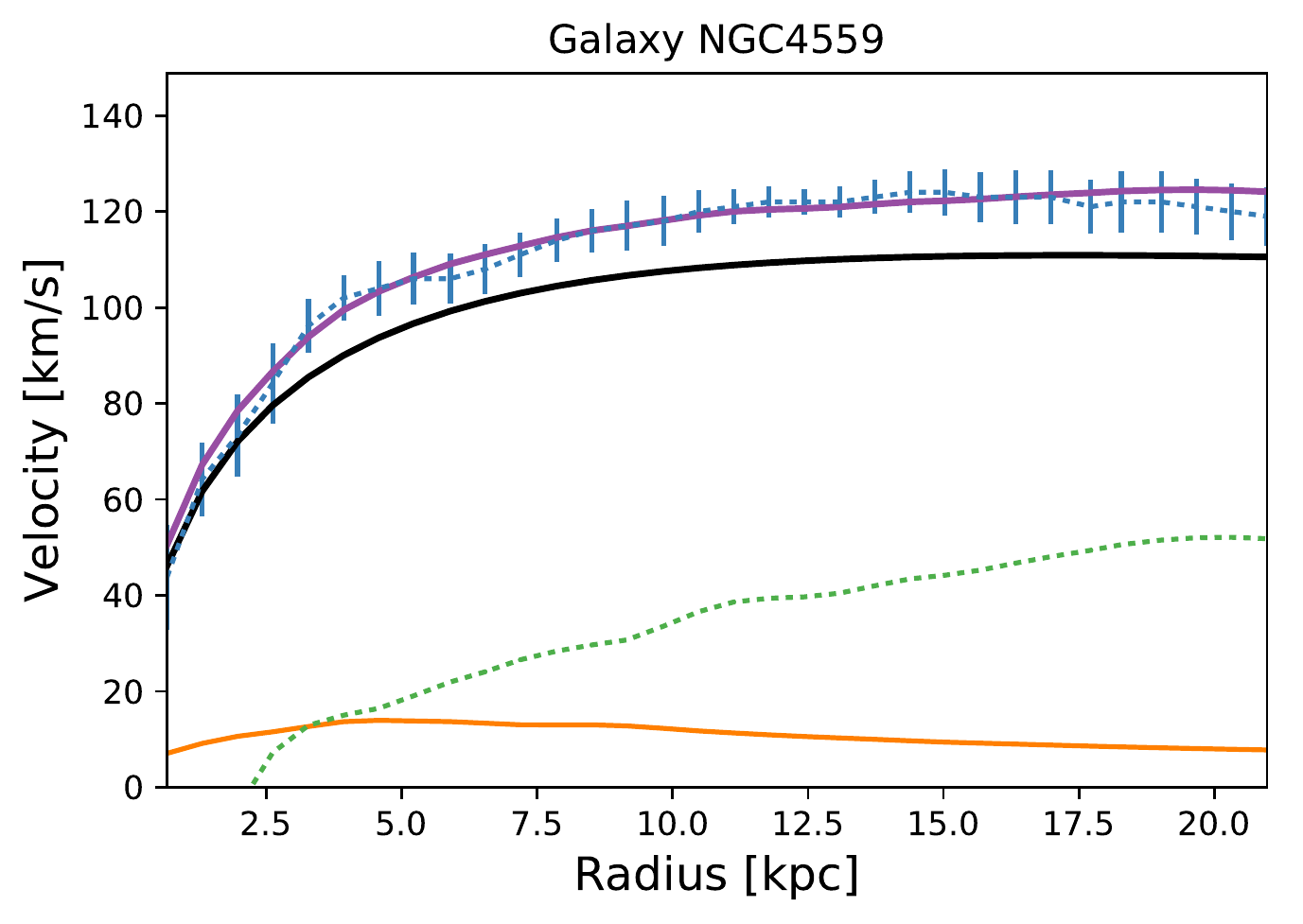}   %&   
  \includegraphics[width=0.45\textwidth]{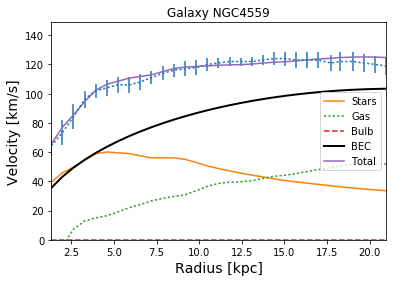}   %\\
\caption{Best--fit rotational curves. NFW profiles on the left. BEC on the right. Legends are like in Fig.\,\,\ref{fig:fit_example}. }
\end{figure}

\begin{figure}
  \includegraphics[width=0.45\textwidth]{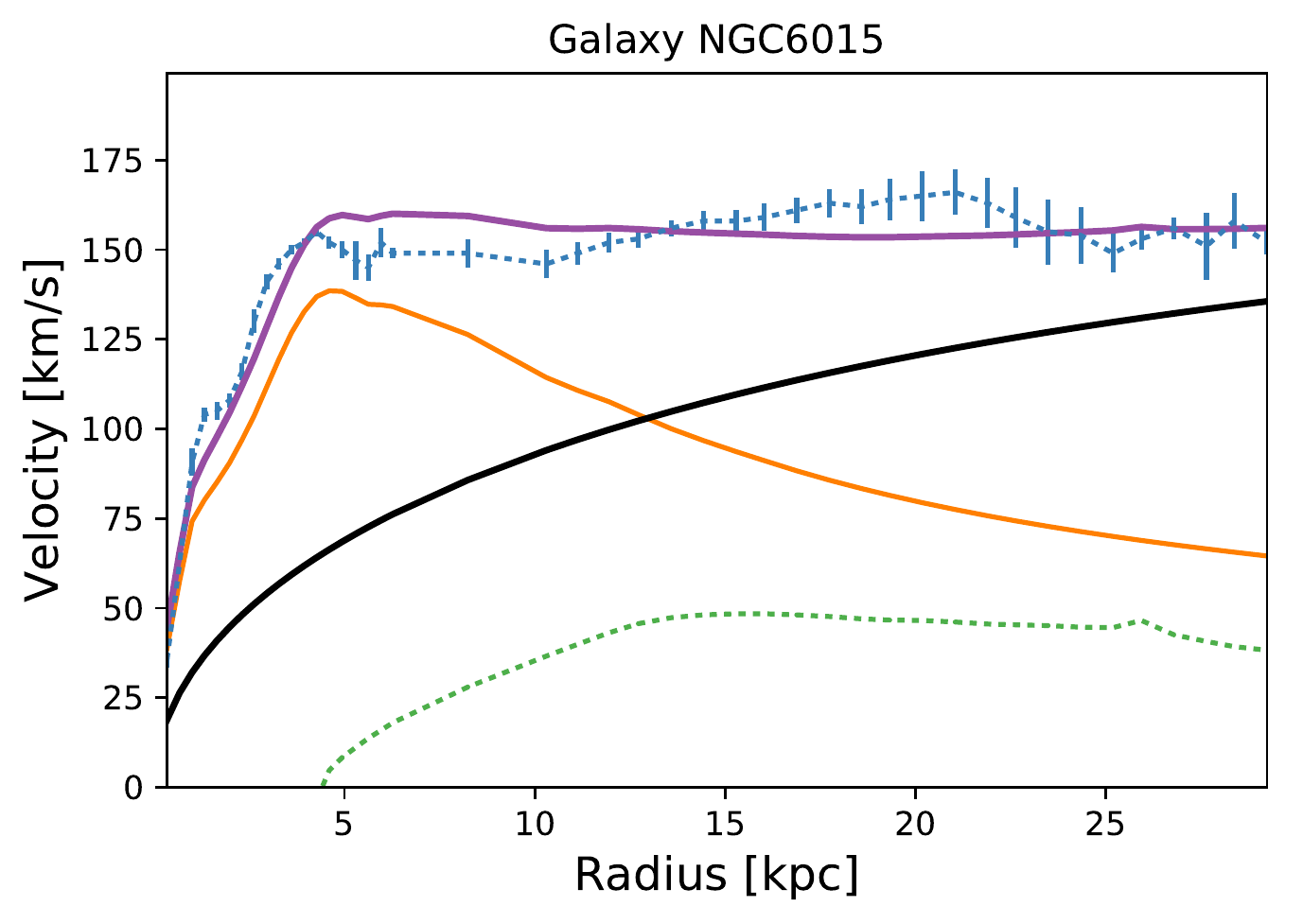}   %&   
  \includegraphics[width=0.45\textwidth]{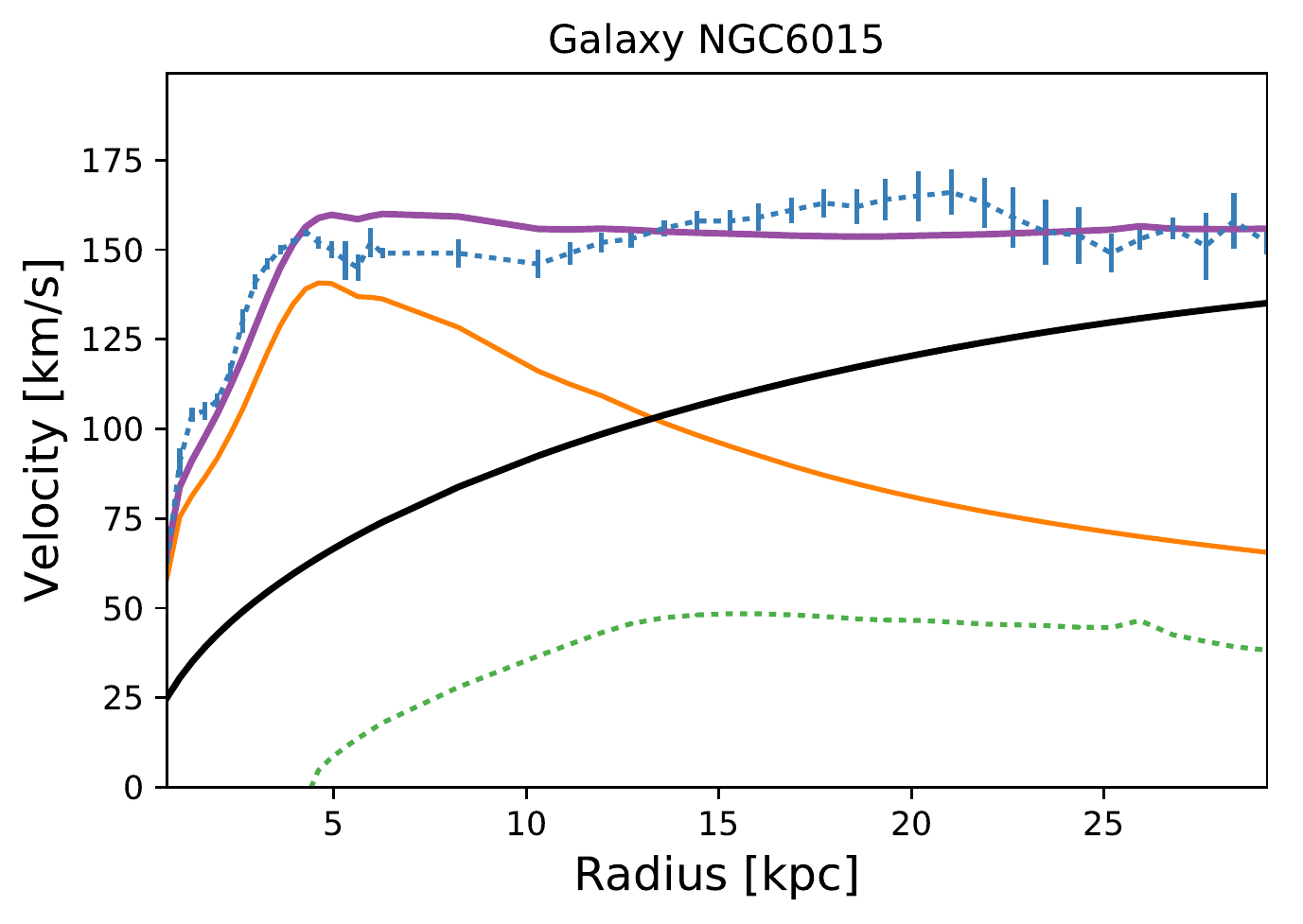}  %\\
\caption{Best--fit rotational curves. NFW profiles on the left. BEC on the right. Legends are like in Fig.\,\,\ref{fig:fit_example}. }
\end{figure}

\begin{figure}
  \includegraphics[width=0.45\textwidth]{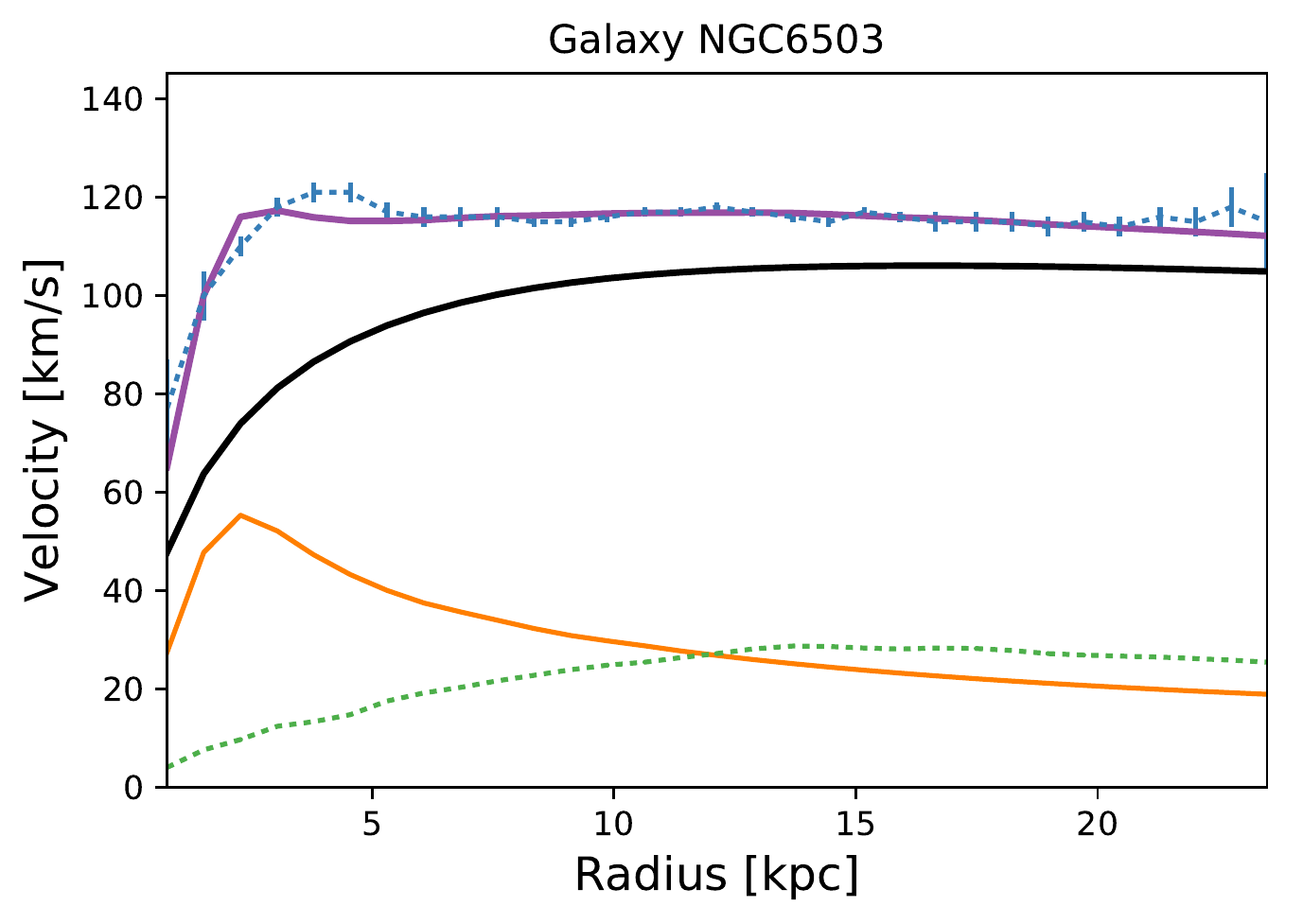}  %&   
  \includegraphics[width=0.45\textwidth]{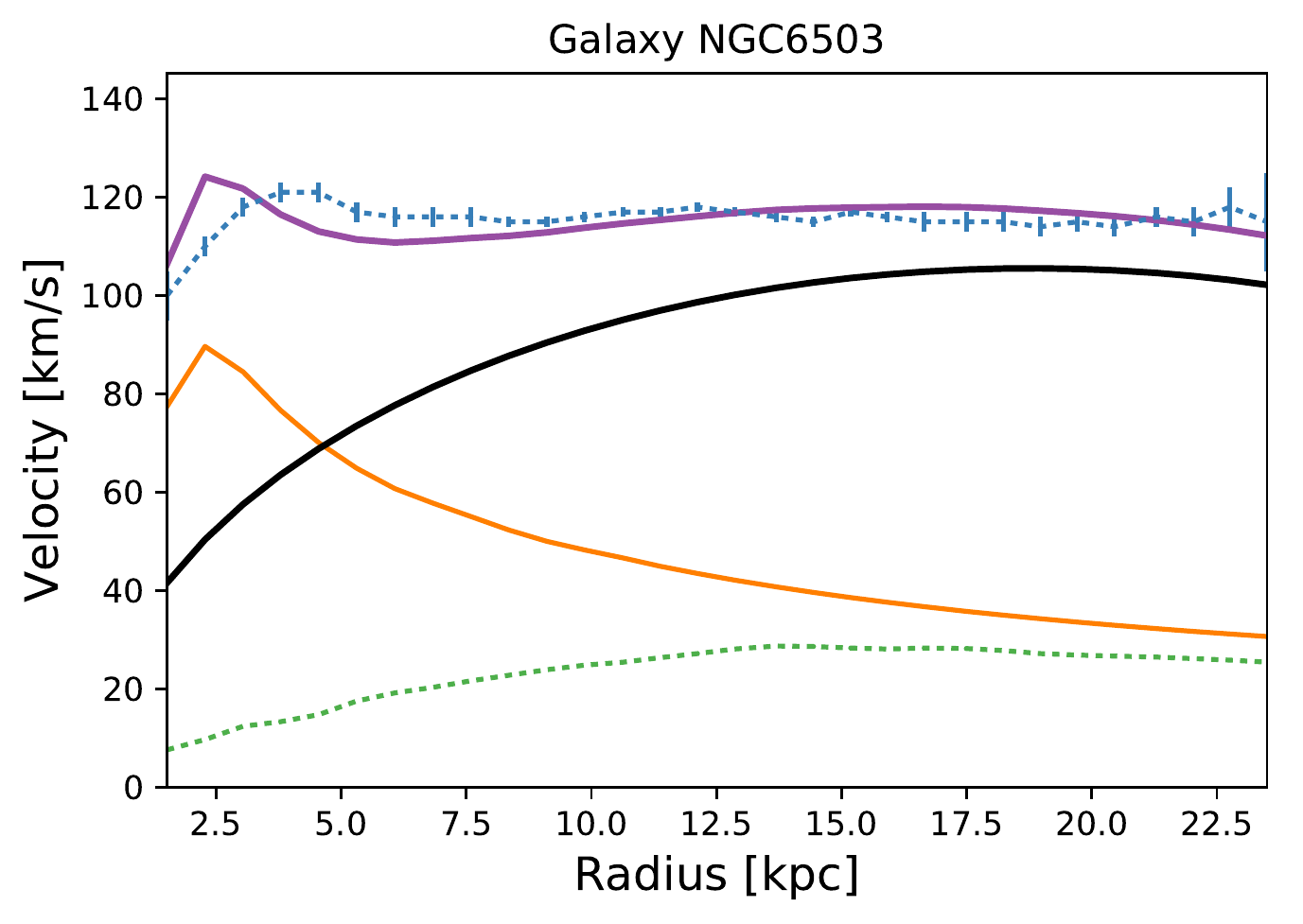}   %\\
  \includegraphics[width=0.45\textwidth]{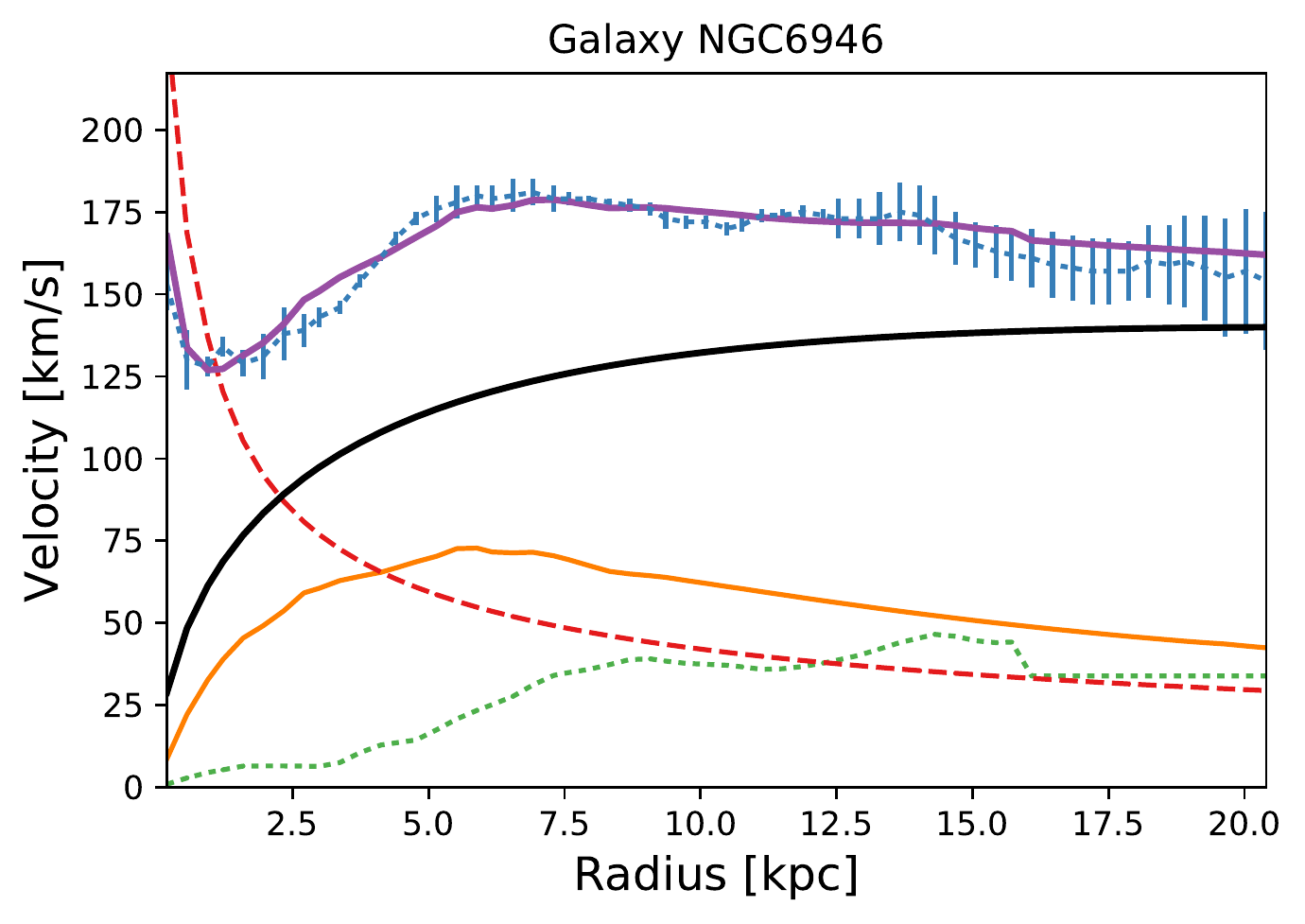} %&   
   \includegraphics[width=0.45\textwidth]{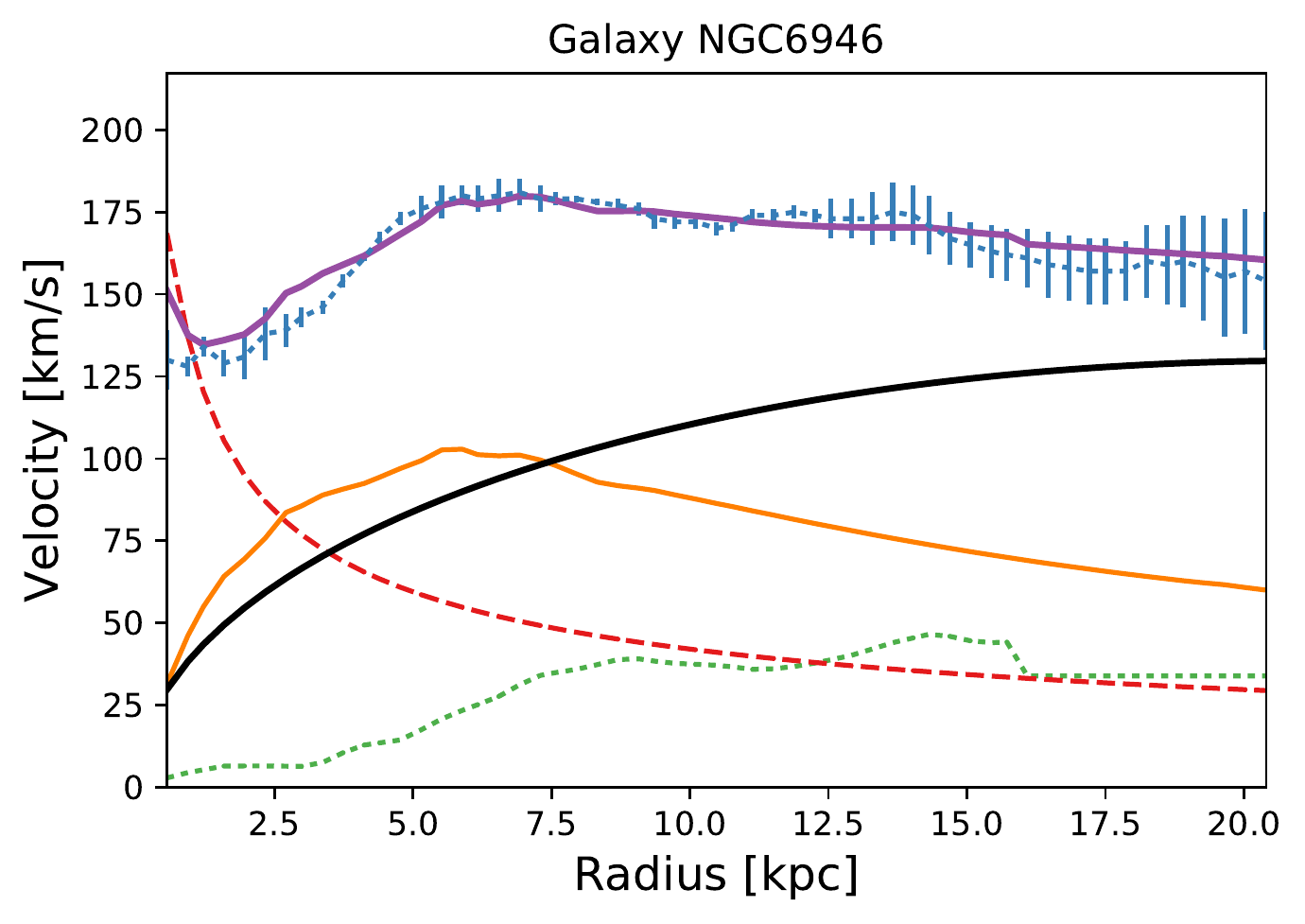}   %\\
\caption{Best--fit rotational curves. NFW profiles on the left. BEC on the right. Legends are like in Fig.\,\,\ref{fig:fit_example}. }
\end{figure}

\begin{figure}
  \includegraphics[width=0.45\textwidth]{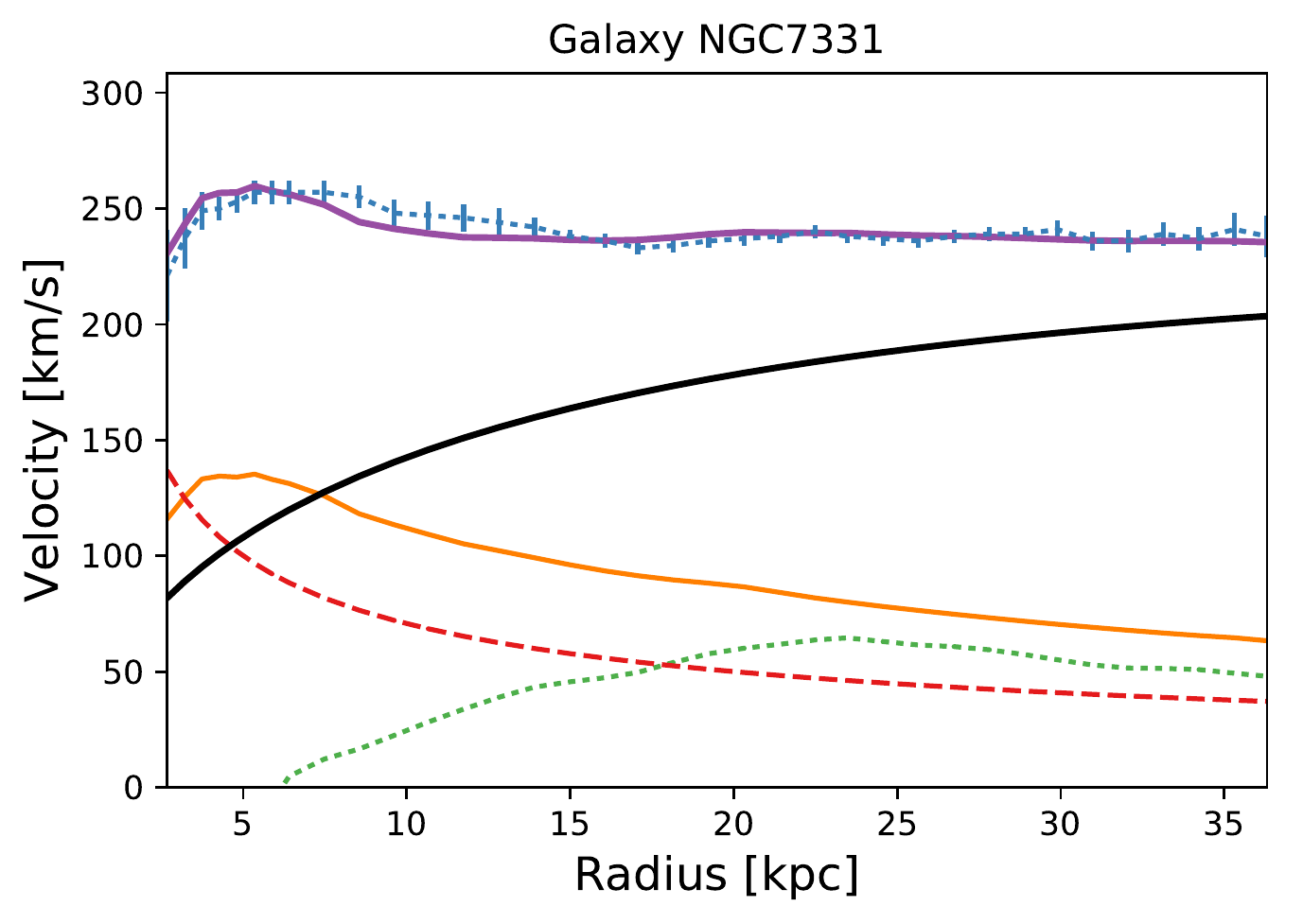}   %&   
   \includegraphics[width=0.45\textwidth]{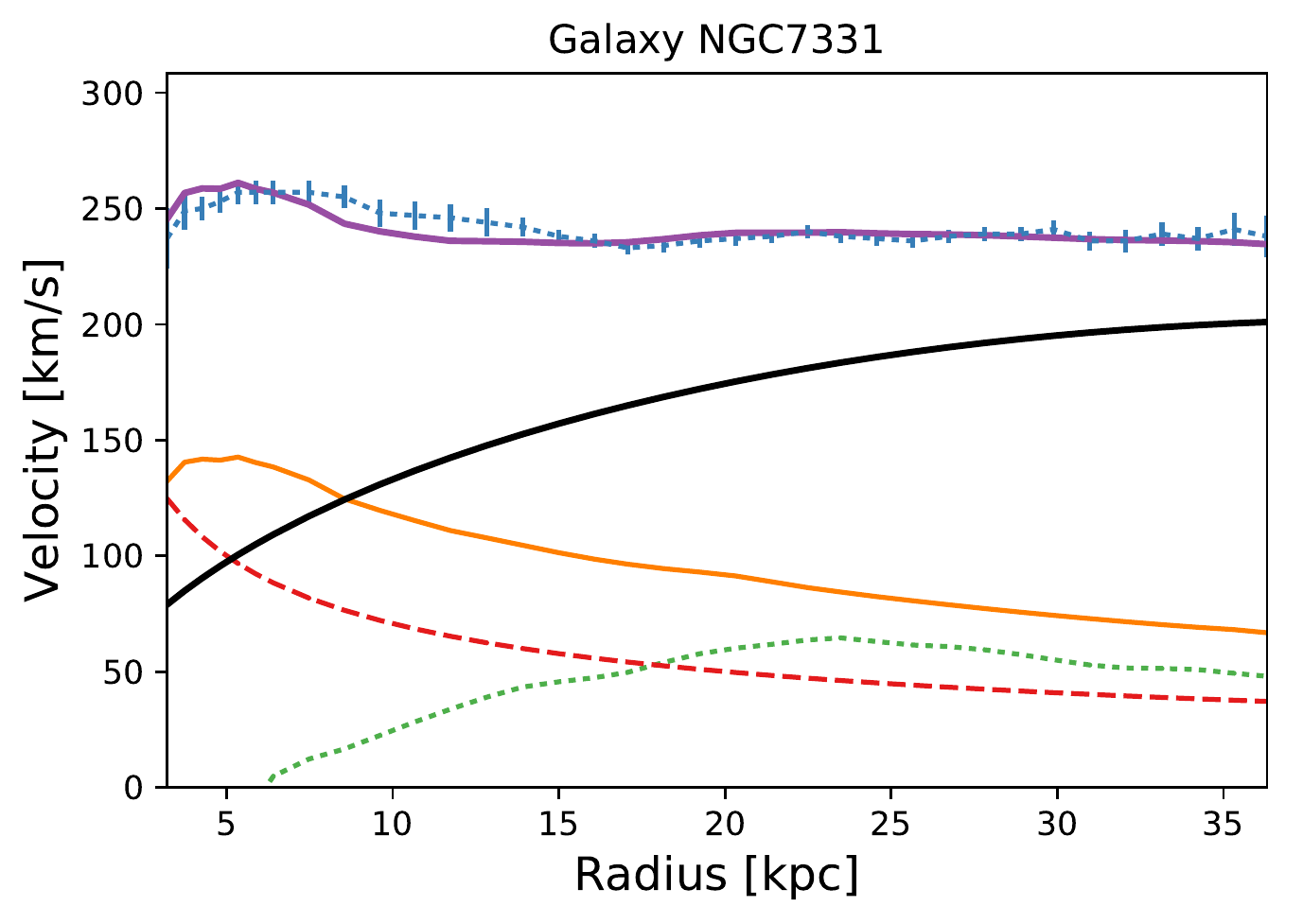}   \\
  \includegraphics[width=0.45\textwidth]{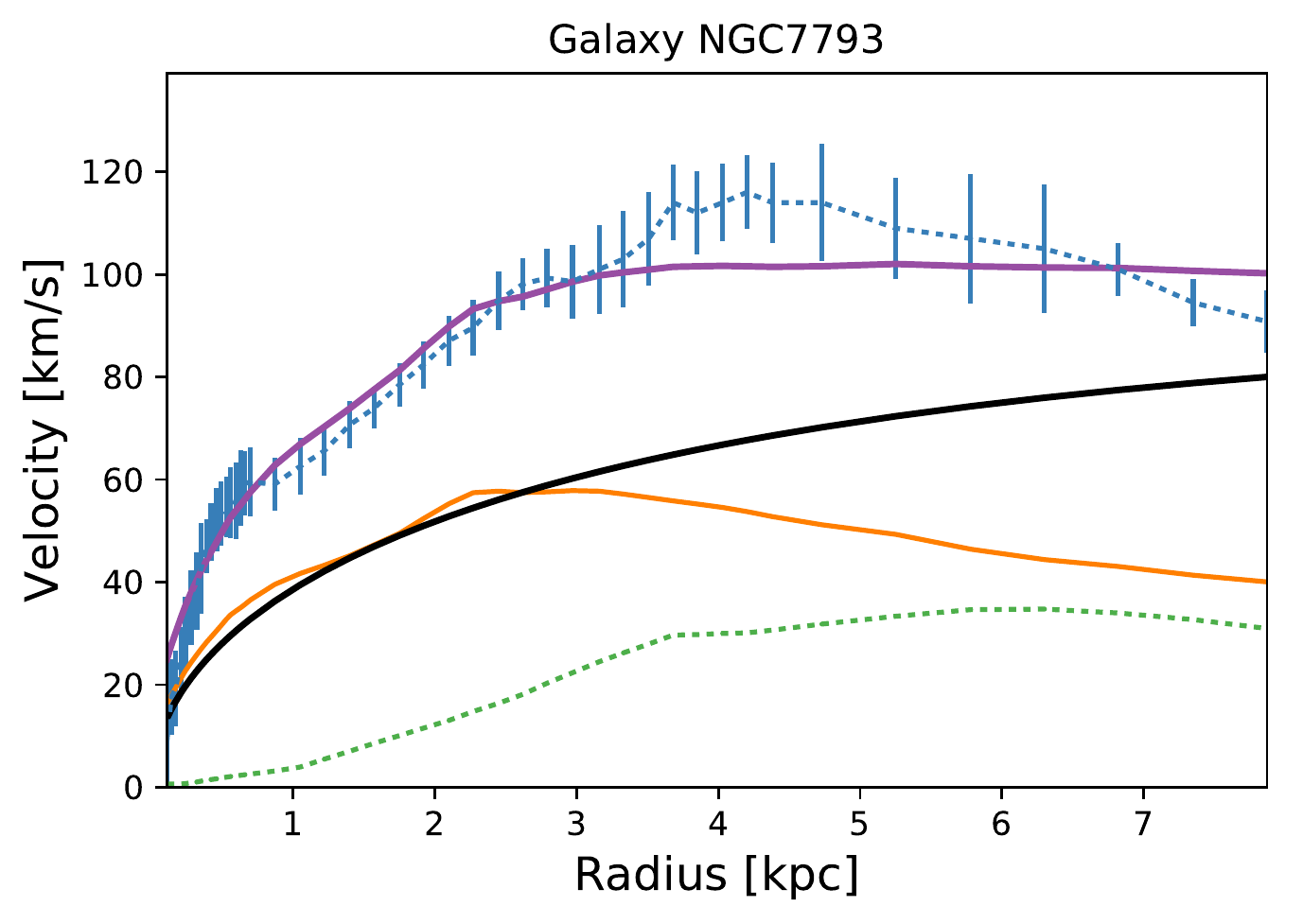}  %&   
  \includegraphics[width=0.45\textwidth]{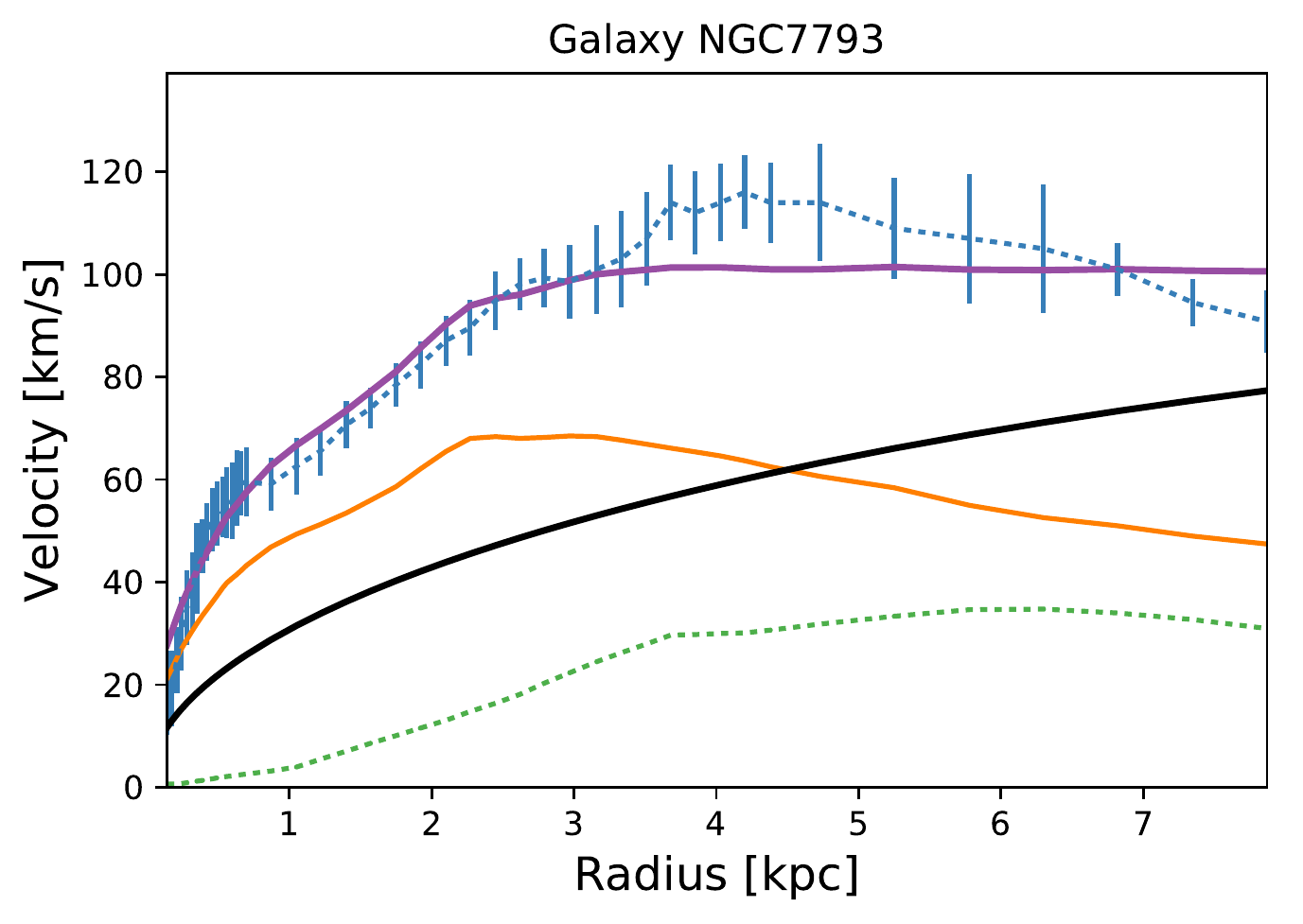}  \\
\caption{Best--fit rotational curves. NFW profiles on the left. BEC on the right. Legends are like in Fig.\,\,\ref{fig:fit_example}. }

\end{figure}

\begin{figure}
 \includegraphics[width=0.45\textwidth]{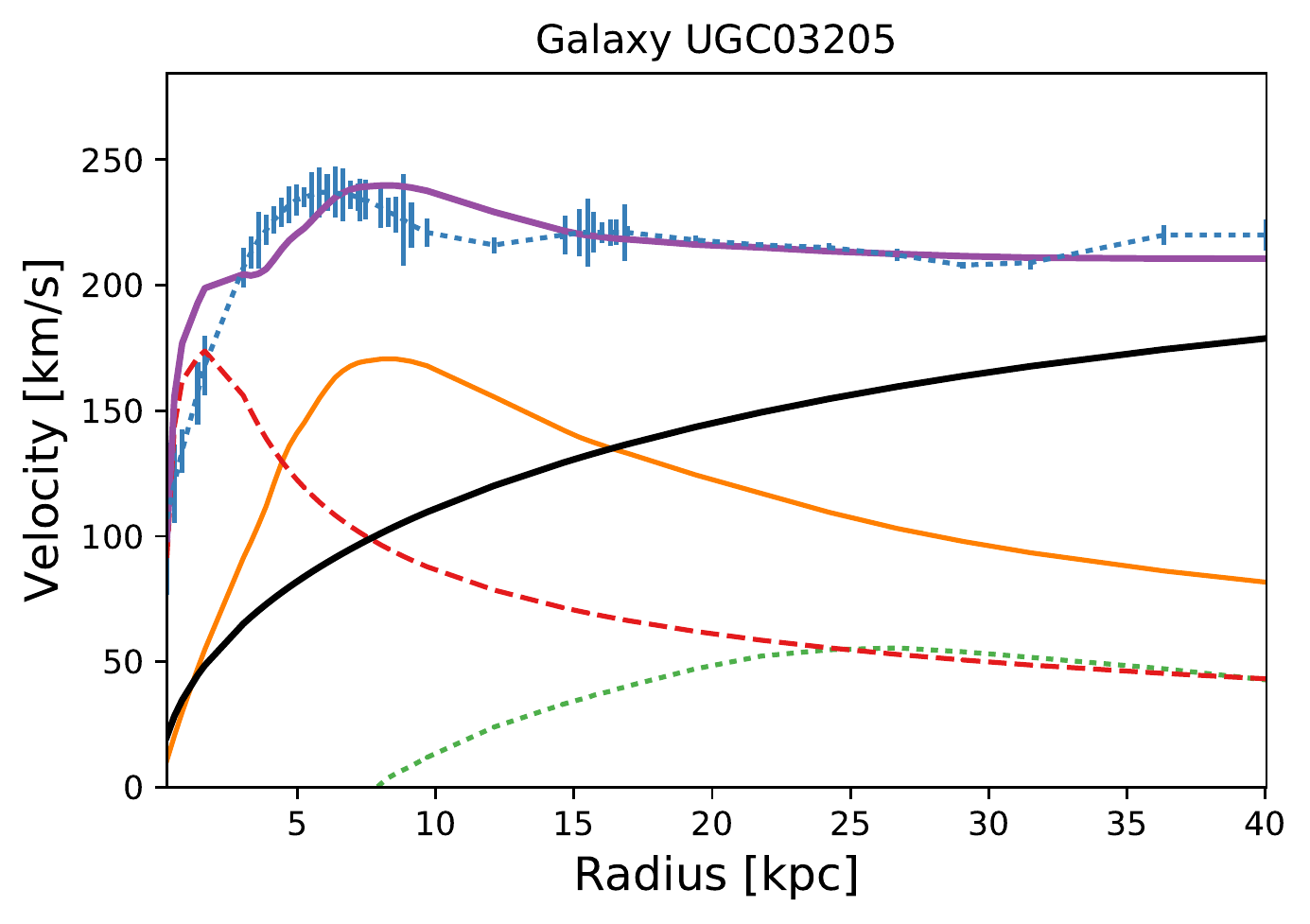}   %&   
  \includegraphics[width=0.45\textwidth]{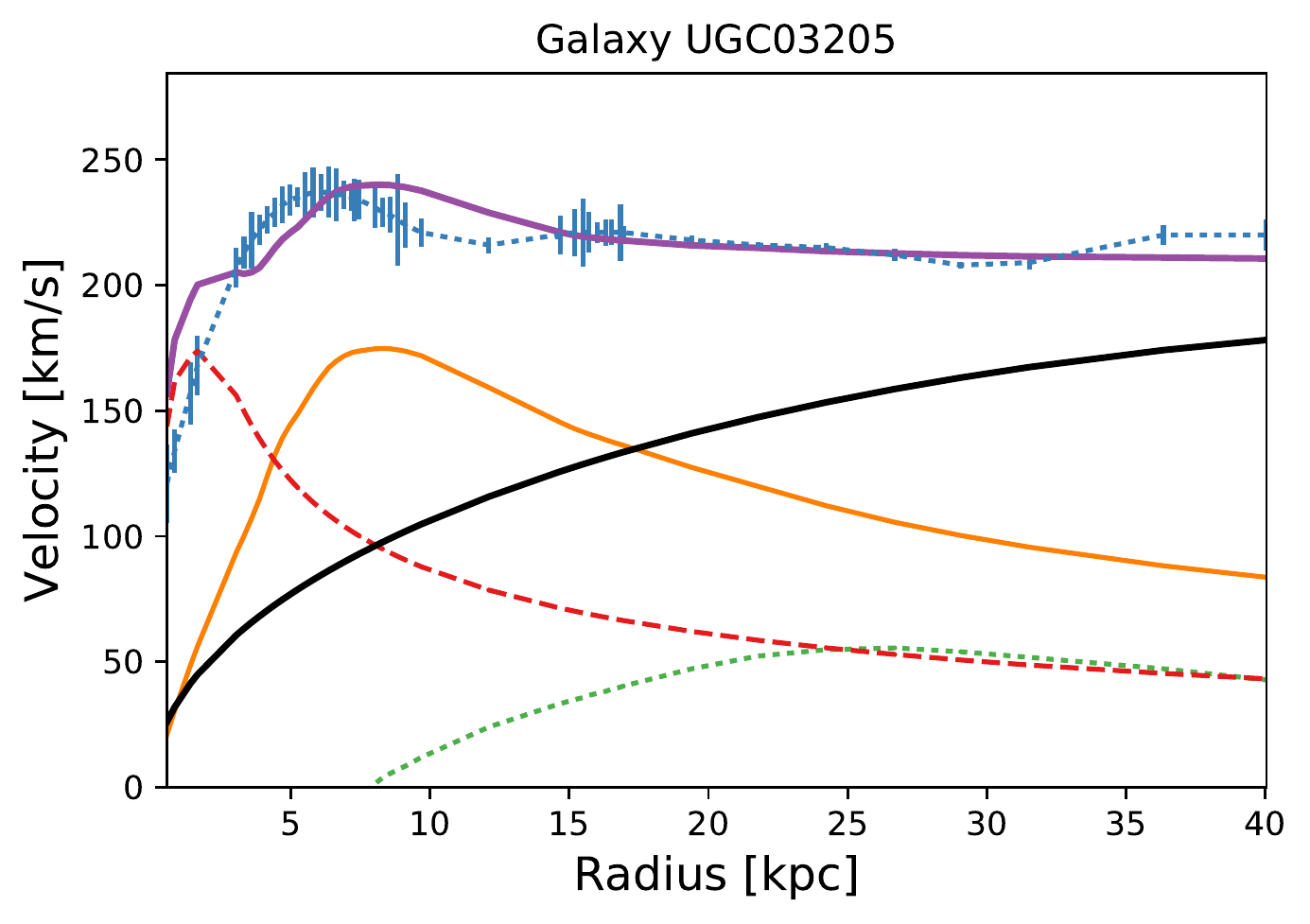} \\
  \includegraphics[width=0.45\textwidth]{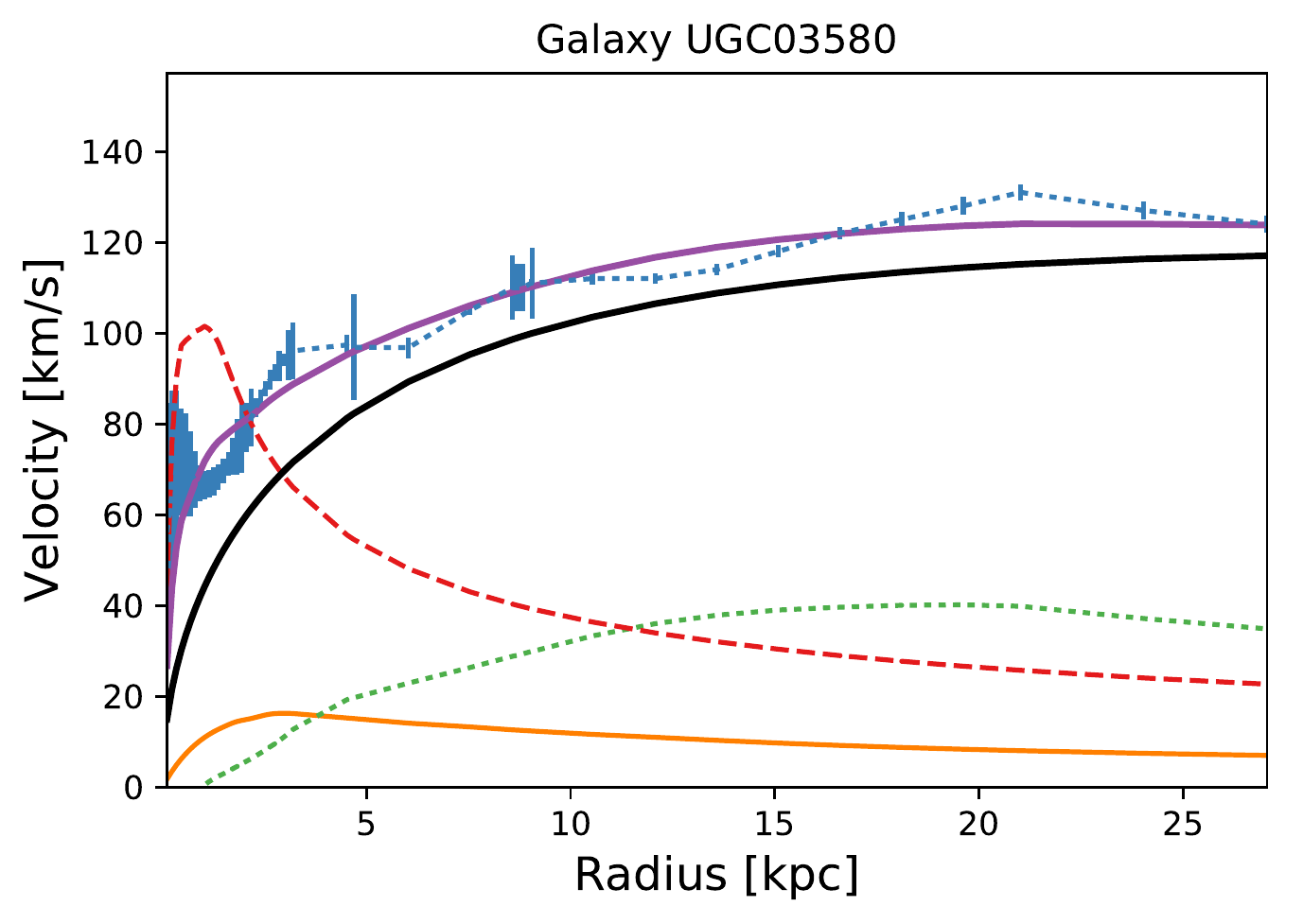}  %&   
  \includegraphics[width=0.45\textwidth]{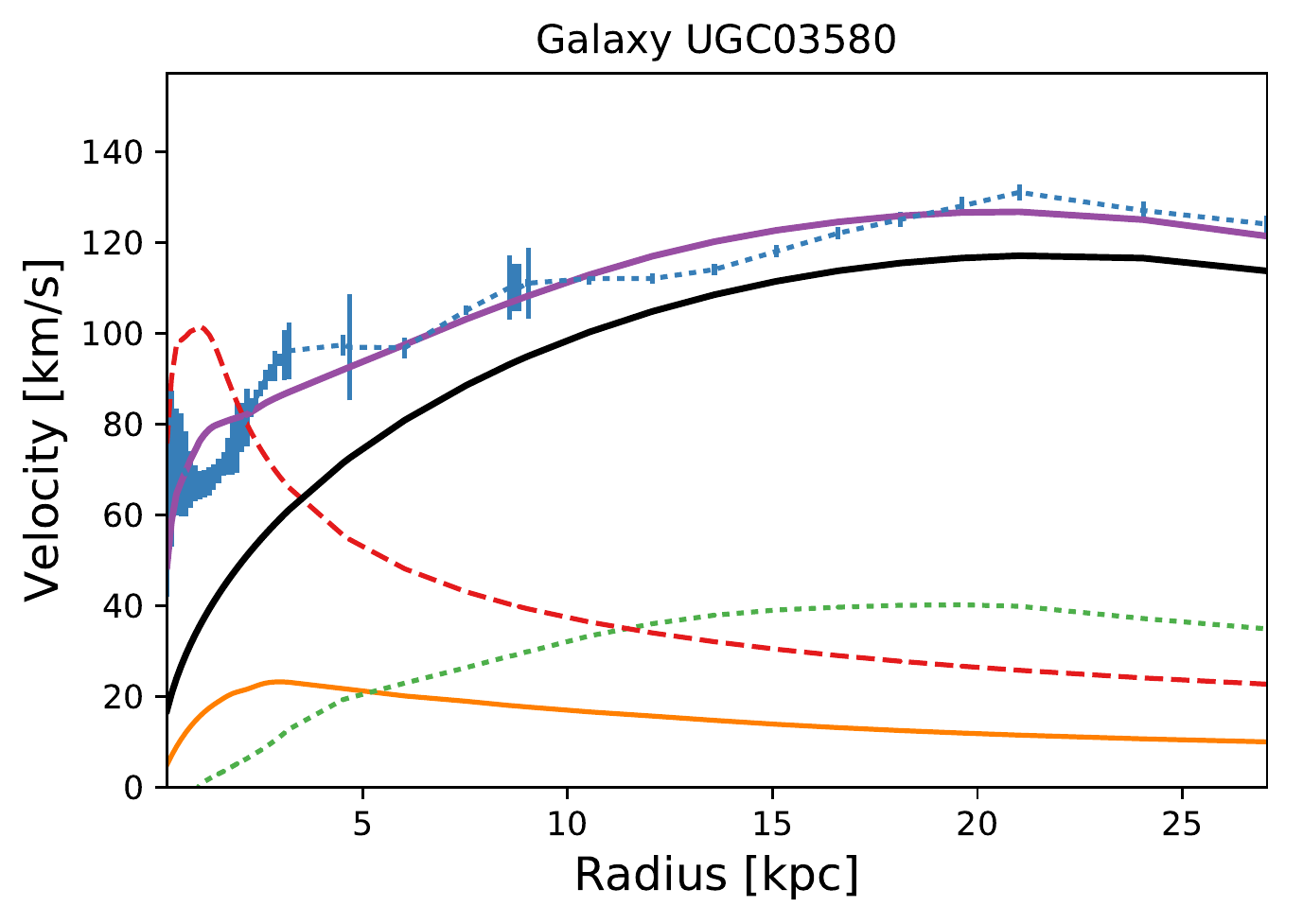}  % \\
  \caption{Best--fit rotational curves. NFW profiles on the left. BEC on the right. Legends are like in Fig.\,\,\ref{fig:fit_example}. }
\end{figure}

\begin{figure}
   \includegraphics[width=0.45\textwidth]{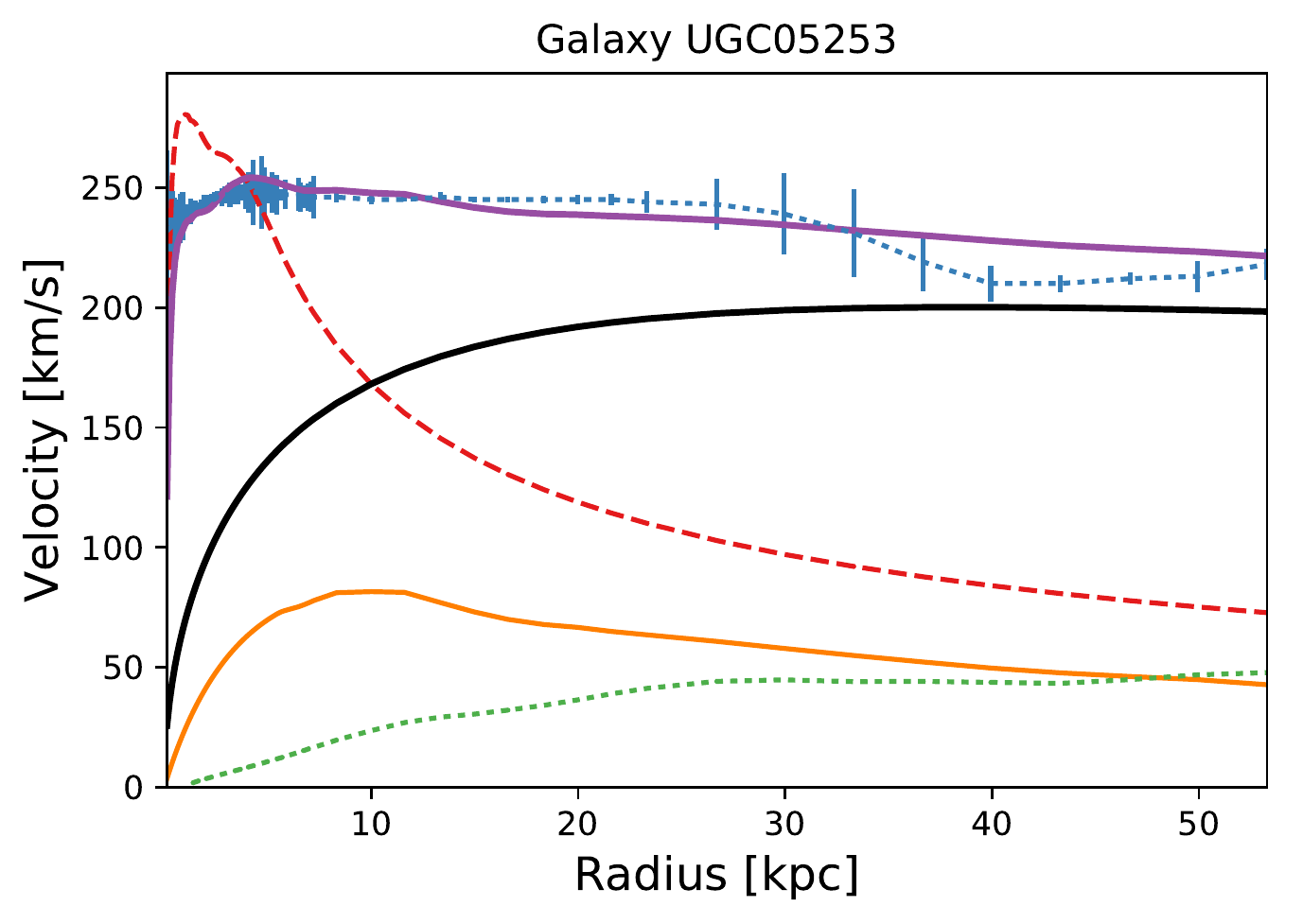} %&   
   \includegraphics[width=0.45\textwidth]{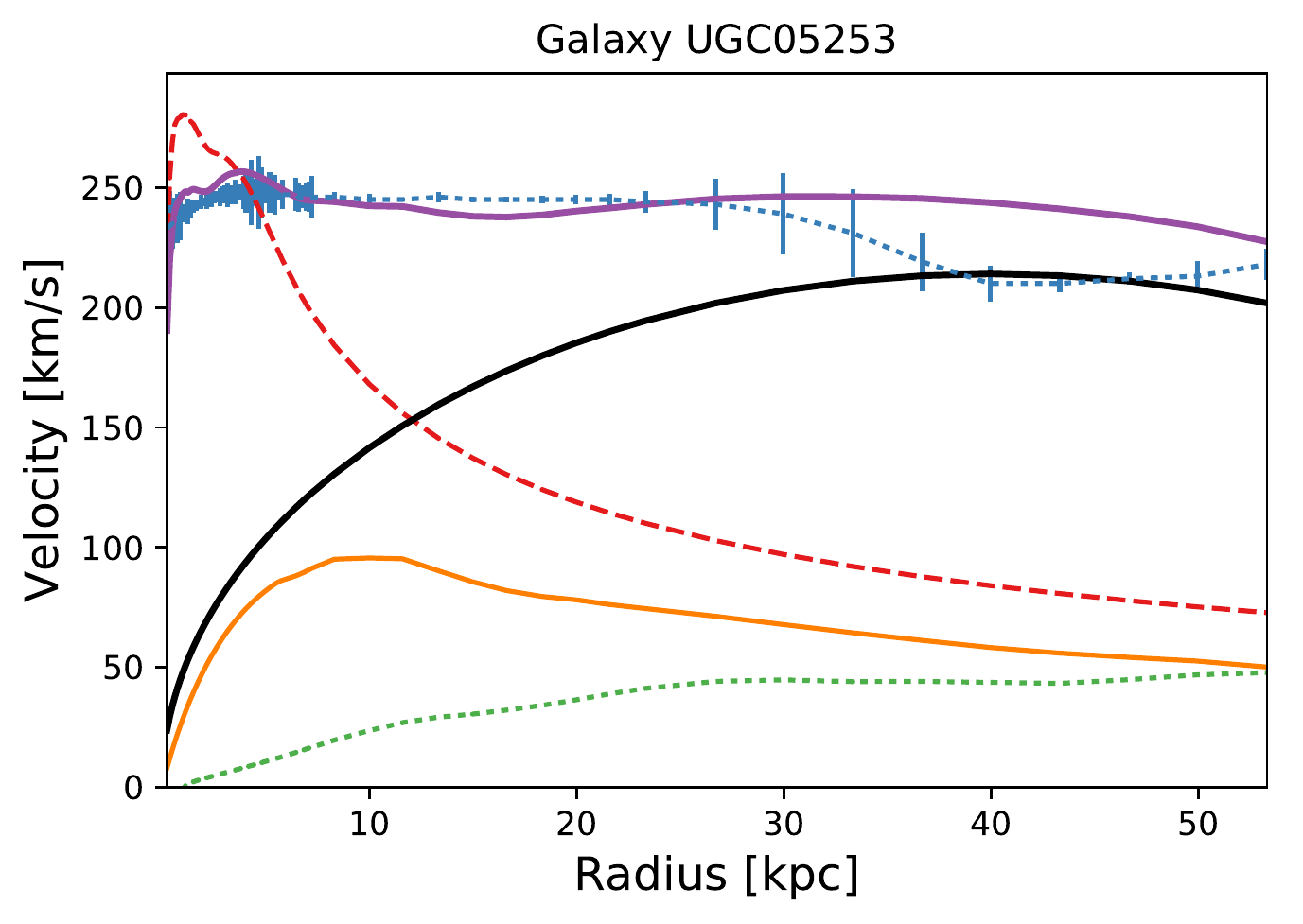}   %\\
\caption{Best--fit rotational curves. NFW profiles on the left. BEC on the right. Legends are like in Fig.\,\,\ref{fig:fit_example}. }
\end{figure}

\begin{figure}
  \includegraphics[width=0.45\textwidth]{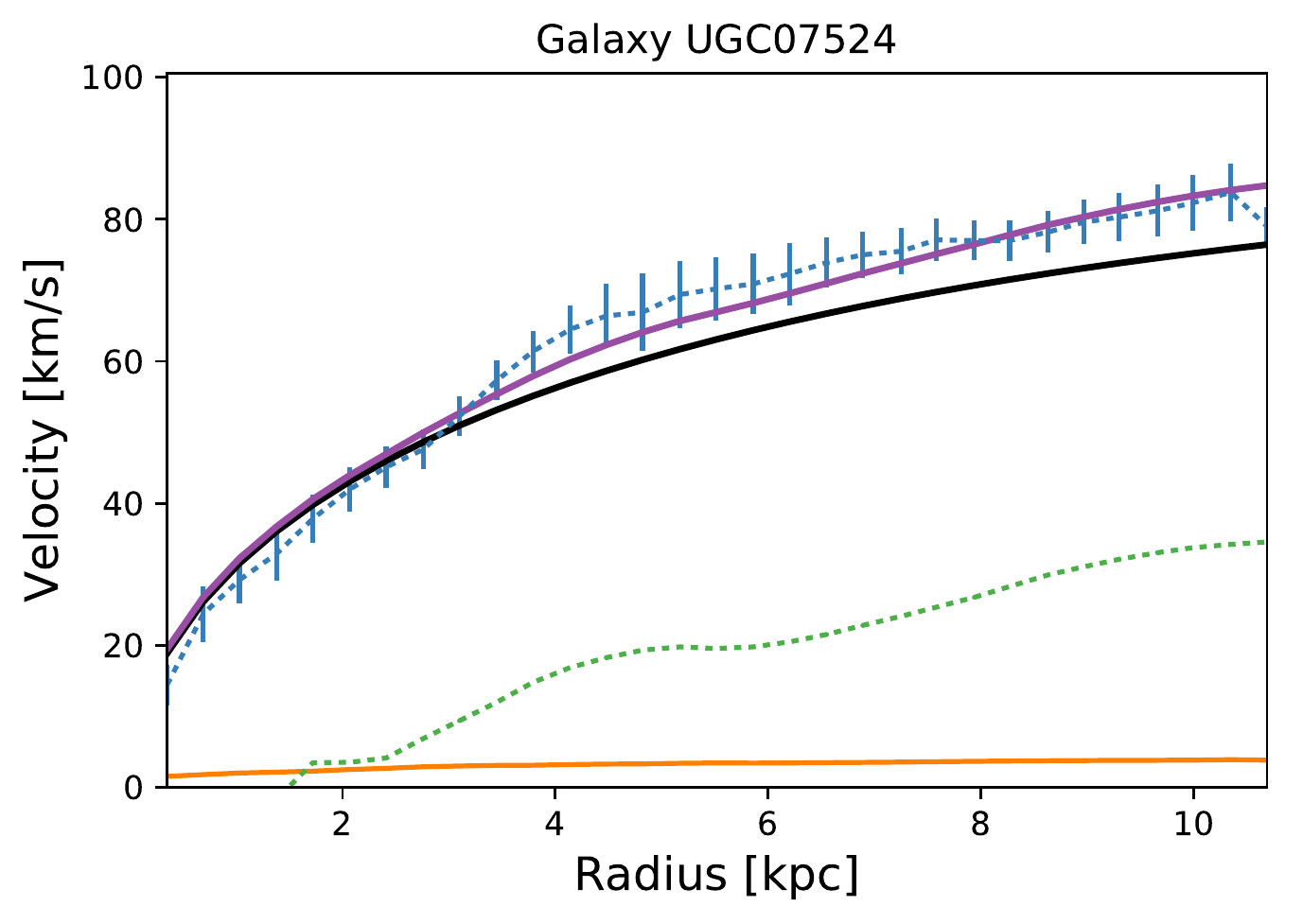}   %&   
   \includegraphics[width=0.45\textwidth]{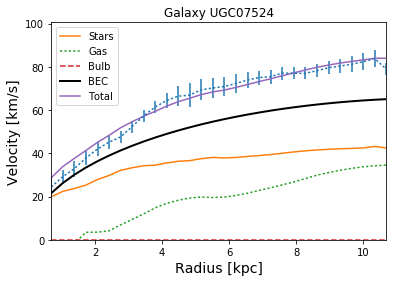}  \\
  \includegraphics[width=0.45\textwidth]{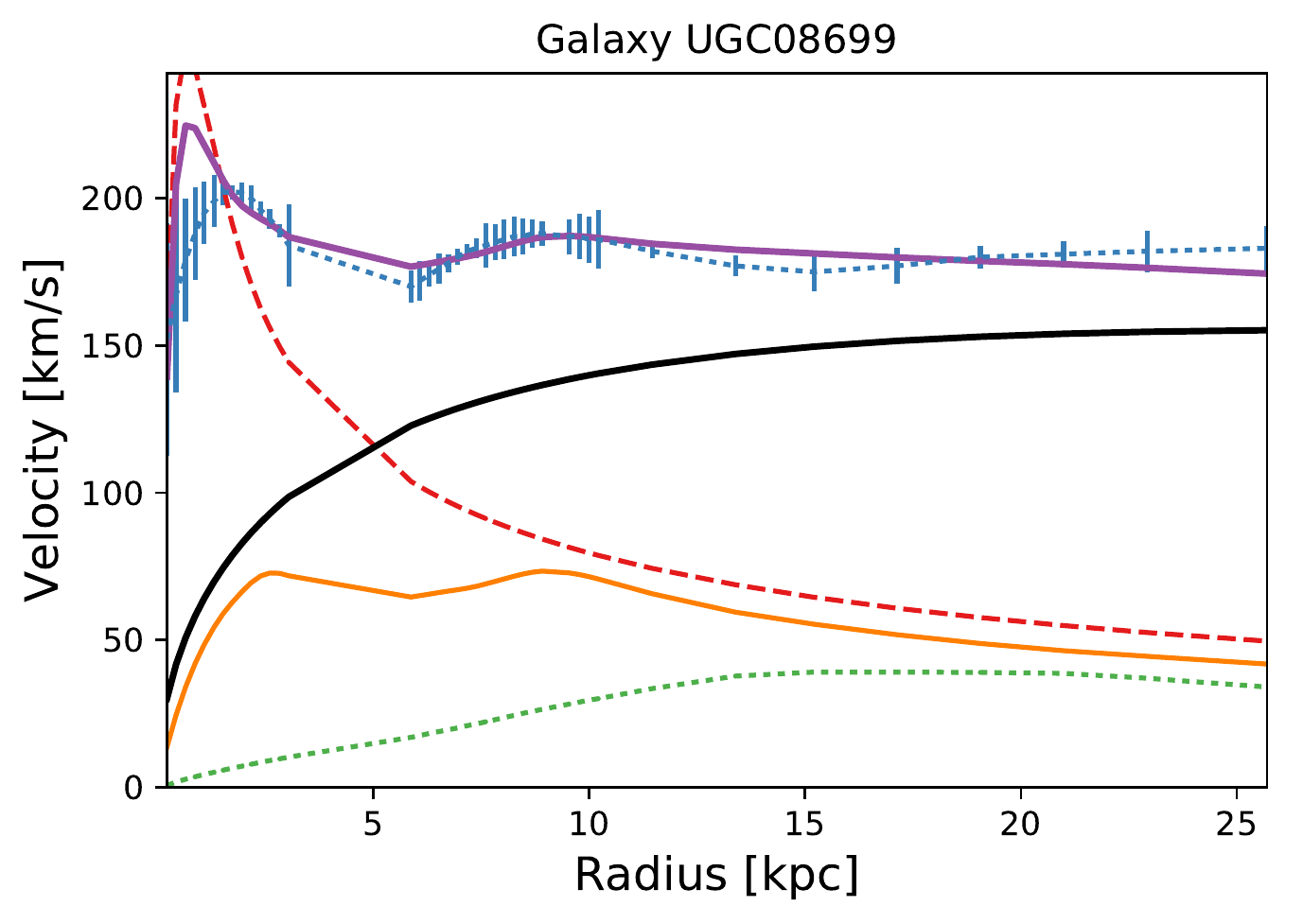} %&   
  \includegraphics[width=0.45\textwidth]{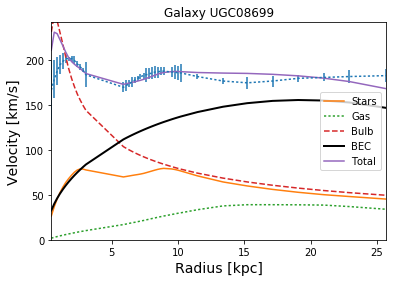}  %\\
  \caption{Best--fit rotational curves. NFW profiles on the left. BEC on the right. Legends are like in Fig.\,\,\ref{fig:fit_example}. }
  \end{figure}
  
  \begin{figure}
 \includegraphics[width=0.45\textwidth]{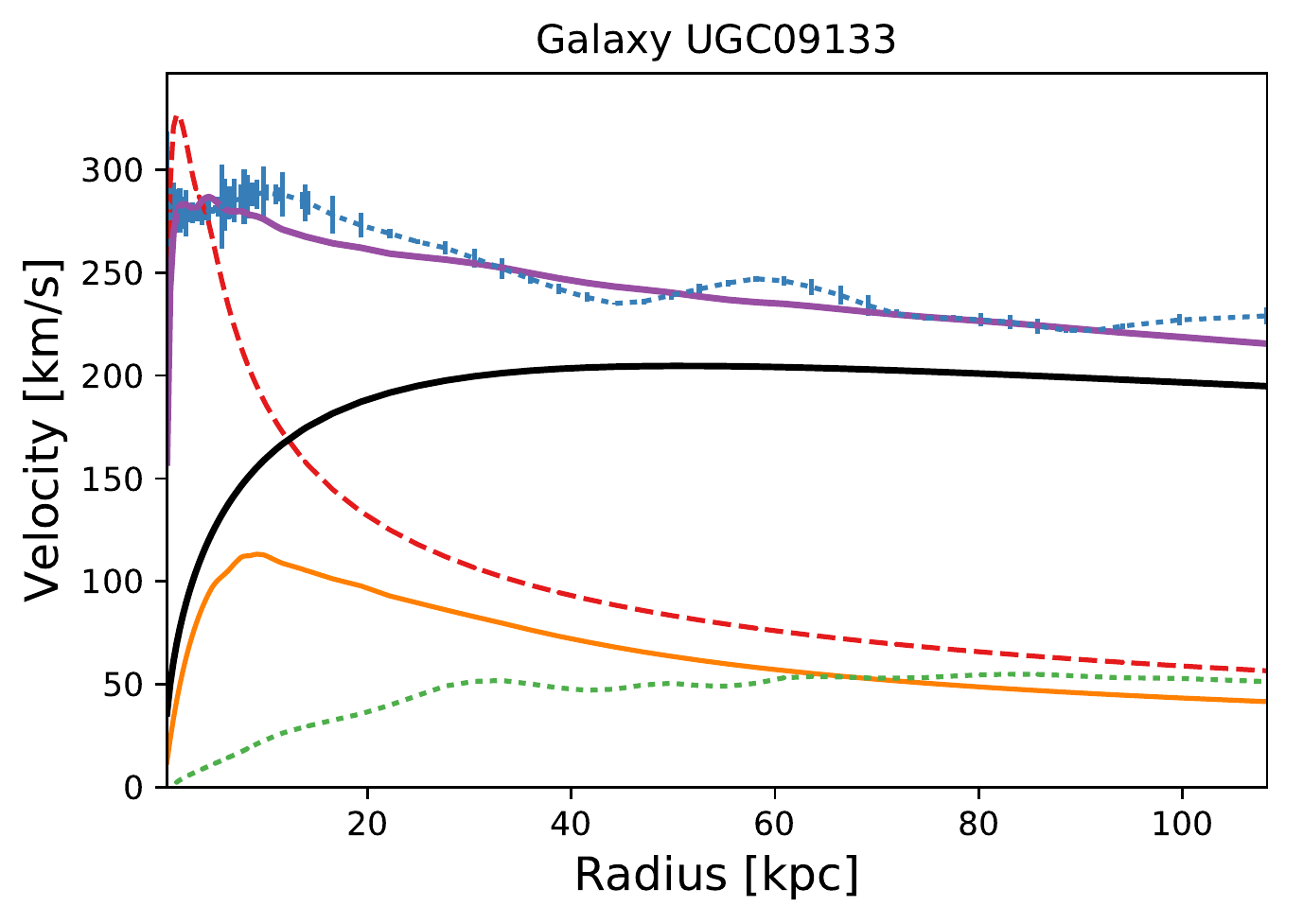}   %&   
  \includegraphics[width=0.45\textwidth]{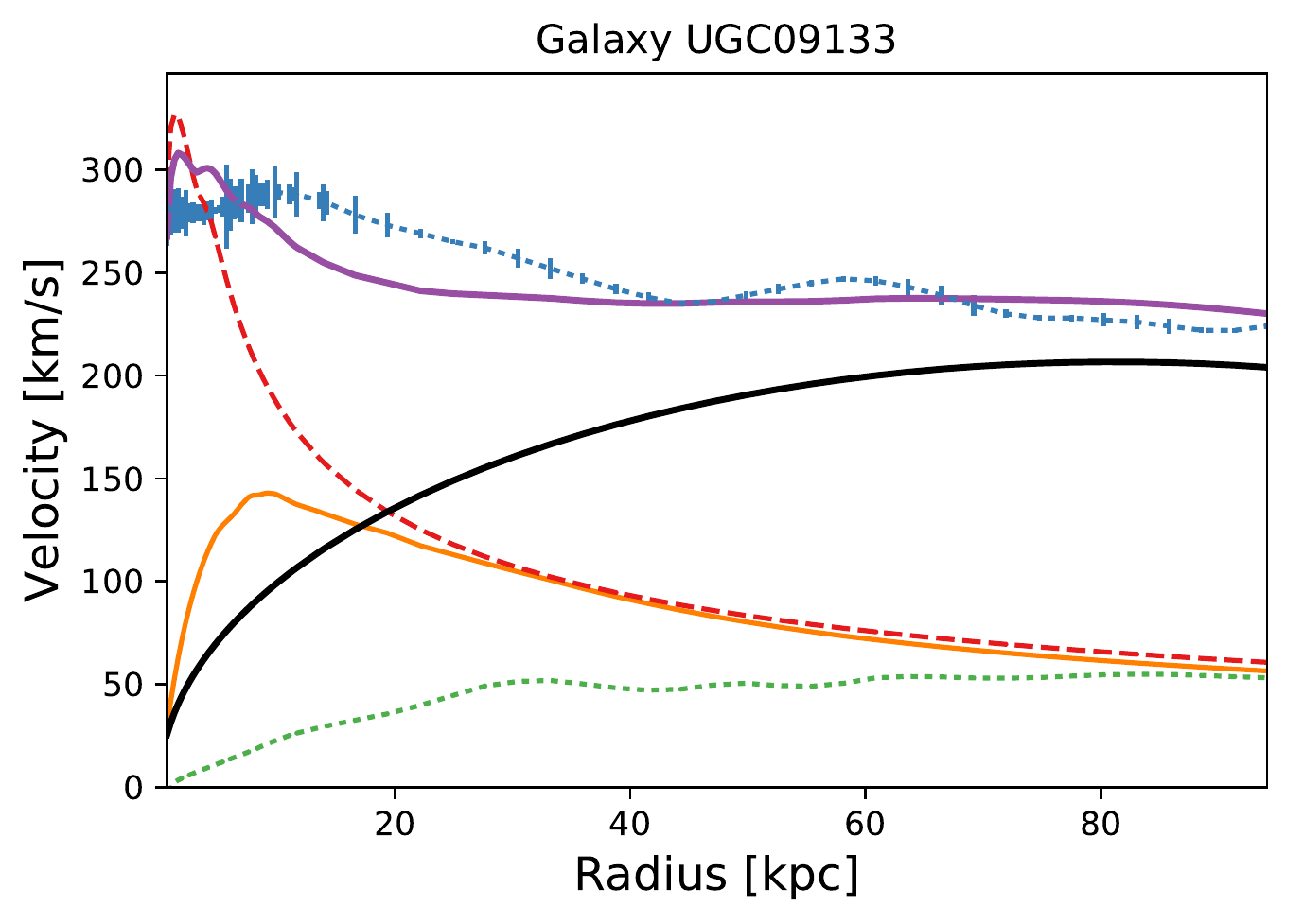} %\\
  \caption{Best--fit rotational curves. NFW profiles on the left. BEC on the right. Legends are like in Fig.\,\,\ref{fig:fit_example}. }
  \end{figure}
  
  \begin{figure}
  \includegraphics[width=0.45\textwidth]{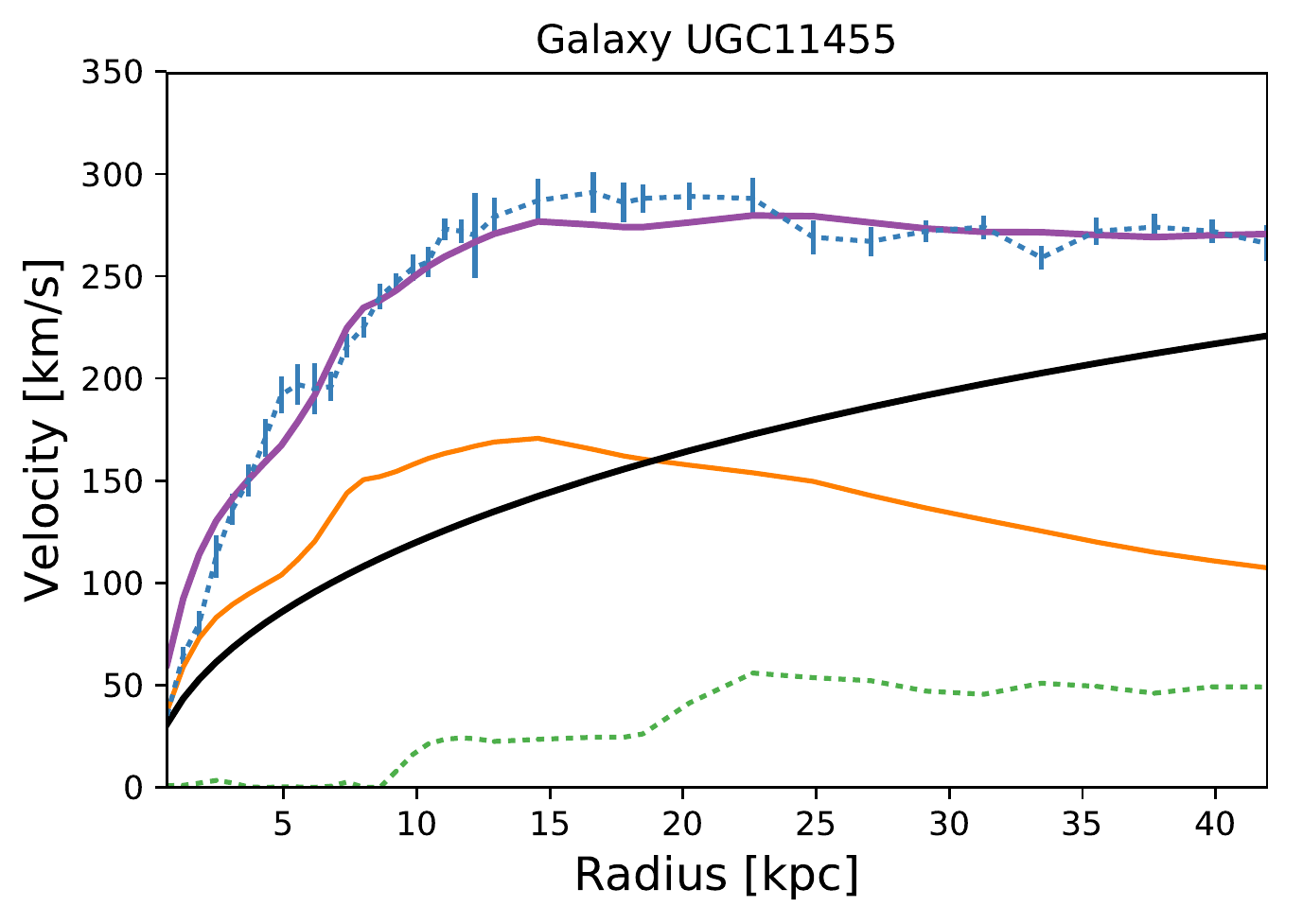} %&   
  \includegraphics[width=0.45\textwidth]{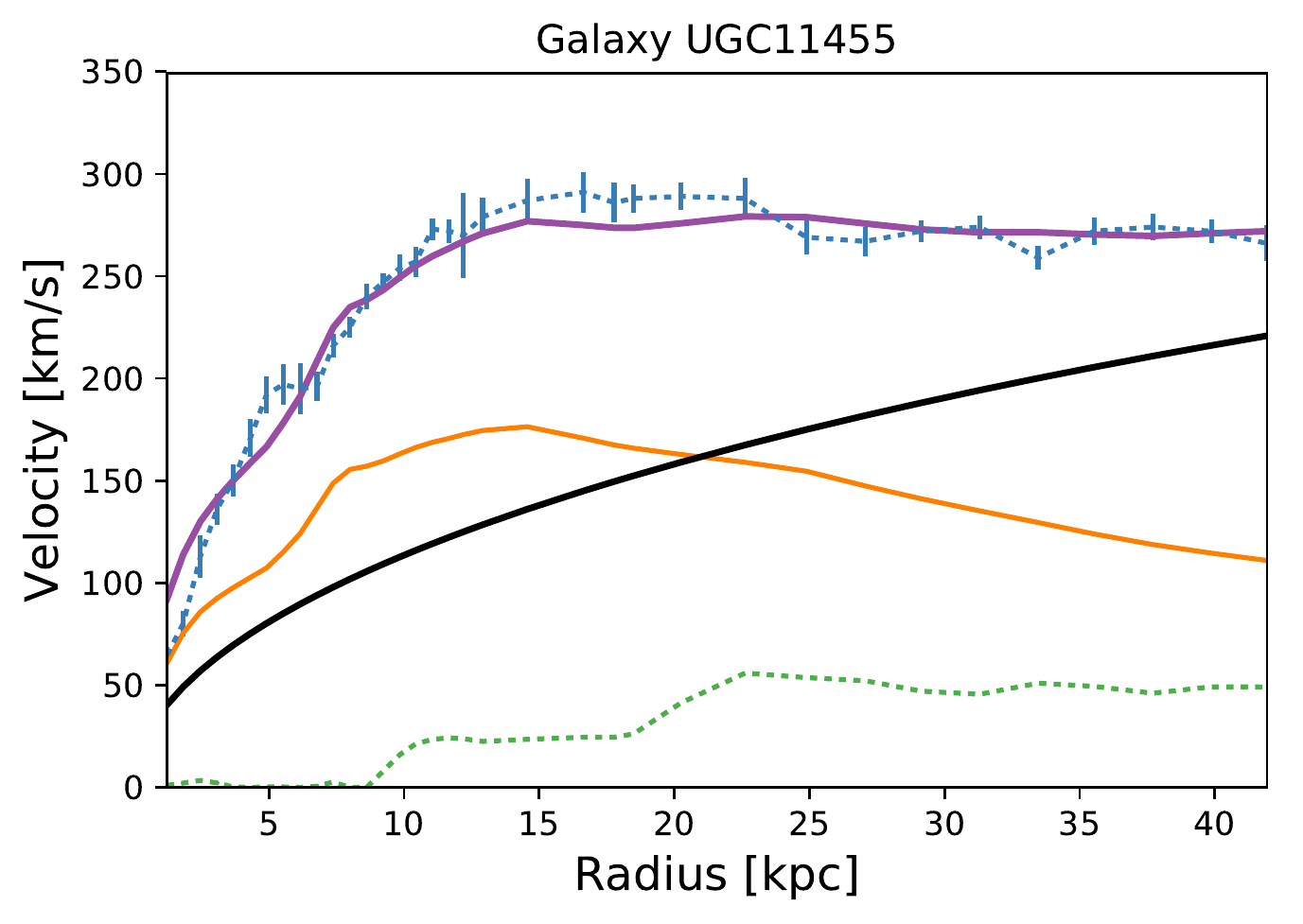}   \\
    \includegraphics[width=0.45\textwidth]{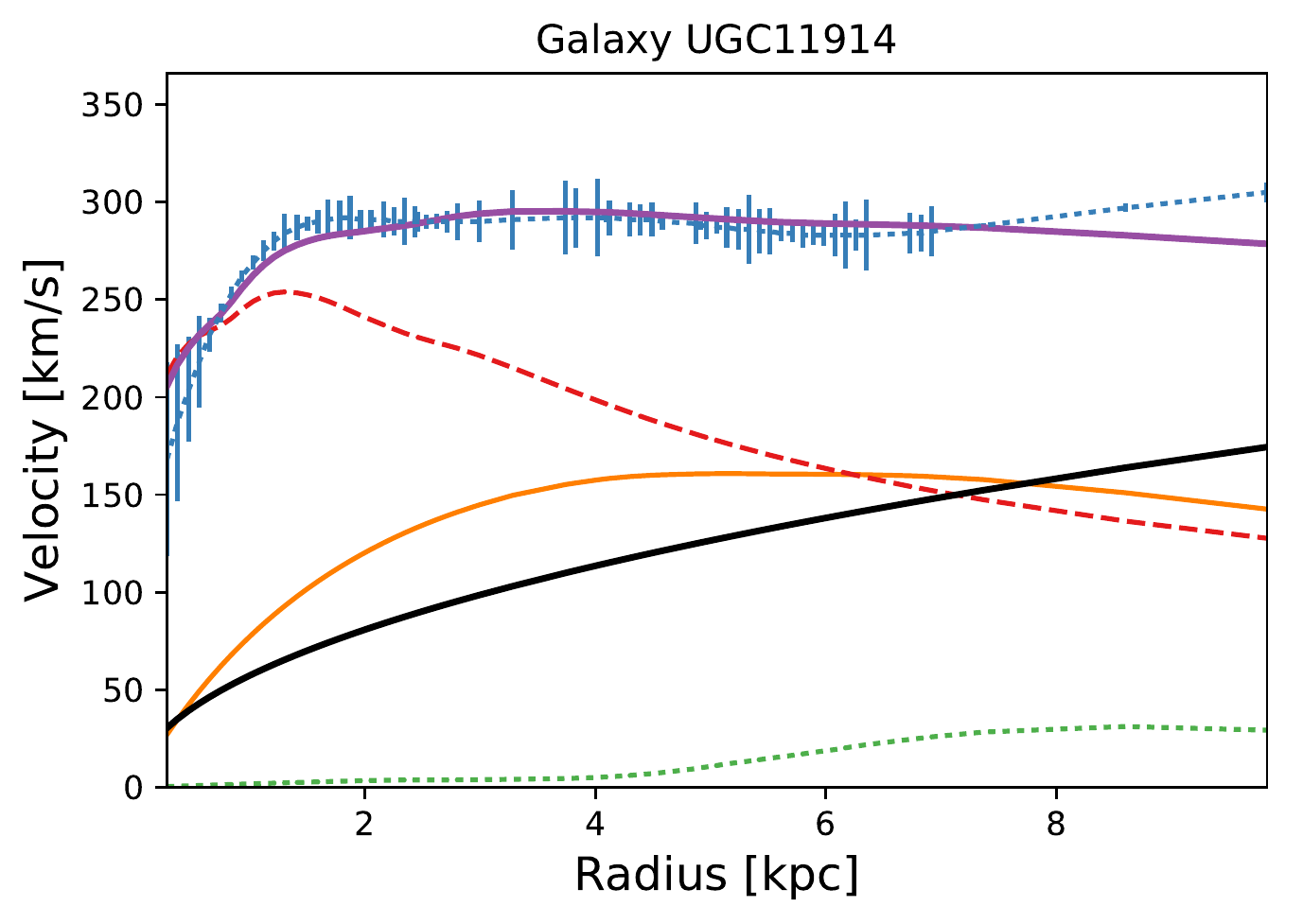} %&   
   \includegraphics[width=0.45\textwidth]{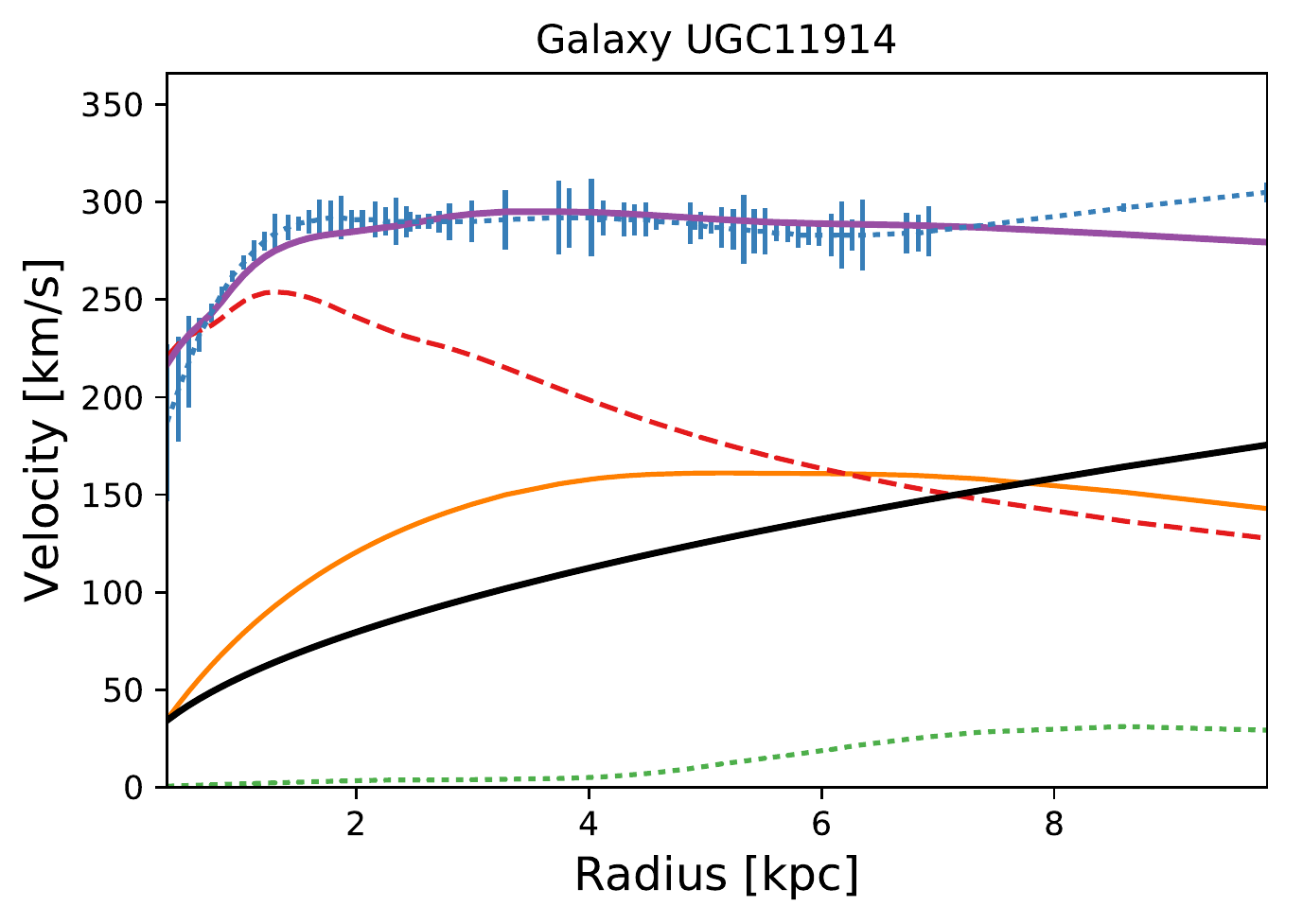}   %\\
\caption{Best--fit rotational curves. NFW profiles on the left. BEC on the right. Legends are like in Fig.\,\,\ref{fig:fit_example}. }
\end{figure}

\end{appendix}
%%%%%%%%%%%%%%%%%%%%%%%%%%%%%%%%%%%%%%%%%%%%
%%%%%%%%%%%%%%%%%%%%%%%%%%%%%%%%%%%%%%%%%%%%

%%%%%%%%%%%%%%%%%%%%%%%%%%%%%%%%%%%%%%%%%%%%
%%%%%%%%%%%%%%%%%%%%%%%%%%%%%%%%%%%%%%%%%%%%
\end{document}